\documentclass[twoside,10pt]{article}
\usepackage{amsmath}
\usepackage{amssymb}
\usepackage{subfigure}
\usepackage{graphicx}

\renewcommand{\theequation}{\thesection.\arabic{equation}}

\newtheorem{satz}{Theorem}[section]
\newtheorem{definition}[satz]{Definition}
\newtheorem{lemma}[satz]{Lemma}
\newtheorem{koro}[satz]{Corollary}
\newtheorem{bemerkung}[satz]{Remark}
\newtheorem{proposition}[satz]{Proposition}

\newtheorem{theorem}{Theorem}[section]

\newtheorem{remark}[theorem]{Remark}
\newtheorem{notation}{Notation}[section]
\newtheorem{assumption}{Assumption}
\renewcommand{\section}{\secdef\sct\sect}
\newcommand{\sct}[2][default]{\refstepcounter{section}
\vspace{0.5cm} \setcounter{equation}{0}
\centerline{ \scshape \arabic{section}.\ #1} \vspace{0.3cm}}
\newcommand{\sect}[1]{
\vspace{0.5cm} \centerline{\large\scshape #1} \vspace{0.3cm}}
\renewcommand{\subsection}{\secdef \subsct\sbsect}
\newcommand{\subsct}[2][default]{\refstepcounter{subsection}
\nopagebreak \vspace{0.5\baselineskip} {\flushleft\bf
\arabic{section}.\arabic{subsection}~\bf #1  } \nopagebreak}
\newcommand{\sbsect}[1]{\vspace{0.1cm}\noindent
{\bf #1}\vspace{0.1cm}}
\renewcommand{\subsubsection}{\secdef \subsubsect\sbsbsect}
\newcommand{\subsubsect}[2][default]{\refstepcounter{subsubsection} \nopagebreak
\vspace{0.1\baselineskip} \nopagebreak {\flushleft
\sffamily\slshape
\arabic{section}.\arabic{subsection}.\arabic{subsubsection}
\ \sffamily #1\/.}\ }
\newcommand{\sbsbsect}[1]{\vspace{0.1cm}\noindent
{\bf #1}\ }
\input{tcilatex}
\begin{document}

\title{Characterization of the Quasi-Stationary State of an Impurity Driven
by Monochromatic Light I - The Effective Theory}
\author{J.-B. Bru, W. de Siqueira Pedra, M. Westrich}
\maketitle

\begin{abstract}
We consider an impurity ($N$--level atom) driven by monochromatic light in a
host environment which is a fermionic thermal reservoir. The external light
source is a time--periodic perturbation of the atomic Hamiltonian
stimulating transitions between two atomic energy levels $E_{1}$ and $E_{N}$
and thus acts as an optical pump. The purpose of the present work is the
analysis of the effective atomic dynamics resulting from the full
microscopic time--evolution of the compound system. We prove, in particular,
that the atomic dynamics of population relaxes for large times to a
quasi-stationary state, that is, a stationary state up to small oscillations
driven by the external light source. This state turns out to be uniquely
determined by a balance condition. The latter is related to
{}\textquotedblleft generalized Einstein relations\textquotedblright {}\ of
spontaneous/stimulated emission/absorption rates, which are conceptually
similar to the phenomenological relations derived by Einstein in 1916. As an
application we show from quantum mechanical first principles how an
inversion of population of energy levels of an impurity in a crystal can
appear. Our results are based on the spectral analysis of the generator of
the evolution semigroup related to a non--autonomous Cauchy problem
effectively describing the atomic dynamics.
\end{abstract}

\section{Introduction\label{Section intro}}

In the present paper and in a companion one \cite{BruPedraWestrich2011b} we
study the dynamics of an impurity (an atom) in a crystal, or host
environment, interacting with an external monochromatic light source serving
as an optical pump. The host environment corresponds to free electrons in
thermal equilibrium within the crystal whereas the atom is described by a $N$%
--level system, the pure states of which are the unit vectors of the finite
dimensional Hilbert space $\mathbb{C}^{d}$ ($d\geq N\geq 1$). The external
monochromatic light source is a time--periodic classical field stimulating
transitions between the two energy levels $E_{1}$ and $E_{N}$. The
microscopic non--autonomous dynamics of the full system, that is, impurity,
host environment and external light source, is then described through a
two--parameter group of automorphisms generated by a time--dependent
symmetric derivation acting on the $C^{\ast }$--algebra of observables of
the compound system. This dynamics is generally non--unitary when restricted
to the atomic subalgebra. The restricted dynamics of the impurity shows a
dissipative behavior, provided the coupling to the host environment (thermal
reservoir) is effective. In the second part \cite{BruPedraWestrich2011b} of
the present work we prove that the dynamics of the impurity is properly
described -- up to small corrections for a moderate reservoir--atom--pump
interaction -- by some effective non--autonomous time--evolution involving
atomic degrees of freedom only. See Theorem \ref{ninja thm cool}.

The detailed study of this effective dynamics is the main goal of this
paper. More precisely, we are interested in an effective description of the
time--evolution of populations of the atomic energy levels, i.e., an
effective atomic \emph{block--diagonal} dynamics. To this end, we use
evolution group techniques transforming the non--autonomous Cauchy problem
into an autonomous one in a suitable Banach space. The same method is also
used in \cite{BruPedraWestrich2011b} to analyze the full microscopic
dynamics. In fact, many key arguments of the present analysis recur in \cite%
{BruPedraWestrich2011b}, albeit technically more involved. We remark that
evolution semigroups to study the long--time behaviour of quantum systems
have also been used by Abou--Salem and Fr\"{o}hlich \cite{Salem}.

Observe that similar models without external light source have been
extensively analyzed, see for instance \cite%
{AlickiLendi2007,AttalJoyePillet2006a,AttalJoyePillet2006b,AttalJoyePillet2006c}
and references therein. One of the most important questions considered
previously (see, e.g., \cite%
{AttalJoyePillet2006a,AttalJoyePillet2006b,AttalJoyePillet2006c,BachFroehlichSigal2000,JaksicPillet1996b}%
) are existence and asymptotic stability of stationary states, especially
asymptotic stability of thermal equilibrium states. The latter refers to the
so--called {}{}``return to equilibrium\textquotedblright{}\ which typically
occurs in models involving one thermal reservoir at fixed temperature weakly
coupled to an isolated atom. Note also that, as soon as there are several
thermal reservoirs at distinct temperatures, the system does not possess a
thermal equilibrium state, but rather a {}``non--equilibrium stationary state%
\textquotedblright{}\ (NESS) \cite{JaksicPillet2001b,MerkliMueckSigal2007}.
In our setting we certainly cannot not expect a thermal equilibrium state to
exist due to the optical pump.

However, it turns out that the atomic block--diagonal dynamics attains for
large times a quasi--stationary state $\varrho=\varrho|_{\mathfrak{D}}$,
that is, a stationary state up to small oscillations driven by the optical
pump. Here, $\varrho|_{\mathfrak{D}}$ is the restriction of the
quasi--stationary state $\varrho$ to the subalgebra $\mathfrak{D}$ of
population observables, that is, block--diagonal observables (i.e.,
observables describing quantum coherences between atomic energy levels are
excluded). We show that the structure of this quasi--stationary state is
uniquely characterized by a rather intuitive balance condition involving
population densities only (Theorem \ref{corollary uniqueness limit density
matrix}). This balance condition reads $(A+B)\varrho=0$. The operator $A$
does not depend on the pump and can be interpreted as a matrix of
spontaneous transition rates, whereas $B$ is the matrix related to the
stimulated transition rates, which are induced by the optical pump. We show
that $A$ and $B$ satisfy some relations which are conceptually related to
the well--known Einstein relations \cite{Einstein1916}.

Einstein considered an atom interacting with a black body (i.e., broad band)
radiation field and derived by phenomenological arguments that the
stimulated transition rates $B$ between atomic energy levels are
proportional to the radiation density as well as to the spontaneous
transition rates $A$. In contrast, we consider the situation of a narrow
band (i.e., monochromatic field) driving an atom in interaction with a
thermal reservoir. In this case, we find that the effective stimulated
processes are proportional to the intensity of the radiation in this model
and that, apart from this simple dependency on the pump intensity, the
natural decay rates of the atom (i.e., the dynamical properties of the atom
as the pump is turned off) uniquely determines the stimulated transition
rates $B$. The relevant dynamical parameters turn out to be the decoherence
times of the atom, which are purely quantum mechanical objects. In a recent
work \cite{BermanMerkliSigal2008}, Berman, Merkli and Sigal show for a
similar class of models (without pump term, however) that quantum coherences
of the atom decay exponentially fast for well--defined time scales $\mathbf{t%
}_{\mathrm{dec}}$ of decoherence. We show that the (pump independent)
parameters $\mathbf{t}_{\mathrm{dec}}$ and the intensity of the pump
uniquely define effective stimulated rates.

As a result of our analysis we find that the time--evolution of the
population density of the atomic energy levels, i.e., the block--diagonal
dynamics, differs from the phenomenological time--evolution usually used in
the physics literature (cf. \cite{AlickiLendi2007}). The block--diagonal
evolution is usually described by the so--called \emph{Pauli master equation}%
, which is a phenomenological first order autonomous linear differential
equation. Our analysis shows that the effective block--diagonal evolution
derived from the microscopic model is well described by a
integro--differential equation (Theorems \ref{lemmalongtime copy(4)} and \ref%
{The pre--master equation}). Nevertheless, the long--time behavior of the
solutions of our equation is in generic situations the same as the one
predicted by the Pauli master equation.

The model we study here is a basic model for pumping schemes of doped
crystals as explained in \cite{Levine1968}. One expects for large times a
steady emission of photons with fixed frequency $E_{N^{\prime\prime}}-E_{N^{%
\prime}}$, where $E_{1}<\cdots<E_{N^{\prime}}<\cdots<E_{N^{\prime\prime}}<%
\cdots<E_{N}$ are the (ordered) eigenvalues of the atomic Hamiltonian, i.e.,
the atomic energy levels. Such a steady emission should result from a
stronger occupation of the energy atomic level $E_{N^{\prime\prime}}$ as
compared to $E_{N^{\prime}}<E_{N^{\prime\prime}}$. This effect, called \emph{%
inversion of population}, is a central mechanism to obtain lasing materials
(cf. \cite{AspectFabreGrynberg2010}) and we demonstrate that the stationary
state $\varrho$ characterized by the balance condition can show such an
inversion of population.

In another recent work \cite{BachMerkliPedraSigal2011}, Bach, Merkli, Pedra
and Sigal investigate the possibility to control the decoherence time(s) $%
\mathbf{t}_{\mathrm{dec}}$ using a different external light source, which
does not act as an optical pump but imposes oscillations of the atomic
energy levels. In certain situations control of decoherence could thus
enhance the inversion of the population. Consequently, measuring the
threshold of the pump intensity needed for inversion of population could
yield a simple experimental test for control of decoherence.

To our knowledge there is only one framework in which some aspect of laser
phenomenology has been rigorously analyzed from first principles, namely for
some versions of the Dicke model \cite{Dic}, see \cite%
{HeppLieb1973,HeppLieb1973-2,HeppLieb1973-3,AlliSewell1995}. In \cite%
{HeppLieb1973,HeppLieb1973-3}, Hepp and Lieb study an interacting
conservative system consisting of a reservoir (a radiation field), a finite
number $\mathrm{N}$ of two--level atoms, and finitely many quantized
oscillators as {}\textquotedblleft radiation--modes\textquotedblright {}. In
\cite{HeppLieb1973-2,HeppLieb1973-3} the reservoir is absent in their study.
In \cite{HeppLieb1973,HeppLieb1973-3} it is shown that this system undergoes
a phase transition in the limit $\mathrm{N}\rightarrow \infty $ with the
appearance of a\ coherent radiation driven by the reservoir, whereas \cite%
{HeppLieb1973-2,HeppLieb1973-3} establish a transition from a normal to a
superradiant phase as $\mathrm{N}\rightarrow \infty $. More than 20 years
later, Alli and Sewell use in \cite{AlliSewell1995} a dissipative (i.e.,
non--conservative) version of the Dicke model to get similar results as $%
\mathrm{N}\rightarrow \infty $. For more details, we recommend for instance
\cite[Chap. 11]{Sewell}. In a more recent paper \cite{Bagarello2002} the
dissipative model of \cite{AlliSewell1995} is proven to be the Markovian
approximation of the model considered in \cite{HeppLieb1973}. On the other
hand, solid state lasers are usually constructed with weakly doped crystals
\cite{Levine1968}. Consequently, an infinite number of impurities is not
fundamental for the inversion of population. Moreover, the phenomenology of
lasers as described in physics textbooks is based on three-- or four--level
atoms \cite{Levine1968}, but Dicke--type models are based on two--level
atoms which cannot explain the inversion of population at finite number of
impurities. The assumption of impurities with three or four levels is also
very realistic from the experimental point of view, as most of lasing
materials used in the praxis are of this type \cite{Levine1968}. We focus on
understanding the pumping scheme in lasers which is (i) coherent with
physics textbooks and experimental facts and (ii) uses first principles of
quantum mechanics only. An important open problem remains however to find a
realistic description of a cavity. A recent work in this direction is due to
Bruneau and Pillet \cite{BruneauPillet2009}.

To resume, we focus here on the derivation and analysis of the structure of
the quasi--stationary state of the corresponding time--dependent master
equation for the impurity, whereas its link to the full microscopic dynamics
is established in \cite{BruPedraWestrich2011b}.

The paper is organized as follows. In Section \ref{Section def model} we
introduce the microscopic model. Then, in Section \ref{Section master
equation} we define the effective master equation and specify its relation
-- proven in \cite{BruPedraWestrich2011b} -- to the microscopic model. In
Section \ref{Section effective howland}, we introduce the evolution
semigroup of the non--autonomous effective master equation on a suitable
enlarged Hilbert space. This leads to a study of an autonomous problem. The
spectral analysis of the generator of such evolution semigroup performed in
Sections \ref{Sectino restriction finite}--\ref{Section The pre--master
equation} yields a pre--master equation and a balance condition, which
characterizes the quasi--stationary state for large times in terms of
generalized Einstein relations as explained in Section \ref%
{sec:The-Generalized-Einstein}. Numerical simulations show in Section \ref%
{section inv of pop} an inversion of population with a dynamics far from the
one described by the usual Pauli equations. Finally, Section \ref{sectino
appendix} is an appendix collecting, for the reader's convenience, some
heuristics and technical proofs as well as some notions and results on
completely positive (CP) semigroups used in our proofs.

\begin{notation}[Generic constants]
\label{remark constant}\mbox{
}\newline
To simplify our notation, we denote by $C,c,c^{\prime}$ any generic positive
and finite constant. Note that these constants do not need be the same from
one statement to another.
\end{notation}

\section{The host environment--impurity--light source microscopic model\label%
{Section def model}}

Laser devices are based on several physical processes. The first one
consists of pumping the electron densities of atomic energy levels to obtain
a so--called inversion of population. It is achieved in many cases by using
an external source of light. This procedure is called here \emph{optical}
pumping. Note however that pumping can also be implemented by means of
chemical reactions, electric currents, and other methods. A further step is
to use this inversion of population to obtain optical amplification through
stimulated emission of photons. These two processes implement a \emph{gain
medium}, which is then put into a resonant optical cavity to obtain a laser,
i.e., \emph{light amplification by stimulated emission of radiation}. Here,
we are only interested in the first process, namely the optical pumping
performed in the gain medium.

Since the invention of lasers in 1958 by the physicists Schawlow, Townes and
Bassov, a significant amount of gain mediums has been found.\ This includes
semiconductors, liquids with dyes, gases such as carbon dioxide, argon,
etc., and solids such as doped crystals and glasses. A large part of gain
mediums are based on a host material, usually a solid or a liquid, to which
impurities or dyes which can be pumped are provided. A typical example is
given by the well--known Nd:YAG laser which uses a crystal doped with
neodymium. Inversion of population is obtained in this family of laser
devices by optical pumping. Such a gain medium is typically what we have in
mind here.

An appropriately chosen impurity is crucial to get inversion of population
by pumping and hence a positive optical gain. In particular, the pumping
step should involve at least three or four atomic energy levels, as
explained in any physics textbook on lasers. For instance, in the case of a
monochromatic pump, neodymium impurities in Nd:YAG lasers are well described
as four level systems.

Surprisingly, also dissipative processes are important in order to get an
efficient optical pumping, in general. For instance, we could analyze the
model for impurities in crystals which will be defined below with $\lambda=0$
(meaning that the impurity does not interact with the surrounding
environment) and $\eta\neq0$ (meaning that the optical pump is turned on).
For such a choice of parameters, the dynamics can be explicitly computed,
showing that the atomic populations undergo Rabi oscillations and one \emph{%
cannot} obtain a (quasi--) \emph{stable} inversion of population and not
even a positive inversion in time average. In solid--state lasers, the
interaction with the host material, i.e., the crystal or the glass, provides
a dissipative component to the (effective) atomic dynamics. The same should
be found in a gain medium constituted of a liquid with a dye to be optically
pumped.

The dissipative mechanism can be quite different from one gain medium to
another. It can be due to many different types of interactions such as
interactions with phonons, electrons in a crystal, etc. Nevertheless, its
only important property for the pumping process is to damp Rabi oscillations
of pumped atoms and to modify, in interplay with the optical pump,
transition rates between atomic energy levels in order to get a (quasi--)
stable inversion of population.

Therefore, we can conclude three things about the optical pumping, i.e., the
stimulated process to get an inversion of population:

\begin{itemize}
\item[(a)] To explain optical pumping, it suffices to use a \emph{generic}
dissipative mechanism. To this end, we consider here an interaction between
a $N$--level atom (serving as a small system in a host environment) and a
fermionic thermal (macroscopic) reservoir whose mathematical setting is
standard, see, e.g., \cite{BratteliRobinson1987,BratteliRobinson1996}. This
choice excludes interactions between impurities and bosonic particles like
phonons. This could also be implemented, but for simplicity we refrain from
considering it as it qualitatively leads to similar results. Note, however,
that more physical models for host environments could be useful to get more
precise quantitative results.

\item[(b)] To obtain an efficient optical pumping, the strengths $\tilde{\eta%
},\tilde{\lambda}>0$ of respectively the optical pump and the interaction
with the host environment have to be of the same order, i.e., $\tilde{\eta}%
\sim\tilde{\lambda}$, see Section \ref{moderate section}. Indeed, if $\tilde{%
\eta}>>\tilde{\lambda}$, then Rabi oscillations with frequency of order $%
\tilde{\eta}$ are the dominating processes governing the dynamics of
populations. This oscillations are then generally damped in a time--scale of
order $\tilde{\lambda}^{-1}$ not depending much on the intensity of the
pump. In the opposite situation, i.e., when $\tilde{\eta}<<\tilde{\lambda}$,
no inversion of population can appear as the system relaxes in this case to
a state near the ground state (or a thermal equilibrium state) of the atom
and all the energy provided by the (weak) optical pump is lost into the
(very large) host environment.

\item[(c)] Finally, the inversion of population described in physics
textbooks makes senses if the {}\textquotedblleft $N$--level
atom\textquotedblright {}\ picture stays valid. In other words, the
interaction with the host environment must be a small perturbation of the
Hamiltonian representing the $N$--level atom. By (b), the optical pump
should also be seen as a perturbation of the $N$--level atom. In view of
this observation it is appropriate to use (Kato's) perturbation theory \cite%
{Kato} of (discrete) eigenvalues of the $N$--level atom.
\end{itemize}

The host environment discussed in (a), i.e., the fermionic thermal
reservoir, is described in Section \ref{sect crystal} in details. We then
define the impurity and the external monochromatic light source in Sections %
\ref{part.syst}--\ref{section pump nija}, respectively. The coupled full
system is set up in Sections \ref{Section dynamics}--\ref{moderate section}.

\subsection{The host environment as a fermionic thermal reservoir\label{sect
crystal}}

Let $\mathfrak{h}_{1}:=L^{2}(\mathbb{R}^{3},\mathbb{C})$ be the separable
Hilbert space representing the one--particle space of the host environment
(reservoir). The one--particle Hamiltonian $h_{1}$ is then defined by using
a dispersion relation. Indeed, we choose some measurable, rotationally
invariant function $\mathbf{E}:\mathbb{R}^{3}\rightarrow\mathbb{R}$, i.e., $%
\mathbf{E}(p)=\mathrm{E}(|p|)$, and define the multiplication operator $%
h_{1}=h_{1}(\mathbf{E})$ by $f(p)\mapsto\mathbf{E}(p)f(p)$ on $\mathfrak{h}%
_{1}$. Physically, $\mathbf{E}(p)$ represents the energy of one particle
with momentum $p$ within the reservoir.

In the case where the host material of the gain medium is a crystal, observe
that $L^{2}$ spaces on Brillouin zones as one--particle spaces are more
realistic than $L^{2}(\mathbb{R}^{3},\mathbb{C})$. As explained above (cf.
Observation (a)), the results obtained would qualitatively be the same. For
instance, the rotation symmetry assumed above for the dispersion relation $%
\mathbf{E}$ is not essential to the analysis performed below. However, to
satisfy the assumptions of Theorem \ref{ninja thm cool}, we require that $%
\mathbf{E}$ behaves (at least near $p=0$) like $|p|$ up to some
diffeomorphism. For instance, the dispersion relation $\mathbf{E}%
(p)\equiv|p|^{2}$ is allowed as $|p|$ and $|p|^{2}$ are clearly the same
function up a diffeomorphism of $\mathbb{R}^{3}\backslash\{0\}$.

The field algebra ${\mathcal{V}}_{\mathcal{R}}$ of the reservoir is the CAR$%
\ C^{\ast}$--algebra generated by the annihilation and creation operators $%
a(f),$ $a^{+}(f):=a(f)^{\ast}$, $f\in\mathfrak{h}_{1}$, acting on the
antisymmetric Fock space ${\mathcal{F}}_{-}(\mathfrak{h}_{1})$ and
fulfilling the canonical anti--commutation relations (CAR):
\begin{equation*}
a(f_{1})a^{+}(f_{2})+a^{+}(f_{2})a(f_{1})=\langle f_{1},f_{2}\rangle\ ,\quad
a(f_{1})a(f_{2})+a(f_{2})a(f_{1})=0
\end{equation*}
for any $f_{1},f_{2}\in\mathfrak{h}_{1}$.

The (unperturbed) dynamics of the reservoir is given by the family $\{\tau
_{t}^{{\mathcal{R}}}\}_{t\in \mathbb{R}}$ of Bogoliubov automorphisms on the
algebra ${\mathcal{V}}_{{\mathcal{R}}}$ uniquely defined by the condition:
\begin{equation}
\tau _{t}^{{\mathcal{R}}}\left( a(f)\right) =a(\mathrm{e}^{ith_{1}}f),\quad
f\in \mathfrak{h}_{1},\quad t\in \mathbb{R}\ .  \label{eq:Bogoliubov}
\end{equation}%
Physically, this means that the fermionic particles of the reservoir do not
interact with each other, i.e., they form an ideal Fermi gas. The group $%
\tau ^{{\mathcal{R}}}:=\{\tau _{t}^{{\mathcal{R}}}\}_{t\in \mathbb{R}}$ of
automorphisms is strongly continuous and hence, $({\mathcal{V}}_{{\mathcal{R}%
}},\tau ^{{\mathcal{R}}})$ is a $C^{\ast }$--dynamical system. We denote its
generator by $\delta _{{\mathcal{R}}}$.

Note that generators of $C^{\ast}$--dynamical systems are symmetric
derivations. This means that the domain $\mathrm{Dom}(\delta_{{\mathcal{R}}%
}) $ of the generator $\delta_{{\mathcal{R}}}$ is a dense sub--$\ast$%
--algebra of ${\mathcal{V}}_{{\mathcal{R}}}$ and, for all $A,B\in\mathrm{Dom}%
(\delta_{{\mathcal{R}}})$,
\begin{equation*}
\delta_{{\mathcal{R}}}(A)^{\ast}=\delta_{{\mathcal{R}}}(A^{\ast}),\quad%
\delta_{{\mathcal{R}}}(AB)=\delta_{{\mathcal{R}}}(A)B+A\delta_{{\mathcal{R}}%
}(B)\ .
\end{equation*}

Thermal equilibrium states of the reservoir are defined through the bounded
positive operators%
\begin{equation*}
d_{\mathcal{R}}:=\frac{1}{1+e^{\beta h_{1}}}
\end{equation*}%
acting on $\mathfrak{h}_{1}$ for all \emph{inverse temperatures} $\beta \in
(0,\infty )$. Indeed, the so--called \emph{symbol} $d_{\mathcal{R}}$
uniquely defines a (faithful) quasi--free state%
\begin{equation}
\omega _{\mathcal{R}}:=\omega _{d_{\mathcal{R}}}  \label{reservoir state}
\end{equation}%
on the fermion algebra ${\mathcal{V}}_{{\mathcal{R}}}$ by the conditions $%
\omega _{d_{\mathcal{R}}}(\mathbf{1}_{{\mathcal{R}}})=1$ and%
\begin{equation*}
\omega _{d_{\mathcal{R}}}\left( a^{+}(f_{1})\ldots
a^{+}(f_{m})a(g_{1})\ldots a(g_{n})\right) =\delta _{m,n}\,\mathrm{det}%
\left( [\langle f_{j},d_{\mathcal{R}}g_{k}\rangle ]_{j,k}\right)
\end{equation*}%
for all $\left\{ f_{j}\right\} _{j=1}^{n},\left\{ g_{j}\right\}
_{j=1}^{n}\subset \mathfrak{h}_{1}$. The positive, normalized linear
functional $\omega _{\mathcal{R}}$ is the unique $\beta $--KMS\ state of$\ $%
the $C^{\ast }$--dynamical system $({\mathcal{V}}_{{\mathcal{R}}},\tau ^{{%
\mathcal{R}}})$ and is called the thermal equilibrium state of the reservoir
at inverse temperature $\beta \in (0,\infty )$.

This definition of thermal states is rather abstract, but it can physically
be motivated as follows: Confining the particles within a box of side length
$L$ corresponds to the replacement of the momentum space $\mathbb{R}^{3}$ by
$\frac{2\pi }{L}\mathbb{Z}^{3}$, i.e., $L^{2}(\mathbb{R}^{3})$ by $\ell ^{2}(%
\frac{2\pi }{L}\mathbb{Z}^{3})$. In particular, the spectrum of $h_{1}$ (and
hence of its fermionic second quantization $\mathrm{d}\Gamma _{-}(h_{1})$)
becomes purely discrete. Additionally, the operators $e^{-\beta \mathrm{d}%
\Gamma _{-}(h_{1})}$ are in this case trace--class for all side lengths $L$.
Hence, we can define Gibbs states
\begin{equation*}
\mathfrak{g}_{\mathcal{R}}^{(L)}(\cdot ):=\frac{\mathrm{Tr}(\;\cdot
\;e^{-\beta \mathrm{d}\Gamma _{-}(h_{1})})}{\mathrm{Tr}(e^{-\beta \mathrm{d}%
\Gamma _{-}(h_{1})})}\ ,
\end{equation*}%
which have the thermal equilibrium state $\omega _{\mathcal{R}}$ as unique
weak--$\ast $ limit\footnote{%
This refers to the weak$^{\ast }$--topology on the locally convex real space
${\mathcal{V}}_{\mathcal{R}}^{\ast }$, the dual space of the separable
Banach space ${\mathcal{V}}_{\mathcal{R}}$.} for $L\rightarrow \infty $.
This follows, for instance, from the results of \cite[Chapters 5.2 and 5.3]%
{BratteliRobinson1996} on KMS states.

\subsection{The impurity as a $N$--level atom\label{part.syst}}

The impurity (atom) is modeled by a finite quantum system, i.e., its
observables are the self--adjoint elements of the finite dimensional $%
C^{\ast}$--algebra $\mathcal{B}(\mathbb{C}^{d})$ of all linear operators on $%
\mathbb{C}^{d}$ for $d\in\mathbb{N}$.

In the sequel it is convenient to define left and right multiplication
operators on $\mathcal{B}(\mathbb{C}^{d})$: For any $A\in\mathcal{B}(\mathbb{%
C}^{d})$ we define the linear operators $\underrightarrow{A}$ and $%
\underleftarrow{A}$ acting on $\mathcal{B}(\mathbb{C}^{d})$ by
\begin{equation}
B\mapsto\underrightarrow{A}B:=AB\quad\text{and}\quad B\mapsto\underleftarrow{%
A}B:=BA\ .  \label{left multi}
\end{equation}

The Hamiltonian of the atom is an arbitrary observable $H_{\mathrm{at}}=H_{%
\mathrm{at}}^{\ast}\in\mathcal{B}(\mathbb{C}^{d})$ representing its total
energy. We denote its eigenvalues and corresponding eigenspaces by $E_{k}\in%
\mathbb{R}$ and ${\mathcal{H}}_{k}\subset\mathbb{C}^{d}$ for $%
k\in\{1,\ldots,N\}$ ($N\geq2$), respectively. $E_{k}$ is chosen such that $%
E_{j}<E_{k}$ whenever $j<k$. In other words, $E_{k}$ is the energy of the $k$%
th atomic level and vectors of ${\mathcal{H}}_{k}$ describe the sub--band
structure of the corresponding energy level. The dimension $n_{k}$ of the
eigenspace ${\mathcal{H}}_{k}$ is the degeneracy of the $k$th atomic level.

As usual, the Hamiltonian $H_{\mathrm{at}}$ defines a free atomic dynamics,
i.e., a continuous one--parameter group of automorphisms $\tau^{\mathrm{at}%
}:=\{\tau_{t}^{\mathrm{at}}\}_{t\in\mathbb{R}}$ of the $C^{\ast}$--algebra $%
\mathcal{B}(\mathbb{C}^{d})$ defined by
\begin{equation}
\tau_{t}^{\mathrm{at}}(A):=\mathrm{e}^{itH_{\mathrm{at}}}A\mathrm{e}^{-itH_{%
\mathrm{at}}}\ ,\quad A\in\mathcal{B}(\mathbb{C}^{d})\ ,
\label{atom atomorph}
\end{equation}
for all $t\in\mathbb{R}$.

Thermal equilibrium states of the free atom are Gibbs states $\mathfrak{g}_{%
\mathrm{at}}$ given by the density matrix
\begin{equation}
\rho_{\mathfrak{g}}:=\frac{e^{-\beta H_{\mathrm{at}}}}{\mathrm{Tr}_{\mathbb{C%
}^{d}}\left(e^{-\beta H_{\mathrm{at}}}\right)}  \label{Gibbs.init}
\end{equation}
for any inverse temperature $\beta\in(0,\infty)$.

In presence of the optical pump and the host environment (the thermal
reservoir), the state of the atom is generally far from the Gibbs state $%
\mathfrak{g}_{\mathrm{at}}$. We thus consider arbitrary atomic states $%
\omega_{\mathrm{at}}$. For any state $\omega_{\mathrm{at}}$ on $\mathcal{B}(%
\mathbb{C}^{d})$, there is a unique trace--one positive operator $\rho_{%
\mathrm{at}}$ on $\mathbb{C}^{d}$, the so--called density matrix of $\omega_{%
\mathrm{at}}$, such that%
\begin{equation*}
\omega_{\mathrm{at}}(A)=\mathrm{Tr}_{\mathbb{C}^{d}}\left(\rho_{\mathrm{at}%
}\ A\right)\ ,\qquad A\in\mathcal{B}(\mathbb{C}^{d})\ .
\end{equation*}

Note that any state $\omega _{\mathrm{at}}$ on $\mathcal{B}(\mathbb{C}^{d})$
can be represented as a vector state via its GNS representation $(\mathfrak{H%
}_{\mathrm{at}},\pi _{\mathrm{at}},\Omega _{\mathrm{at}})$, see, e.g., \cite[%
Theorem 2.3.16]{BratteliRobinson1987}. If $\omega _{\mathrm{at}}$ is
faithful then $(\mathfrak{H}_{\mathrm{at}},\pi _{\mathrm{at}},\Omega _{%
\mathrm{at}})$ is explicitly given as follows. The Hilbert space $\mathfrak{H%
}_{\mathrm{at}}$ corresponds to the linear space $\mathcal{B}(\mathbb{C}%
^{d}) $ endowed with the Hilbert--Schmidt scalar product
\begin{equation}
\left\langle A,B\right\rangle _{\mathrm{at}}:=\mathrm{Tr}_{\mathbb{C}%
^{d}}(A^{\ast }B)\ ,\qquad A,B\in \mathcal{B}(\mathbb{C}^{d})\ .
\label{trace h at}
\end{equation}%
The representation $\pi _{\mathrm{at}}$ is the left multiplication explained
above in (\ref{left multi}), i.e.,
\begin{equation*}
\pi _{\mathrm{at}}\left( A\right) =\underrightarrow{A}\ ,\qquad A\in
\mathcal{B}(\mathbb{C}^{d})\ .
\end{equation*}%
The cyclic vector of the GNS representation of $\omega _{\mathrm{at}}$ is
defined by using the density matrix $\rho _{\mathrm{at}}\in \mathcal{B}(%
\mathbb{C}^{d})$ of $\omega _{\mathrm{at}}$ as
\begin{equation}
\Omega _{\mathrm{at}}:=\rho _{\mathrm{at}}^{1/2}\in \mathfrak{H}_{\mathrm{at}%
}\ .  \label{defnition vaccum at}
\end{equation}%
Using the cyclicity of the trace we obtain that
\begin{equation*}
\omega _{\mathrm{at}}\left( A\right) =\langle \Omega _{\mathrm{at}},%
\underrightarrow{A}\Omega _{\mathrm{at}}\rangle _{\mathrm{at}}\ ,\qquad A\in
\mathcal{B}(\mathbb{C}^{d})\ .
\end{equation*}%
This GNS representation $(\mathfrak{H}_{\mathrm{at}},\pi _{\mathrm{at}%
},\Omega _{\mathrm{at}})$ is known in the literature as the \emph{standard
representation} of the state $\omega _{\mathrm{at}}$. See \cite[Section 5.4]%
{DerezinskiFruboes2006}.

The dynamics given by the continuous one--parameter group $\tau^{\mathrm{at}%
} $ of automorphisms of the $C^{\ast}$--algebra $\mathcal{B}(\mathbb{C}^{d})$
defined by (\ref{atom atomorph}) can be represented in the Schr{\"{o}}dinger
picture of Quantum Mechanics through the so--called (standard) \emph{%
Liouvillean} operator%
\begin{equation}
L_{\mathrm{at}}:=\underrightarrow{H_{\mathrm{at}}}-\,\underleftarrow{H_{%
\mathrm{at}}}=[H_{\mathrm{at}},\ \cdot\ ]=L_{\mathrm{at}}^{\ast}
\label{liouvillean atom}
\end{equation}
acting on the Hilbert space $\mathfrak{H}_{\mathrm{at}}$. Indeed, it is easy
to check that:

\begin{lemma}[Schr{\"{o}}dinger picture of $\protect\tau^{\mathrm{at}}$]
\label{liouvillean atom lemma}\mbox{ }\newline
For all $t\in\mathbb{R}$,
\begin{equation*}
\omega_{\mathrm{at}}\left(\tau_{t}^{\mathrm{at}}(A)\right)=\langle\Omega_{%
\mathrm{at}}\left(t\right),\,\pi_{\mathrm{at}}(A)\,\Omega_{\mathrm{at}%
}\left(t\right)\rangle_{\mathrm{at}}\ ,\qquad A\in\mathcal{B}(\mathbb{C}%
^{d})\ ,
\end{equation*}
where $\Omega_{\mathrm{at}}\left(t\right):=e^{-itL_{\mathrm{at}}}\Omega_{%
\mathrm{at}}$.
\end{lemma}

Finally, for $k\in\left\{ 1,\ldots,N\right\} $, note that the population of
the $k$th atomic level in the state $\omega_{\mathrm{at}}$ is defined by the
expectation
\begin{equation}
p_{k}(\rho_{\mathrm{at}}):=\omega_{\mathrm{at}}\left(\mathbf{1}\left[H_{%
\mathrm{at}}=E_{k}\right]\right)=\mathrm{Tr}_{\mathbb{C}^{d}}\left(\mathbf{1}%
\left[H_{\mathrm{at}}=E_{k}\right]\rho_{\mathrm{at}}\right)\geq0\ ,
\label{population}
\end{equation}
where $\mathbf{1}\left[H_{\mathrm{at}}=E_{k}\right]\in\mathcal{B}(\mathbb{C}%
^{d})$ is the orthogonal projection onto the eigenspace ${\mathcal{H}}_{k}$.
If $\omega_{\mathrm{at}}=\mathfrak{g}_{\mathrm{at}}$ is the Gibbs state of
the atom then, for any inverse temperature $\beta\in\left(0,\infty\right)$,
\begin{equation*}
\forall j,k\in\left\{ 1,\ldots,N\right\} ,\ j<k:\qquad p_{j}(\rho_{\mathfrak{%
g}})>p_{k}(\rho_{\mathfrak{g}})\ .
\end{equation*}
In contrast, we say that a state $\omega$ or a density matrix $\rho$ shows
\emph{inversion of population} if there are $j,k\in\left\{
1,\ldots,N\right\} $\ such that $j<k$, i.e., $E_{j}<E_{k}$, and $%
p_{j}(\rho)<p_{k}(\rho)$. In other words, inversion of population requires a
higher energy level more populated than a lower one. Of course, this
phenomenon can only appear in a state out of equilibrium and one usually
uses external light sources to artificially pump electrons from a low energy
level of the atom to a higher one.

\subsection{The external monochromatic light source as a classical optical
pump\label{section pump nija}}

The optical pump, i.e., the monochromatic photon field interacting with the
atom, is described by the following time--periodic perturbation of the
atomic Hamiltonian $H_{\mathrm{at}}$:
\begin{equation}
\eta\cos(\varpi t)H_{\mathrm{p}}\ ,\qquad\varpi:=E_{N}-E_{1}>0\ ,\qquad t\in%
\mathbb{R}\ .  \label{pump ninja1}
\end{equation}
Recall that $E_{1}<\cdots<E_{N}$ denote the $N$ eigenvalues of $H_{\mathrm{at%
}}=H_{\mathrm{at}}^{\ast}\in\mathcal{B}(\mathbb{C}^{d})$. Here,
\begin{equation}
H_{\mathrm{p}}:=h_{\mathrm{p}}+h_{\mathrm{p}}^{\ast}\in\mathcal{B}(\mathbb{C}%
^{d})  \label{pump ninja1bis}
\end{equation}
for some $h_{\mathrm{p}}\in\mathcal{B}(\mathbb{C}^{d})$ satisfying
\begin{eqnarray}
\ker\left(h_{\mathrm{p}}\right)^{\perp} & \subseteq & {\mathcal{H}}_{1}:=%
\mathrm{ran}\left(\mathbf{1}\left[H_{\mathrm{at}}=E_{1}\right]\right)\ ,
\label{pump ninja1bisbis} \\
\mathrm{ran}\left(h_{\mathrm{p}}\right) & \subseteq & {\mathcal{H}}_{N}:=%
\mathrm{ran}\left(\mathbf{1}\left[H_{\mathrm{at}}=E_{N}\right]\right)\ .
\end{eqnarray}
In other words, the\ optical pump produces only transitions between the
lowest and the highest atomic levels $1$ and $N$, as described in standard
textbooks on the physics of lasers.

\noindent From the physical point of view, the time--dependent optical pump
may be regarded as a partial classical limit of a closed (autonomous)
physical system involving a quantized pump modeled by a quantum harmonic
oscillator. The corresponding initial state for this quantized pump should
be chosen as being a coherent state. See, e.g., \cite{Hepp1974,Westrich2008}.

\begin{remark}[Non monochromatic light sources as classical optical pumps]
\mbox{ }\newline
Results of this paper can easily be extended to non--monochromatic light
sources as classical optical pumps. This case corresponds here to replace
the cosine in (\ref{pump ninja1}) by some time--periodic and continuous
function. However, in order to keep technical aspects as simple as possible,
we refrain from considering this more general case.
\end{remark}

\subsection{The uncoupled reservoir--atom system\label{Section dynamics}}

Define the $C^{\ast }$--algebra ${\mathcal{V}}:=\mathcal{B}(\mathbb{C}%
^{d})\otimes {\mathcal{V}}_{\mathcal{R}}$. As both $C^{\ast }$--algebras $%
\mathcal{B}(\mathbb{C}^{d})$ and ${\mathcal{V}}_{\mathcal{R}}$ are already
realized as algebras of bounded operators on Hilbert spaces and since $%
\mathcal{B}(\mathbb{C}^{d})$ is finite dimensional, we do not have to
specify the meaning of the tensor product. Observables of the
reservoir--atom system are self--adjoint elements of ${\mathcal{V}}$. Its
free dynamics is described by the strongly continuous one--parameter group $%
\tau :=\{\tau _{t}\}_{t\in \mathbb{R}}$ of automorphisms of ${\mathcal{V}}$
defined by
\begin{equation}
\tau _{t}:=\tau _{t}^{\mathrm{at}}\otimes \tau _{t}^{{\mathcal{R}}}\ ,\quad
t\in \mathbb{R}\ .  \label{free dynamics}
\end{equation}%
This tensor product is well--defined and unique because the atomic algebra $%
\mathcal{B}(\mathbb{C}^{d})$ is finite dimensional. The generator of the
free dynamics defined by $\tau $ is a symmetric derivation, denoted by $%
\delta $, which acts on a dense sub--$\ast $--algebra $\mathrm{Dom}(\delta )$
of ${\mathcal{V}}$.

Let $\omega_{\mathrm{at}}$ be any initial (not necessarily Gibbs) state of
the atom and define the initial state of the atom--reservoir system by%
\begin{equation}
\omega_{0}:=\omega_{\mathrm{at}}\otimes\omega_{{\mathcal{R}}}\ .
\label{initial state}
\end{equation}
Again, the latter is well--defined and unique, by finite dimensionality of $%
\mathcal{B}(\mathbb{C}^{d})$. If $\omega^{\mathrm{at}}=\mathfrak{g}_{\mathrm{%
at}}$ is the Gibbs state then $\omega_{0}$ is clearly a $(\beta,\tau)$--KMS
state. Observe also that $\mathfrak{g}_{\mathrm{at}}$ is a faithful state
and we assume without loss of generality that $\omega^{\mathrm{at}}$ is also
a faithful state. Indeed, the set of faithful states is dense in the set of
all states of the atom. Since the quasi--free state $\omega_{{\mathcal{R}}}$
of the reservoir is also faithful, this property carries over to the initial
state $\omega_{0}$ of the composite system.

\begin{remark}[Coupled initial state of the atom--reservoir system]
\label{remark coupled initial state}\mbox{ }\newline
Considering impurities interacting with the host environment long before the
pump is turned on, the initial state of the atom--reservoir system should,
in principle, not be a product state as in (\ref{initial state}). Instead,
it should be a thermal equilibrium state of the coupled atom--reservoir
system. However, in rather generic situations it can be shown that at small
atom--reservoir couplings this thermal state is near the product state $%
\omega _{0}$ with $\omega ^{\mathrm{at}}=\mathfrak{g}_{\mathrm{at}}$ and the
results would be the same up to sub--leading corrections. As the KMS states
of the model considered here are unique (KMS states of bounded perturbations
of a free fermion gas are unique), the latter follows from standard results
on the stablity of KMS\ states, see for instance \cite[Section 5.4.1]%
{BratteliRobinson1996}. Indeed, we can even treat, by the same methods, any
initial state of the composite system as soon as its relative entropy with
respect to the product state (\ref{initial state}) is finite. This feature
is verified for the equilibrium thermal state of the composite system at
weak coupling. In order to keep technical aspects as simple as possible, we
will not consider this case.
\end{remark}

\subsection{The atom--reservoir interaction\label{section atom--reservoir
interaction}}

The interaction between the atom and the fermionic thermal reservoir
involves the so--called \emph{fermionic field operators} defined, for all $%
f\in\mathfrak{h}_{1}$, by%
\begin{equation*}
\Phi(f):=\frac{1}{\sqrt{2}}(a^{+}(f)+a(f))=\Phi(f)^{\ast}\in\mathcal{B}({%
\mathcal{F}}_{-}(\mathfrak{h}_{1}))\ .
\end{equation*}
Choose now a finite collection $\left\{ Q_{_{\ell}}\right\}
_{\ell=1}^{m}\subset\mathcal{B}(\mathbb{C}^{d})$ of self--adjoint operators
and an orthonormal (finite) system $\left\{ f_{\ell}\right\}
_{\ell=1}^{m}\subset\mathfrak{h}_{1}$. Then, the atom--reservoir interaction
is implemented by the bounded symmetric derivation
\begin{equation*}
\delta_{\mathrm{at},\mathcal{R}}:=i\sum_{\ell=1}^{m}\left[%
Q_{_{\ell}}\otimes\Phi(f_{\ell}),\;\cdot\;\right]\ .
\end{equation*}

Note that the orthonormality of the family $\left\{ f_{\ell}\right\}
_{\ell=1}^{m}$ does not inflict loss of generality. Indeed, for an arbitrary
finite set $\{\tilde{Q}_{_{\ell}}\}_{\ell=1}^{\tilde{m}}\subset\mathcal{B}(%
\mathbb{C}^{d})$ of self--adjoint operators and (possibly not orthonormal)
family $\{\tilde{f}_{\ell}\}_{\ell=1}^{\tilde{m}}\subset\mathfrak{h}_{1}$,
there are $m\in\mathbb{N}$, a finite collection $\left\{ Q_{_{\ell}}\right\}
_{\ell=1}^{m}\subset\mathcal{B}(\mathbb{C}^{d})$ of self--adjoint operators
and an orthonormal system $\left\{ f_{\ell}\right\} _{\ell=1}^{m}\subset%
\mathfrak{h}_{1}$ such that%
\begin{equation*}
\sum_{\ell=1}^{\tilde{m}}\tilde{Q}_{_{\ell}}\otimes\Phi(\tilde{f}%
_{\ell})=\sum_{\ell=1}^{m}Q_{_{\ell}}\otimes\Phi(f_{\ell})\ .
\end{equation*}

As we seek to maintain mathematical rigor while keeping technical aspects as
simple as possible, we assume some technically useful conditions on the
family $\left\{ f_{\ell }\right\} _{\ell =1}^{m}\subset \mathfrak{h}_{1}$
guaranteeing the assumptions of Theorem \ref{ninja thm cool} to be
satisfied. Note that these conditions will by no means restrict the range of
our analysis to rather physically meaningless submodels. First, $\left\{
f_{\ell }\right\} _{\ell =1}^{m}\subset \mathfrak{h}_{1}$ is taken as a
family of rotationally invariant functions, i.e., $f_{\ell }(p)=\mathrm{f}%
_{\ell }(|p|)$ for all $p\in \mathbb{R}^{3}$ and $\ell \in \{1,\ldots ,m\}$
with $\mathrm{f}_{\ell }:\mathbb{R}_{0}^{+}\rightarrow \mathbb{C}$. The
rotational invariance of $f_{\ell }$ for all $\ell \in \{1,\ldots ,m\}$ is
only assumed for technical simplicity as more general choices of such
functions would yield similar results, at least qualitatively. The second,
rather technical assumption is an analyticity condition which is only
required to prove Theorem \ref{ninja thm cool}. It is given here for
completeness, but it can clearly be omitted as no proof of this paper uses
it. This analyticity condition reads as follows: For all $\ell \in
\{1,\ldots ,m\}$, the complex valued functions $\mathrm{g}_{\ell }$ and $%
\mathrm{g}_{\ell }^{\#}$ respectively defined by%
\begin{equation}
\forall x\in \mathbb{R}:\quad \mathrm{g}_{\ell }(x):=|x|\left( 1+\mathrm{e}%
^{-\beta x}\right) ^{-1/2}\left\{
\begin{array}{lcr}
\mathrm{f}_{\ell }(x) & , & x\geq 0\ , \\
&  &  \\
\overline{\mathrm{f}_{\ell }(-x)} & , & x<0\ ,%
\end{array}%
\right.  \label{function gl}
\end{equation}%
and $\mathrm{g}_{\ell }^{\#}(x):=i\overline{\mathrm{g}_{\ell }(-x)}$ on $%
\mathbb{R}$ have an analytic continuation to the strip $\mathbb{R+}i(-C,C)$,
and satisfy%
\begin{equation*}
\underset{\vartheta \in (-C,C)}{\sup }\left\{ \int_{\mathbb{R}}(|\mathrm{g}%
_{\ell }(x+i\vartheta )|+|\mathrm{g}_{\ell }^{\#}(x+i\vartheta )|)^{2}%
\mathrm{d}x\right\} <\infty
\end{equation*}%
for all $\ell \in \{1,\ldots ,m\}$. For instance, to satisfy these
conditions one may choose for any $\ell \in \{1,\ldots ,m\}$, the function $%
\mathrm{f}_{\ell }$ as linear combinations of terms of the form $%
|x|^{2k-1}\exp (-cx^{2})$ with $k\in \mathbb{N}_{0}$.

\subsection{Dynamics of the coupled atom--reservoir--pump system\label%
{Section full dynamaics}}

The full dynamics of the system involves the classical pump described in
Section \ref{section pump nija}, which is implemented as a periodic
perturbation of the dynamics of the atom--reservoir system described in
Section \ref{Section dynamics}. Indeed, let
\begin{equation*}
\delta_{\mathrm{at},\mathrm{p}}:=i[H_{\mathrm{p}}\otimes\mathbf{1}_{{%
\mathcal{R}}},\;\cdot\;]
\end{equation*}
and $\eta\in\mathbb{R}$. The coupled atom--reservoir--pump dynamics is then
generated by the time--dependent symmetric derivation
\begin{equation}
\delta_{t}^{(\lambda,\eta)}:=\delta+\eta\cos(\varpi t)\delta_{\mathrm{at},%
\mathrm{p}}+\lambda\delta_{\mathrm{at},\mathcal{R}},\quad t\in\mathbb{R}\ .
\label{derivation full nija}
\end{equation}
Here, $\lambda,\eta\in\mathbb{R}$ are the atom--reservoir and atom--pump
coupling constants, respectively.

Observe that $\delta_{t}^{(\lambda,\eta)}$ acts on a dense sub--$\ast$%
--algebra $\mathrm{Dom}(\delta_{t})=\mathrm{Dom}(\delta)$ of ${\mathcal{V}}$
which does not depend on $t\in\mathbb{R}$. Indeed, $\delta_{\mathrm{at},%
\mathcal{R}}$ and $\delta_{\mathrm{at},\mathrm{p}}$ are bounded symmetric
derivations and $\delta_{t}^{(\lambda,\eta)}$ is the generator of a strongly
continuous one--parameter group of automorphisms of $\mathcal{V}$. As the
map
\begin{equation*}
t\mapsto\delta_{t}^{(\lambda,\eta)}-\delta_{0}^{(\lambda,\eta)}
\end{equation*}
is norm--continuous, $\delta_{t}^{(\lambda,\eta)}$ generates a strongly
continuous two--parameter family $\{\tau_{t,s}^{(\lambda,\eta)}\}_{t\geq s}$
of automorphisms of $\mathcal{V}$ corresponding to the non--autonomous
dynamics of the (coupled) atom--reservoir--pump system. The operator $%
\tau_{t,s}^{(\lambda,\eta)}$ can even be explicitly constructed as a Dyson
series because the operator $\eta\cos(\varpi t)\delta_{\mathrm{at},\mathrm{p}%
}$ is bounded and $\cos$ is a smooth function.

The time--evolution of the state of the full system is then given by
\begin{equation*}
\omega_{t}:=\omega_{0}\circ\tau_{t,0}^{(\lambda,\eta)}=\left(\omega_{\mathrm{%
at}}\otimes\omega_{{\mathcal{R}}}\right)\circ\tau_{t,0}^{(\lambda,\eta)},%
\quad t\in\mathbb{R}_{0}^{+}\ .
\end{equation*}
The reduction of this state onto the atomic degrees of freedom yields a
time--dependent atomic state defined by%
\begin{equation}
\omega_{\mathrm{at}}\left(t\right)\left(A\right):=\omega_{t}(A\otimes\mathbf{%
1}_{{\mathcal{R}}}),\qquad A\in\mathcal{B}(\mathbb{C}^{d})\ ,  \label{ninja1}
\end{equation}
for all $t\in\mathbb{R}_{0}^{+}$.

\subsection{Moderate optical pump and atom--reservoir interaction\label%
{moderate section}}

As explained at the beginning of Section \ref{Section def model} (cf.
Observation (c)), we are interested in the regime where $|\lambda|,|\eta|<<1$%
. In other words, we take the atom--reservoir and atom--pump interactions as
being small, but non--vanishing, perturbations of the free dynamics
generated by the symmetric derivation $\delta$. Moreover, we assume that the
pump is moderate with respect to the atom--reservoir interaction in the
following sense:

\begin{assumption}[Moderate optical pump]
\label{assumption important}\mbox{
}\newline
For any $\lambda\in\mathbb{R}$ and some fixed constant $C\in(0,\infty)$: $%
|\eta|\leq C\lambda^{2}$.
\end{assumption}

Actually, in all our proofs, it would suffice to impose $|\eta |\leq
C|\lambda |$ for some sufficiently small constant $C\in (0,\infty )$.
However, as it will be shown below (cf. Remark \ref{remark Moderate optical
pump}), the contribution of the pump to the final atomic state is of order $%
\eta ^{2}/\lambda ^{4}$ whereas the contribution of the interaction with the
reservoir is of order one (in the parameter $\eta ^{2}/\lambda ^{4}$). Thus,
imposing $|\eta |\sim \lambda ^{2}$ means physically that both the pump and
the reservoir contribute in an essential way to the final state of the atom.
The (relative) strengths $\tilde{\eta},\tilde{\lambda}>0$ of respectively
the optical pump and the interaction with the host environment turn out to
be equal to $\tilde{\eta}:=\eta ^{2}/\lambda ^{2}$ and $\tilde{\lambda}%
:=\lambda ^{2}$. In particular, Assumption \ref{assumption important} means
that $\tilde{\eta}\sim \tilde{\lambda}$. See Observation (b) above.
Consequently, we say in this context that the pump is weak whenever $|\eta
|<<\lambda ^{2}$, i.e., $\tilde{\eta}<<\tilde{\lambda}$.

In the opposite situation when for small $(\eta,\lambda)\in\mathbb{R}^{2}$
one has $|\eta|>>\lambda^{2}$, i.e., $\tilde{\eta}>>\tilde{\lambda}$, Rabi
oscillations are the dominant processes driving the populations of the
atomic energy levels. Indeed, a similar version of Corollary \ref%
{lemmalongtime copy(3)} is still valid in the \emph{strong pump regime} $%
|\eta|>>\lambda^{2}$ if $|\eta|<<|\lambda|$. Using this result one can show
that in general Rabi oscillations dominate the dynamics of populations at
time--scales comparable to $|\eta|^{-1}$ and that there is a global
relaxation of populations at time--scale $\lambda^{-2}$ not depending much
on the pump intensity. By this last property we could also call the regime $%
|\eta|>>\lambda^{2}$ \emph{saturated pumping}. However, we are rather
interested in the situation where pump and natural relaxations of the atom
compete in a non trivial way with each other and the evolution of the full
system is well described by some relaxing dynamics driving the atom to a
state with a persisting inversion of population. See again the discussions
at the beginning of Section \ref{Section def model}, in particular
Observation (b).

\section{The effective atomic master equation\label{Section master equation}}

The aim of this paper is to analyze the atomic dynamics resulting from the
restriction on $\mathcal{B}(\mathbb{C}^{d})$ of the full dynamics generated
by the symmetric derivation $\delta _{t}^{(\lambda ,\eta )}$. See Equation (%
\ref{derivation full nija}). This corresponds to the family $\{\omega _{%
\mathrm{at}}\left( t\right) \}_{t\in \mathbb{R}_{0}^{+}}$ of states defined
by (\ref{ninja1}) or, equivalently, to the family $\{\rho _{\mathrm{at}%
}\left( t\right) \}_{t\in \mathbb{R}_{0}^{+}}$ of density matrices. More
precisely, we are interested in the time--behavior of observables related to
atomic levels only, and not to correlations between different levels. This
amounts to study the orthogonal projection $P_{\mathfrak{D}}\left( \rho _{%
\mathrm{at}}\left( t\right) \right) $ of the atomic density matrix $\rho _{%
\mathrm{at}}\left( t\right) $ on the subspace
\begin{equation}
\mathfrak{D}\equiv \mathfrak{D}(H_{\mathrm{at}}):=\mathcal{B}({\mathcal{H}}_{%
\mathrm{1}})\oplus \cdots \oplus \mathcal{B}({\mathcal{H}}_{\mathrm{N}%
})\subset \mathfrak{H}_{\mathrm{at}}  \label{ninja0}
\end{equation}%
of block--diagonal matrices. In other words, we analyze the density matrix
\begin{equation}
P_{\mathfrak{D}}\left( \rho _{\mathrm{at}}\left( t\right) \right) =\overset{N%
}{\underset{k=1}{\sum }}\ \mathbf{1}\left[ H_{\mathrm{at}}=E_{k}\right] \
\rho _{\mathrm{at}}\left( t\right) \ \mathbf{1}\left[ H_{\mathrm{at}}=E_{k}%
\right]  \label{ninja0bis}
\end{equation}%
for any $t\in \mathbb{R}_{0}^{+}$.

As proven in \cite{BruPedraWestrich2011b}, the density matrix $\rho _{%
\mathrm{at}}\left( t\right) $ can be well approximated, uniformly in time,
on the subspace $\mathfrak{D}$ by the solution of an effective
non--autonomous initial value problem in $\mathcal{B}(\mathbb{C}^{d})$,
\emph{the effective atomic master (or Kossakowski--Lindblad) equation}
described in Section \ref{effective subsection}. The corresponding time
evolution is generated by the time--dependent Lindbladian $\mathfrak{L}%
_{t}^{\left( \lambda ,\eta \right) }$ defined in Section \ref{Section M eq
defs}. In Section \ref{section Spohn's assumption} we impose a condition,
introduced in \cite[Theorem 2]{Spohn1977} in the context of relaxing CP
semigroups, on the dissipative part of this generator which is a
non--commutative version of the irreducibility of classical Markov chains
and ensures the existence of the (quasi--) stationary state for populations
at large times.

\subsection{The atomic time--dependent Lindbladian\label{Section M eq defs}}

Lindblad operators (or Lindbladians) are standard objects used to describe
(generally dissipative) dynamics compatible with some phenomenologically
reasonable prescriptions like complete positivity. As explained in \cite%
{AlickiLendi2007}, the typical example of application of such operators is
related to the time--evolution of composite systems made of a small (open)
quantum system interacting with a macroscopic one (a reservoir). In this
context, the dynamics restricted to the small system is, in many situations,
well--described by a semigroup $\left\{ \mathrm{e}^{t\mathfrak{L}}\right\}
_{t\geq0}$ which is trace--preserving and completely positive, see Section %
\ref{section compl positive}. Generators $\mathfrak{L}$ of such completely
positive (CP) semigroups are called Lindblad operators or Lindbladians.

A first rigorous result in this direction is due to Davies \cite%
{Davies1974b,Davies1975,Davies1976a} in certain scaling limit, the
so--called weak coupling limit for similar interacting systems as ours with $%
\eta =0$, i.e., without the optical pump or any other time dependent term.
See also \cite{DerezinskiFruboes2006} and references therein. For more
details on CP semigroups, we also recommend Section \ref{section compl
positive}. Note however that, in contrast to Davies' approach, we never take
the limit $\lambda \rightarrow 0$. It suffices in our analysis to have a
sufficiently small coupling constant $\lambda \neq 0$. Observe also that we
have to control at the same time the (non--trivial) influence of two
microscopic couplings of different orders ($|\eta |\thicksim \lambda ^{2}$,
by the moderate pump assumption) and the precise meaning of a weak--coupling
limit is not clear from the beginning. The dynamic laws we deal with are
moreover non--autonomous.

The physical systems considered here yield a non--autonomous effective
time--evolution with a \emph{time--dependent},\ generally non
anti--self--adjoint generator $\mathfrak{L}_{t}^{\left( \lambda ,\eta
\right) }\in \mathcal{B}(\mathfrak{H}_{\mathrm{at}})$ \cite%
{BruPedraWestrich2011b} given, for any $t\in \mathbb{R}$, by:
\begin{equation}
\mathfrak{L}_{t}^{\left( \lambda ,\eta \right) }(\rho ):=\mathfrak{L}_{%
\mathrm{at}}(\rho )+\eta \cos (\varpi t)\mathfrak{L}_{\mathrm{p}}(\rho
)+\lambda ^{2}\mathfrak{L}_{\mathcal{R}}(\rho ),\qquad \rho \in \mathfrak{H}%
_{\mathrm{at}}\ .  \label{lindblad ninja1}
\end{equation}

The first term defining $\mathfrak{L}_{t}^{\left( \lambda ,\eta \right) }$
is the Lindbladian of the free atomic dynamics. It is the
anti--self--adjoint operator
\begin{equation}
\mathfrak{L}_{\mathrm{at}}(\rho ):=-iL_{\mathrm{at}}(\rho )=-i[H_{\mathrm{at}%
},\rho ]=-\mathfrak{L}_{\mathrm{at}}^{\ast }(\rho ),\qquad \rho \in
\mathfrak{H}_{\mathrm{at}}\ .  \label{Lat}
\end{equation}%
See Lemma \ref{liouvillean atom lemma}. The second term of (\ref{lindblad
ninja1}) encodes the influence of the optical pump. It is defined via the
Lindbladian%
\begin{equation}
\mathfrak{L}_{\mathrm{p}}(\rho ):=-i[H_{\mathrm{p}},\rho ]=-\mathfrak{L}_{%
\mathrm{p}}^{\ast }(\rho ),\qquad \rho \in \mathfrak{H}_{\mathrm{at}}\ .
\label{L pump}
\end{equation}%
The third term includes a dissipative part $\mathfrak{L}_{d}\in \mathcal{B}(%
\mathfrak{H}_{\mathrm{at}})$ corresponding to the non--unitary character of
the effective dynamics and so, $\mathfrak{L}_{\mathcal{R}}\in \mathcal{B}(%
\mathfrak{H}_{\mathrm{at}})$ is \emph{not} anti--self--adjoint. In fact, the
Lindbladian $\mathfrak{L}_{\mathcal{R}}$ is related to the second--order
perturbation term coming from the atom--reservoir (electron field--impurity)
interaction in a suitable representation and equals
\begin{equation}
\mathfrak{L}_{\mathcal{R}}(\rho ):=-i[H_{\mathrm{Lamb}},\rho ]+\mathfrak{L}%
_{d}(\rho ),\qquad \rho \in \mathfrak{H}_{\mathrm{at}}\ .  \label{L R}
\end{equation}

In order to define the so--called \emph{atomic Lamb shift} $H_{\mathrm{Lamb}%
} $, which encodes the shift of the atomic energy levels due to the presence
of the reservoir, and the \emph{effective atomic dissipation} $\mathfrak{L}%
_{d}$ some preliminary definitions are necessary: We denote the spectrum of
any operator $A$ by $\sigma \left( A\right) $, its positive part by $\sigma
^{+}\left( A\right) :=\sigma \left( A\right) \cap \mathbb{R}_{0}^{+}$, and
define the sets%
\begin{equation}
\mathfrak{t}_{\epsilon }:=\{(j,k):E_{j}-E_{k}=\epsilon \}\subset
\{1,2,\ldots ,N\}\times \{1,2,\ldots ,N\}  \label{t eps}
\end{equation}%
for each eigenvalue%
\begin{equation}
\epsilon \in \sigma (L_{\mathrm{at}})=\sigma ([H_{\mathrm{at}},\ \cdot \
])=\left\{ E_{j}-E_{k}:j,k\in \{1,2,\ldots ,N\}\right\} \ .
\label{definition spectre atomique}
\end{equation}%
Let $\{V_{j,k}^{(\ell )}\}_{j,k,\ell }\subset \mathcal{B}(\mathbb{C}^{d})$
be the family of operators defined by
\begin{equation}
V_{j,k}^{(\ell )}:=\mathbf{1}\left[ H_{\mathrm{at}}=E_{j}\right] \ Q_{\ell
}\ \mathbf{1}\left[ H_{\mathrm{at}}=E_{k}\right]  \label{definition V1}
\end{equation}%
for $j,k\in \{1,2,\ldots ,N\}$ and $\ell \in \{1,2,\ldots ,m\}$, and let $%
\{f_{\ell }^{(\beta )}\}_{\ell =1}^{m}$ be the family of functions $\mathbb{R%
}\rightarrow \mathbb{R}_{0}^{+}$ defined by%
\begin{equation}
f_{\ell }^{(\beta )}(x):=4\pi \frac{\left\vert x\text{ }\mathrm{f}_{\ell
}\left( \left\vert x\right\vert \right) \right\vert ^{2}}{1+\mathrm{e}%
^{-\beta x}}=4\pi \left\vert \mathrm{g}_{\ell }(x)\right\vert ^{2}
\label{definition V1bis}
\end{equation}%
at any fixed inverse temperature $\beta \in (0,\infty )$ of the fermionic
reservoir, see (\ref{function gl}).

\begin{bemerkung}[Self--adjointness of the space spanned by $%
\{V_{j,k}^{(\ell)}\}_{j,k,\ell} $]
\label{remark projection}\mbox{ }\newline
Since $Q_{\ell}=Q_{\ell}^{\ast}$, one has $(V_{j,k}^{(\ell)})^{%
\ast}=V_{k,j}^{(\ell)}$ for any $j,k\in\{1,2,\ldots,N\}$ and $%
\ell\in\{1,2,\ldots,m\}$, and obviously,
\begin{equation*}
\mathrm{span}\{V_{j,k}^{(\ell)}\}_{j,k,\ell}=\mathrm{span}%
\{(V_{j,k}^{(\ell)})^{\ast}\}_{j,k,\ell}\subset\mathfrak{H}_{\mathrm{at}}\ .
\end{equation*}
This fact is important when using Theorem \ref{Spohn}.
\end{bemerkung}

Then, the atomic Lamb shift $H_{\mathrm{Lamb}}\in\mathcal{B}(\mathbb{C}^{d})$
is the self--adjoint operator defined by
\begin{equation}
H_{\mathrm{Lamb}}:=-\frac{1}{2}\sum_{\epsilon\in\sigma([H_{\mathrm{at}%
},\cdot])\backslash\{0\}}\ \sum_{(j,k)\in\mathfrak{t}_{\epsilon}}\sum_{%
\ell=1}^{m}d_{j,k}^{(\ell)}V_{j,k}^{(\ell)\ast}V_{j,k}^{(\ell)}
\label{Atomic Lamb shift}
\end{equation}
with the real coefficients%
\begin{equation*}
d_{j,k}^{(\ell)}:=\mathcal{PP}\left[f_{\ell}^{(\beta)}\left(%
\cdot+(E_{k}-E_{j})\right)\right]
\end{equation*}
being the principal part $\mathcal{PP}[f]$ of functions $f\equiv
f_{\ell}^{(\beta)}\left(\cdot+(E_{k}-E_{j})\right)$.

Meanwhile, the non anti--self--adjoint operator $\mathfrak{L}_{d}\in\mathcal{%
B}(\mathfrak{H}_{\mathrm{at}})$ describing the effective atomic dissipation
is defined by%
\begin{equation}
\mathfrak{L}_{d}:=\frac{1}{2}\sum_{\epsilon\in\sigma([H_{\mathrm{at}%
},\cdot])}\ \sum_{(j,k)\in\mathfrak{t}_{\epsilon}}\sum_{%
\ell=1}^{m}c_{j,k}^{(\ell)}\mathfrak{L}_{j,k}^{\left(\ell\right)}\ ,
\label{Effective atomic dissipation}
\end{equation}
where $c_{j,k}^{(\ell)}:=\pi f_{\ell}^{(\beta)}(E_{k}-E_{j})$ and%
\begin{equation}
\mathfrak{L}_{j,k}^{\left(\ell\right)}\left(\rho\right):=2V_{j,k}^{(\ell)}%
\rho V_{j,k}^{(\ell)\ast}-V_{j,k}^{(\ell)\ast}V_{j,k}^{(\ell)}\rho-\rho
V_{j,k}^{(\ell)\ast}V_{j,k}^{(\ell)}\ ,\qquad\rho\in\mathfrak{H}_{\mathrm{at}%
}\ .  \label{eq:Effective atomic dissipation 2}
\end{equation}
The terms $V_{j,k}^{(\ell)}\rho V_{j,k}^{(\ell)\ast}$ in these definitions
correspond to transitions from the $k$th to the $j$th atomic levels, whereas
the other terms guarantee the Markov property of the dynamics, i.e., the
preservation of the trace of the density matrix.

Note that the functions $f_{\ell }^{(\beta )}$ satisfy the equality
\begin{equation*}
f_{\ell }^{(\beta )}(-x)=\mathrm{e}^{-\beta x}f_{\ell }^{(\beta )}(x),\qquad
x\in \mathbb{R}\ ,
\end{equation*}%
whereas $V_{j,k}^{(\ell )}=V_{k,j}^{(\ell )\ast }$ because $Q_{_{\ell }}$ is
self--adjoint, by assumption. Using these properties, the effective atomic
dissipation equals
\begin{equation}
\mathfrak{L}_{d}=\frac{1}{2}\sum_{\epsilon \in \sigma ^{+}\left( [H_{\mathrm{%
at}},\cdot ]\right) }\ \sum_{(j,k)\in \mathfrak{t}_{-\epsilon }}\sum_{\ell
=1}^{m}c_{j,k}^{(\ell )}\left( \mathfrak{L}_{j,k}^{\left( \ell \right)
}+(1-\delta _{\epsilon ,0})\mathrm{e}^{-\beta \epsilon }\mathfrak{L}%
_{k,j}^{\left( \ell \right) }\right) \ .
\label{Effective atomic dissipationbis}
\end{equation}%
Observe further that $[H_{\mathrm{Lamb}},\rho _{\mathfrak{g}}]=0$ and%
\begin{equation*}
\mathrm{e}^{itH_{\mathrm{at}}}\mathfrak{L}_{j,k}^{\left( \ell \right) }%
\mathrm{e}^{-itH_{\mathrm{at}}}=\mathrm{e}^{it\epsilon }\mathfrak{L}%
_{j,k}^{\left( \ell \right) }
\end{equation*}%
for all $(j,k)\in \mathfrak{t}_{\epsilon }$, $\epsilon \in \sigma ^{+}\left(
[H_{\mathrm{at}},\cdot ]\right) $. This is the standard form of a
Lindbladian fulfilling the so--called (quantum) \emph{detailed balance
condition in the sense of Alicki--Frigerio--Gorini--Kossakowski--Verri }with
respect to the atomic Gibbs state $\mathfrak{g}_{\mathrm{at}}$. See \cite%
{A,FGKV} which is reviewed in \cite[Section 4.5]{DerezinskiFruboes2006}. On
important consequence of this fact is that the atomic Gibbs state, for all $%
\lambda ,t$, satisfies
\begin{equation*}
\mathfrak{L}_{t}^{\left( \lambda ,0\right) }(\rho _{\mathfrak{g}})=0.
\end{equation*}%
See, for instance, \cite[Section III.4]{AlickiLendi2007}. Here, the
parameter $\beta $ of the density matrix $\rho _{\mathfrak{g}}$ (see (\ref%
{Gibbs.init})) is chosen to be the inverse temperature of the reservoir.

\begin{bemerkung}[Lindbladians as generators of Markov CP semigroups]
\label{markov CP semigroup}\mbox{ }\newline
The operators $H_{\mathrm{at}},H_{\mathrm{p}},H_{\mathrm{Lamb}}\in \mathcal{B%
}(\mathbb{C}^{d})$ are self--adjoint and $\{c_{j,k}^{(\ell )}\}_{j,k,\ell }$
are non--negative numbers. Thus, at any fixed time $t\in \mathbb{R}$ and for
all $(\lambda ,\eta )\in \mathbb{R}^{2}$, the Lindbladians $\mathfrak{L}%
_{t}^{\left( \lambda ,\eta \right) }$ and $\frac{\eta }{2}\mathfrak{L}_{%
\mathrm{p}}+\lambda ^{2}\mathfrak{L}_{\mathcal{R}}\ $are generators of
Markov CP semigroups, by Theorem \ref{Theorem generator CP}.
\end{bemerkung}

\subsection{Irreducibility of quantum Markov chains\label{section Spohn's
assumption}}

In principle, even after having extracted (by some averaging procedure, for
instance) the oscillations of frequency $\varpi$ caused by the presence of
the pump, the family $\{P_{\mathfrak{D}}\left(\rho_{\mathrm{at}%
}\left(t\right)\right)\}_{t\in\mathbb{R}_{0}^{+}}$ of density matrices could
have several accumulation points, limits depending on the initial state, or
even be oscillating (Rabi oscillations) as $t\rightarrow\infty$. We would
like to avoid this situation and spectral properties of the Lindbladian $%
\frac{\eta}{2}\mathfrak{L}_{\mathrm{p}}+\lambda^{2}\mathfrak{L}_{\mathcal{R}%
} $ turn out to be important in this sense, see Section \ref{Theorem blabla}%
. To this end, we require that $0$ is a non--degenerated eigenvalue of $%
\frac{\eta}{2}\mathfrak{L}_{\mathrm{p}}+\lambda^{2}\mathfrak{L}_{\mathcal{R}%
} $ with some non--trivial \emph{real} spectral gap, that is,
\begin{equation*}
\max\left\{ \mathrm{\mathop{\rm Re}}\left\{ w\right\} \,|\, w\in\sigma\left(%
\frac{\eta}{2}\mathfrak{L}_{\mathrm{p}}+\lambda^{2}\mathfrak{L}_{\mathcal{R}%
}\right)\backslash\{0\}\right\} \leq-\lambda^{2}C<0
\end{equation*}
with $C\in(0,\infty)$ being some fixed constant not depending on $\lambda$
and $\eta$.

This is useful (and very likely even essential) to prove Theorem \ref{ninja
thm cool} because it yields uniform bounds in time and allows the study of
the asymptotic dynamics of the atom. The following assumption on the
dissipative part $\mathfrak{L}_{d}$ suffices to ensure the spectral
properties mentioned above (cf. Lemma \ref{g(0,eta)}).

\begin{assumption}[Irreducibility of quantum Markov chains]
\label{assumption3}\mbox{
}\newline
The family $\{V_{j,k}^{(\ell)}\}_{j,k,\ell}\subset\mathcal{B}(\mathbb{C}%
^{d}) $ of operators defined by (\ref{definition V1}) satisfies
\begin{equation*}
\Big(\bigcup\limits _{\{(j,k,\ell)\ :\ c_{j,k}^{(\ell)}\neq0\}}\left\{
V_{j,k}^{(\ell)}\right\} \Big)^{\prime\prime}=\mathcal{B}(\mathbb{C}^{d})
\end{equation*}
with $M^{\prime\prime}$ being the bicommutant of $M\subset\mathcal{B}(%
\mathbb{C}^{d})$. Recall that $c_{j,k}^{(\ell)}:=\pi
f_{\ell}^{(\beta)}(E_{k}-E_{j})$, see also (\ref{definition V1bis}).
\end{assumption}

\noindent The assumption above highlights the role played by dissipative
effects of the fermionic reservoir on the atom in order to get an
appropriate asymptotic evolution of populations of atomic levels. Actually,
the existence and uniqueness of the final ($t\rightarrow\infty$) density
matrix projected on the subspace $\mathfrak{D}$ of block--diagonal matrices
follows from this hypothesis (cf. Theorem \ref{corollary uniqueness limit
density matrix}). See also Observation (a) at the beginning of Section \ref%
{Section def model}.

Assumption \ref{assumption3} is a non--commutative version of the
irreducibility of classical Markov chains. To illustrate this, we consider
the following example: Assume for simplicity that $m=1$ and the degeneracy $%
n_{k}$ of the $k$th atomic level equals $n_{k}=1$ for all $k\in \{1,\ldots
,N=d\}$. Let $\left\{ \varphi _{k}\right\} _{k=1}^{d}\subset \mathbb{C}^{d}$
be an orthonormal basis of eigenvectors of $H_{\mathrm{at}}$ with $H_{%
\mathrm{at}}\varphi _{k}=E_{k}\varphi _{k}$. If the self--adjoint operator $%
Q_{1}$ of the atom--reservoir interaction is defined by
\begin{equation*}
Q_{1}\varphi _{k}\equiv \underset{j=1}{\overset{d}{\sum }}\varphi
_{j},\qquad k\in \{1,\ldots ,d\}\ ,
\end{equation*}%
then the family $\{V_{j,k}^{(1)}\}_{j,k=1}^{d}$ satisfies $%
V_{j,n}^{(1)}V_{n,k}^{(1)}=V_{j,k}^{(1)}$ for all $j,k,n$ and forms an
orthonormal basis of $\mathfrak{H}_{\mathrm{at}}$. In the orthonormal basis $%
\left\{ \varphi _{k}\right\} _{k=1}^{d}$, $V_{j,k}^{(1)}$ is the elementary
matrix made of zeros except at the intersection of the $j$th row with the $k$%
th column where its matrix coefficient is $1$. We assume the \emph{%
irreducibility} of the family $\{c_{j,k}^{(1)}\}_{j,k=1}^{d}\subset \mathbb{R%
}_{0}^{+}$ of non--negative numbers in the sense that, for all $j\neq k$,
there is a finite sequence $(j_{1},k_{1}),\ldots ,(j_{n},k_{n})$ such that $%
c_{j_{1},k_{1}}^{(1)},\ldots ,c_{j_{n},k_{n}}^{(1)}\neq 0$, $j_{1}=j$, $%
k_{n}=k$, and $k_{l}=j_{l+1}$ for $l\in \{1,2,\ldots ,n-1\}$. Physically
speaking it means that any arbitrary pair of atomic levels is connected by
non--vanishing transitions. By using the commutator identity%
\begin{equation*}
\lbrack
A,V_{j,k}^{(1)}]=[A,V_{j,n}^{(1)}V_{n,k}^{(1)}]=V_{j,n}^{(1)}[A,V_{n,k}^{(1)}]+[A,V_{j,n}^{(1)}]V_{n,k}^{(1)}
\end{equation*}%
for all $j,n,k$ and the irreducibility of the family $\{c_{j,k}^{(1)}%
\}_{j,k=1}^{d}$ one can compute the commutant
\begin{equation*}
\Big(\bigcup\limits_{\{(j,k)\ :\ c_{j,k}^{(1)}\neq 0\}}\left\{
V_{j,k}^{(1)}\right\} \Big)^{\prime }=\mathbb{C}\cdot \mathbf{1}_{\mathbb{C}%
^{d}}\ ,
\end{equation*}%
from which Assumption \ref{assumption3} follows. This is in perfect analogy
to well--known results about uniqueness of invariant states of (aperiodic
irreducible) discrete Markov chains. See for instance \cite[Chapter 18]%
{Klenke}.

Assumption \ref{assumption3} concludes the list of required conditions and
from now on, we assume Assumptions \ref{assumption important}--\ref%
{assumption3} to be satisfied.

\subsection{The effective master equation\label{effective subsection}}

We define now the effective atomic master equation on the Hilbert space $%
\mathfrak{H}_{\mathrm{at}}$ as the initial value problem
\begin{equation}
\forall t\geq 0:\qquad \dfrac{d}{dt}\rho (t)=\mathfrak{L}_{t}^{\left(
\lambda ,\eta \right) }(\rho (t)),\qquad \rho (0)=\rho _{\mathrm{at}}\left(
0\right) \equiv \rho _{\mathrm{at}}\in \mathfrak{H}_{\mathrm{at}}\ .
\label{effective atomic master equation}
\end{equation}%
Recall that $\rho _{\mathrm{at}}$ is the density matrix of the initial
atomic state $\omega _{\mathrm{at}}$ of the atom and $\rho _{\mathrm{at}%
}\left( t\right) $ is the density matrix of the time--dependent state $%
\omega _{\mathrm{at}}\left( t\right) $ defined by (\ref{ninja1}) for any $%
t\in \mathbb{R}_{0}^{+}$. Even if one imposes the condition $\rho _{\mathrm{%
at}}\in \mathfrak{D}$, note that $\rho _{\mathrm{at}}\left( t\right) $ is
generally not block--diagonal, i.e., $\rho _{\mathrm{at}}\left( t\right)
\notin \mathfrak{D}$ for all $t\geq 0$. The same is true for the solution of
(\ref{effective atomic master equation}).

The effective atomic master equation has a unique solution which, by finite
dimensionality of $\mathfrak{H}_{\mathrm{at}}$, is explicitly given by a
Dyson series. In particular, this initial value problem defines a
two--parameter family denoted by $\{\hat{\tau}_{t,s}^{(\lambda ,\eta
)}\}_{t\geq s}$. Since the Lindbladian $\mathfrak{L}_{t}^{\left( \lambda
,\eta \right) }$ is continuous and generates a Markov CP semigroup at any
fixed time $t\in \mathbb{R}$ (cf. Remark \ref{markov CP semigroup}), the
two--parameter family $\{\hat{\tau}_{t,s}^{(\lambda ,\eta )}\}_{t\geq
s}\subset \mathcal{B}(\mathfrak{H}_{\mathrm{at}})$ is continuous, completely
positive and preserves the trace. The positivity and trace preservation
imply that this family is uniformly norm bounded\footnote{%
This can be seen by using a decomposition of any $A\in \mathfrak{H}_{\mathrm{%
at}}$ in imaginary and real parts, each of them being also decomposed in
positive and negative parts. Use then the trace--norm, which is equivalent
to the norm on $\mathfrak{H}_{\mathrm{at}}\equiv \mathcal{B}(\mathbb{C}^{d})$%
.}:%
\begin{equation}
\forall \lambda ,\eta ,\varpi ,s,t\in \mathbb{R}{,\ }t\geq s:\qquad
\parallel \hat{\tau}_{t,s}^{(\lambda ,\eta )}\parallel \leq C
\label{gnackidiot}
\end{equation}%
for some finite constant $C\in \left( 0,\infty \right) $ not depending on $%
\lambda $, $\eta $, $\varpi $, $s$, and $t$. When the optical pump is
absent, the dynamics becomes autonomous and the family $\{\hat{\tau}%
_{t,s}^{(\lambda ,0)}\}_{t\geq s}$ corresponds to an one--parameter
semigroup denoted for simplicity by
\begin{equation}
\hat{\tau}_{t}^{(\lambda ,0)}:=\hat{\tau}_{t,0}^{(\lambda ,0)}\ .
\label{semigroup}
\end{equation}

The main interest of the initial value problem (\ref{effective atomic master
equation}) is that its (unique) solution $\rho(t)$ accurately approximates
at small couplings the true density matrix $\rho_{\mathrm{at}}\left(t\right)$
of the time--dependent state $\omega_{\mathrm{at}}\left(t\right)$ on the
subspace $\mathfrak{D}$ (\ref{ninja0}) of block--diagonal matrices for all $%
t\in\mathbb{R}_{0}^{+}$. Indeed, we prove in \cite{BruPedraWestrich2011b}
the following assertion:

\begin{satz}[Validity of the effective atomic master equation]
\label{ninja thm cool}\mbox{ }\newline
Assume that $\rho _{\mathrm{at}}\in \mathfrak{D}$. The unique solution $%
\{\rho (t)\}_{t\geq 0}$ of the effective atomic master equation (\ref%
{effective atomic master equation}) and the atomic density matrix $\{\rho _{%
\mathrm{at}}\left( t\right) \}_{t\geq 0}$ satisfy the bound
\begin{equation*}
\left\Vert P_{\mathfrak{D}}\left( \rho _{\mathrm{at}}(t)-\rho (t)\right)
\right\Vert \leq C_{\varpi }\left\vert \lambda \right\vert
\end{equation*}%
for some constant $C_{\varpi }\in \left( 0,\infty \right) $ depending on $%
\varpi $, but not on the initial state $\omega _{\mathrm{at}}$ of the atom
and the parameters $t$, $\lambda $, and $\eta $, provided $\lambda $ is
sufficiently small.
\end{satz}

\noindent \textit{Sketch of the proof}. The proof of Theorem \ref{ninja thm
cool} is conceptually similar to what is done in Section \ref%
{sec:The-effective-atomic}, but technically much more involved:

\begin{itemize}
\item Similar to the one--parameter semigroup $\{\mathcal{T}_{\alpha
}\}_{\alpha \geq 0}$ defined below, we represent the non--autonomous
evolution $\{U_{t,s}^{(\lambda ,\eta )}\}_{t\geq s}$ as an autonomous
dynamics $\{\mathrm{e}^{\alpha \mathcal{G}}\}_{\alpha \geq 0}$ on an
enlarged Hilbert space $\mathfrak{H}_{\mathrm{evo}}\supset \mathfrak{H}$ of
periodic $\mathfrak{H}$--valued functions (vectors of $\mathfrak{H}$ are
naturally identified with the constant functions in this case). The
generator $\mathcal{G}$ of the new time--evolution is occasionally referred
to as \emph{Howland} or \emph{Floquet} operator. $\mathfrak{H}$ is the
GNS--space of the initial state $\omega _{0\text{ }}$ and $%
\{U_{t,s}^{(\lambda ,\eta )}\}_{t\geq s}$ is a suitable representation of
the full microscopic dynamics $\tau _{t,0}^{(\lambda ,\eta )}$ through a
family of bounded operators on $\mathfrak{H}$.

\item Then, we perform an analytic deformation $\mathcal{G}(\theta )$ (more
precisely, analytic translation) of the unbounded closed operator $\mathcal{G%
}$ and prove that the dynamics driven by $\{\mathrm{e}^{\alpha \mathcal{G}%
}\}_{\alpha \geq 0}$ and $\{\mathrm{e}^{\alpha \mathcal{G}(\theta
)}\}_{\alpha \geq 0}$ are the same on the atomic subspace $\mathfrak{H}_{%
\mathrm{at}}\equiv \mathfrak{H}_{\mathrm{at}}\otimes \{\Omega _{\mathcal{R}%
}\}\subset \mathfrak{H}_{\mathrm{evo}}$. The use of analytic deformations is
the reason for the analyticity condition stated at the end of Section \ref%
{section atom--reservoir interaction}.

\item In contrast to $\mathcal{G}$, whose eigenvalues are all imbedded in
the continuous spectrum, $\mathcal{G}(\theta )$ has discrete spectrum. It
turns out that the discrete eigenspace of $\mathcal{G}(\theta )$ is the
relevant one for the atomic dynamics. We then analyze the discrete spectrum
and eigenspace of $\mathcal{G}(\theta )$ through Kato's perturbation theory
\cite{Kato} for closed operators. If Assumption \ref{assumption important}
holds to the leading order in $\lambda $ and $\eta $, that is, second order
in $\lambda $ and first order in $\eta $, the operator $\mathcal{G}(\theta )$
is -- up to purely imaginary constants -- unitarily equivalent to $\frac{%
\eta }{2}\mathfrak{L}_{\mathrm{p}}+\lambda ^{2}\mathfrak{L}_{\mathcal{R}}$
in finite dimensional invariant subspaces spanning the whole discrete
subspace of $\mathcal{G}(\theta )$.

\item We are then in position, by using the inverse Laplace transform for $%
C_{0}$--semigroups together with Kato projections, to analyze the action of
the semigroup $\{\mathrm{e}^{\alpha \mathcal{G}(\theta )}\}_{\alpha \geq 0}$
on vectors of the atomic subspace. This analysis leads to a version of
Corollary \ref{lemmalongtime copy(3)} for the full microscopic dynamics and
Theorem \ref{ninja thm cool} follows.
\end{itemize}

\hfill{}{}$\Box$

\section{Effective atomic dynamics\label{sec:The-effective-atomic}}

In this Section we study the behavior of the solutions of the effective
atomic master equation, that is, the initial value problem (\ref{effective
atomic master equation}). Since an important issue of laser technology is to
obtain an optical pumping of atomic energy levels, we want to understand the
time behavior of its solution in relation with the phenomenological \emph{%
Pauli master equation} found in standard textbooks on lasers. The small
(more precisely, of order $\lambda ^{2}\varpi ^{-1}$) fast oscillations of
the populations due to the cosine in the master equation prevent from
obtaining a perfect steady behavior at large times. It is thus convenient to
remove them by averaging the non--autonomous dynamics over a moving period.
This leads to a \emph{pre--master equation} proven in Theorem \ref{The
pre--master equation}, whereas the Pauli master equation only extracts the
limiting behavior of populations at large times (see Section \ref%
{sec:The-Generalized-Einstein}).

\subsection{Methodology and Numerical Illustrations\label{section inv of pop}%
}

To this end, we first represent this non--autonomous evolution as an \emph{%
autonomous} dynamics on an enlarged, infinite dimensional Hilbert space
\begin{equation*}
\mathfrak{H}_{\mathrm{evo}}\supsetneq \mathfrak{H}_{\mathrm{at}}\equiv
\mathcal{B}(\mathbb{C}^{d})\ .
\end{equation*}%
See (\ref{H evo}) below. The latter emerges through an additional degree of
freedom which is a new time variable denoted by $\alpha \geq 0$.

By iterating the{}\textquotedblleft variation of constants
formula\textquotedblright , i.e., the integral equation%
\begin{equation}
\forall s,t\in \mathbb{R},\ t\geq s:\qquad \hat{\tau}_{t,s}^{(\lambda ,\eta
)}=\hat{\tau}_{t-s}^{(0,0)}+\int_{s}^{t}\hat{\tau}_{t-v}^{(0,0)}\mathfrak{W}%
_{v}^{\left( \lambda ,\eta \right) }\hat{\tau}_{v,s}^{(\lambda ,\eta )}%
\mathrm{d}v\,  \label{ninja intergal eq}
\end{equation}%
with the $2\pi \varpi ^{-1}$--periodic operator
\begin{equation}
\mathfrak{W}_{t}^{\left( \lambda ,\eta \right) }:=\mathfrak{L}_{t}^{\left(
\lambda ,\eta \right) }-\mathfrak{L}_{t}^{\left( 0,0\right) }=\eta \cos
(\varpi t)\mathfrak{L}_{\mathrm{p}}+\lambda ^{2}\mathfrak{L}_{\mathcal{R}%
}\in \mathcal{B}\left( \mathfrak{H}_{\mathrm{at}}\right) \ ,
\label{ninja intergal eqbis}
\end{equation}%
we get a representation of $\hat{\tau}_{t,s}^{(\lambda ,\eta )}$ as an
absolutely convergent (Dyson) series which shows that $(t,s)\mapsto \hat{\tau%
}_{t,s}^{(\lambda ,\eta )}$ is continuous and
\begin{equation}
\forall k\in \mathbb{Z},\ s,t\in \mathbb{R},\ t\geq s:\qquad \hat{\tau}%
_{t,s}^{(\lambda ,\eta )}=\hat{\tau}_{t+2\pi \varpi ^{-1}k,s+2\pi \varpi
^{-1}k}^{(\lambda ,\eta )}\ .  \label{periodic t effective}
\end{equation}%
In other words, the dynamics between times $(\alpha +2\pi \varpi ^{-1}k)$
and $(t+\alpha +2\pi \varpi ^{-1}k)$ does not depend on $k\in \mathbb{Z}$.
As $\hat{\tau}_{t-s}^{(0,0)}$ does not affect populations, (\ref{Section
effective howland}) also shows that they do not change much within a period
of the pump. As already explained, it is thus natural to average the
non--autonomous dynamics over a moving period of length $2\pi \varpi ^{-1}$
to extract the leading dynamical behavior of populations, in particular the
inversion of population. The latter is described in Section \ref{Section
effective howland}.

This first step of the analysis of the solutions of the atomic master
equation is quite useful because it enables the analysis of the
non--autonomous dynamics via an associated (evolution) semigroup denoted by $%
\{\mathcal{T}_{\alpha }\}_{\alpha \geq 0}$ corresponding to an autonomous
dynamics. Observe that $\{\mathcal{T}_{\alpha }\}_{\alpha \geq 0}$ acts on
an infinite dimensional Hilbert space $\mathfrak{H}_{\mathrm{evo}%
}\varsupsetneq \mathfrak{H}_{\mathrm{at}}$ although the initial
non--autonomous dynamics was finite dimensional. Nevertheless, since the
Hilbert space $\mathfrak{D}\subset \mathfrak{H}_{\mathrm{at}}$ is an
invariant subspace of $\{\mathcal{T}_{\alpha }\}_{\alpha \geq 0}$ when $%
\lambda =\eta =0$, Kato's perturbation theory \cite{Kato} shows, for
sufficiently small coupling constants $\lambda $ and $\eta $, the existence
of a finite dimensional invariant Hilbert space $\mathfrak{H}_{0}^{\left(
\lambda ,\eta \right) }$ of $\{\mathcal{T}_{\alpha }\}_{\alpha \geq 0}$
almost parallel to $\mathfrak{D}$. In particular, concerning the dynamics of
populations, we can finally pass to an autonomous finite dimensional
dynamics. Note however that $\dim \mathfrak{H}_{0}^{\left( \lambda ,\eta
\right) }\mathfrak{>}\dim \mathfrak{D}$, as $\mathfrak{D}$ is a subspace of
the larger invariant space $\mathfrak{H}_{0}^{\left( 0,0\right) }$ of $\{%
\mathcal{T}_{\alpha }\}_{\alpha \geq 0}$ when $\lambda =\eta =0$. This
second step is performed in Sections \ref{Sectino restriction finite}--\ref%
{Section average block--diagonal dynamics}.

As a first application, these results are then used at the end of Section %
\ref{Section average block--diagonal dynamics} to study the large time
behavior of $P_{\mathfrak{D}}\left( \rho \left( t\right) \right) $. We show
in particular the existence of a density matrix $\rho _{\infty }$ (Theorem %
\ref{lemmalongtime copy(4)}), uniquely determined by a balance condition
(Theorem \ref{corollary uniqueness limit density matrix}),
well--approximating $P_{\mathfrak{D}}\left( \rho \left( t\right) \right) $
when $t\rightarrow \infty $ for small enough coupling constants $\lambda $, $%
\eta $. In Section \ref{Section The pre--master equation} we derive an
integro--differential equation (\emph{pre--master equation}) on the subspace
of block--diagonal density matrices $\mathfrak{D}\subset $ $\mathfrak{H}_{%
\mathrm{at}}$ effectively describing the physical evolution of the
populations. The dynamics of population is properly described by an
integro--differential equation and not by an effective differential equation
like a Pauli equation. This is due to the strict inequality $\dim \mathfrak{H%
}_{0}^{\left( \lambda ,\eta \right) }\mathfrak{>}\dim \mathfrak{D}$.

The rigorous proofs can be tedious sometimes and require a number of
definitions and notations. Therefore, we outline our study by giving now
some numerical illustrations of these three master equations, that is:

\begin{enumerate}
\item[(i)] The effective atomic master equation (\ref{effective atomic
master equation}).

\item[(ii)] The pre--master equation, see Theorem \ref{The pre--master
equation}.

\item[(iii)] The phenomenological Pauli master equation (\ref{Pauli master
equation}) as explained in standard textbooks on lasers. See Section \ref%
{sec:The-Generalized-Einstein} for more details.
\end{enumerate}

As explained in Sections \ref{Section intro}--\ref{Section def model}, an
important step of laser technology is to obtain an optical pumping of atomic
energy levels. So, we focus here on the so--called inversion of population%
\footnote{%
Note that analytical studies concerning the inversion of population can
easily be done, at least for $d=N=4$, by using the balance equation (\ref%
{balance conditoin}).}. As described in textbooks on laser physics, optical
pumping is in many situations based on three-- or four--level atoms \cite%
{Levine1968}, the second case being the more efficient of both. Therefore,
we restrict our study on a non--degenerated four--level atom, i.e., $d=N=4$.
In this case, the effective atomic master equation (\ref{effective atomic
master equation}) is a non--autonomous evolution equation on a $16$%
--dimensional Hilbert space which can easily be treated by standard
numerical methods. One can then more clearly understand the different
approximations performed in this section which lead to the pre--master and
Pauli master equations.

In our example, the atomic Hamiltonian depends on the parameter $\varpi $
and equals%
\begin{equation}
H_{\mathrm{at}}=\left(
\begin{array}{cccc}
0 & 0 & 0 & 0 \\
0 & \frac{1}{5}\varpi & 0 & 0 \\
0 & 0 & \frac{5}{6}\varpi & 0 \\
0 & 0 & 0 & \varpi%
\end{array}%
\right) \ .  \label{atomic Hamiltonian example}
\end{equation}%
The atom--reservoir interaction is fixed by $m=1$, the self--adjoint matrix%
\begin{equation*}
Q_{_{1}}=\left(
\begin{array}{cccc}
0 & \frac{5}{6} & \frac{1}{4} & \frac{1}{5} \\
\frac{5}{6} & 0 & \frac{5}{14} & \frac{5}{19} \\
\frac{1}{4} & \frac{5}{14} & 0 & 1 \\
\frac{1}{5} & \frac{5}{19} & 1 & 0%
\end{array}%
\right)
\end{equation*}%
and the coupling function $f_{_{1}}$ defined on $\mathbb{R}^{3}$ by%
\begin{equation*}
f_{_{1}}\left( p\right) =\mathrm{f}_{\ell }(|p|)=\frac{1}{2\pi |p|}\exp
\left( -|p|/2\right) \ .
\end{equation*}%
See Section \ref{section atom--reservoir interaction}. The optical pump is
modeled here by the time--periodic perturbation%
\begin{equation*}
\eta \cos (\varpi \alpha )\left(
\begin{array}{cccc}
0 & 0 & 0 & 1 \\
0 & 0 & 0 & 0 \\
0 & 0 & 0 & 0 \\
1 & 0 & 0 & 0%
\end{array}%
\right) ,\qquad \alpha \equiv t\in \mathbb{R},
\end{equation*}%
of the atomic Hamiltonian $H_{\mathrm{at}}$, see Section \ref{section pump
nija}. The initial state of the atom $\rho _{\mathrm{at}}\in \mathfrak{D}%
\subset \mathfrak{H}_{\mathrm{at}}$ is the Gibbs state $\rho _{\mathfrak{g}%
}=P_{\mathfrak{D}}\left( \rho _{\mathfrak{g}}\right) $ (\ref{Gibbs.init})
taken at the same inverse temperature $\beta $ as the fermionic reservoir.
We set the frequency $\varpi $, the inverse temperature $\beta $ and the
coupling constants $\lambda ,\eta $ respectively equal to $\varpi =3$, $%
\beta =0.5$, $\lambda \simeq 0.385$, and $\eta =\lambda ^{2}\simeq 0.148$.

\begin{figure*}[hbtp]
\begin{center}
\mbox{
\leavevmode
\subfigure
{ \includegraphics[angle=0,scale=1,clip=true,width=5.5cm]{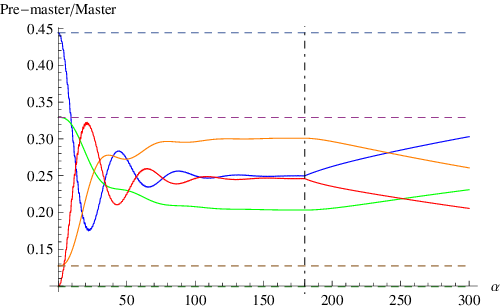} }
\leavevmode
\subfigure
{ \includegraphics[angle=0,scale=1,clip=true,width=5.5cm]{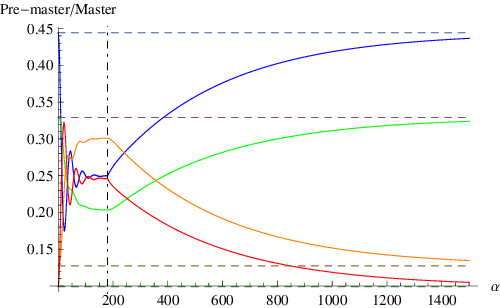} }
\leavevmode
\subfigure
{ \includegraphics[angle=0,scale=1,clip=true,width=5.5cm]{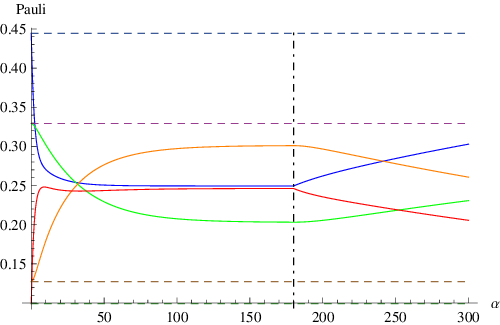} }
}
\end{center}
\caption{\emph{Illustration of the populations as  functions of $\alpha \equiv
t\in \mathbb{R}_{0}^{+}$ at $\beta =0.5$, $\lambda \simeq 0.385$, $\eta
=\lambda ^{2}\simeq 0.148$ and $\varpi =3$. Blue, green, orange, and red
lines correspond to the populations of the 1st, the 2nd, the 3rd
and the 4th atomic energy levels, respectively. The dashed dotted line
marks the time $\alpha=180$ when the pump is turned off, i.e., $\eta =0$ for all times
beyond this line. The four dashed lines mark the populations of the
thermal equilibrium state, i.e., the Gibbs state $\rho _{\mathfrak{g}}$ at
inverse temperature $\beta =0.5$ of the thermal reservoir. The two left
figures are computed from the master equation. The pre--master equation
gives exactly the same picture at this time scale. The right figure is
computed from the Pauli master equation.}}
\label{inversion}
\end{figure*}

The population density of the $k$th atomic level, at any fixed time $\alpha
\in \mathbb{R}_{0}^{+}$ and for any $k\in \left\{ 1,\ldots ,4\right\} $, is
given by
\begin{equation*}
\mathrm{d}_{k}(\alpha ):=p_{k}(\rho \left( \alpha \right) )=\mathrm{Tr}_{%
\mathbb{C}^{4}}(\mathbf{1}\left[ H_{\mathrm{at}}=E_{k}\right] \ \rho \left(
\alpha \right) \ \mathbf{1}\left[ H_{\mathrm{at}}=E_{k}\right] )\geq 0\ .
\end{equation*}%
Here, $\rho (\alpha )$ is the solution of the effective atomic master
equation (\ref{effective atomic master equation}) and $\mathrm{d}_{k}(\alpha
)$ is directly related to the atomic populations because of Theorem \ref%
{ninja thm cool}. Similarly, populations can be defined for the solutions of
the pre--master equation and the Pauli master equation. The plots of all
these densities are given in figure \ref{inversion} and clearly show a
stable inversion of population as
\begin{equation*}
\mathrm{d}_{3}(\alpha )>\mathrm{d}_{1}(\alpha )>\mathrm{d}_{4}(\alpha )>%
\mathrm{d}_{2}(\alpha )
\end{equation*}%
for large enough $\alpha \leq 180$. As soon as the optical pump is turned
off, the systems relaxes to the Gibbs state $\rho _{\mathfrak{g}}$.
%TCIMACRO{%
%\TeXButton{inversion}{\begin{figure*}[hbtp]
%\begin{center}
%\mbox{
%\leavevmode
%\subfigure
%{ \includegraphics[angle=0,scale=1,clip=true,width=5.5cm]{master1.eps} }
%\leavevmode
%\subfigure
%{ \includegraphics[angle=0,scale=1,clip=true,width=5.5cm]{master1bis.eps} }
%\leavevmode
%\subfigure
%{ \includegraphics[angle=0,scale=1,clip=true,width=5.5cm]{pauli.eps} }
%}
%\end{center}
%\caption{\emph{Illustration of the populations as  functions of $\alpha \equiv
%t\in \mathbb{R}_{0}^{+}$ at $\beta =0.5$, $\lambda \simeq 0.385$, $\eta
%=\lambda ^{2}\simeq 0.148$ and $\varpi =3$. Blue, green, orange, and red
%lines correspond to the populations of the 1st, the 2nd, the 3rd
%and the 4th atomic energy levels, respectively. The dashed dotted line
%marks the time $\alpha=180$ when the pump is turned off, i.e., $\eta =0$ for all times
%beyond this line. The four dashed lines mark the populations of the
%thermal equilibrium state, i.e., the Gibbs state $\rho _{\mathfrak{g}}$ at
%inverse temperature $\beta =0.5$ of the thermal reservoir. The two left
%figures are computed from the master equation. The pre--master equation
%gives exactly the same picture at this time scale. The right figure is
%computed from the Pauli master equation.}}
%\label{inversion}
%\end{figure*}}}%
%BeginExpansion
%EndExpansion

\begin{figure*}[hbtp]
\begin{center}
\mbox{
\leavevmode
\subfigure
{ \includegraphics[angle=0,scale=1,clip=true,width=4.1cm]{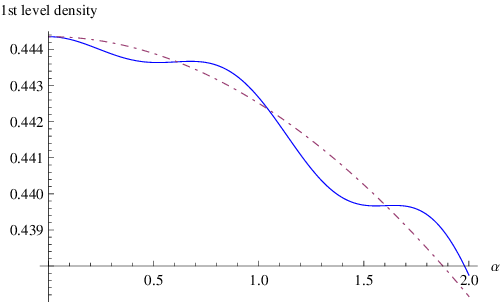} }
\leavevmode
\subfigure
{ \includegraphics[angle=0,scale=1,clip=true,width=4.1cm]{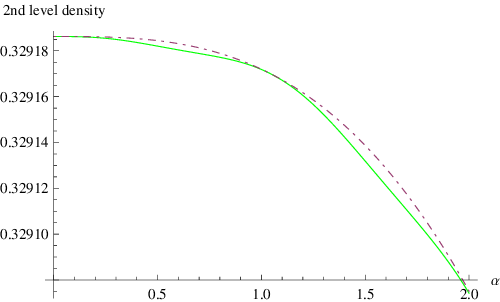} }
\leavevmode
\subfigure
{ \includegraphics[angle=0,scale=1,clip=true,width=4.1cm]{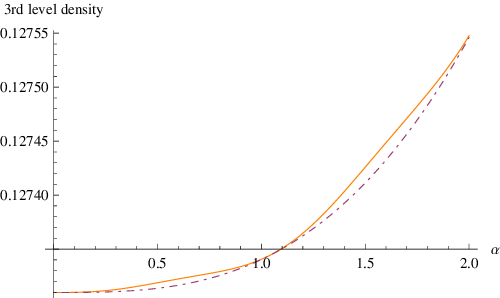} }
\leavevmode
\subfigure
{ \includegraphics[angle=0,scale=1,clip=true,width=4.1cm]{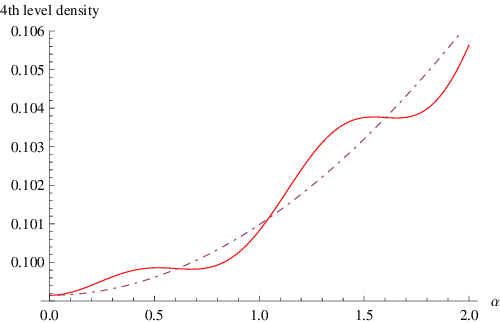} }
}
\end{center}
\caption{\emph{Illustration of the populations as functions of $\alpha \equiv
t\in \left[ 0,2\right] $ at $\beta =0.5$, $\lambda \simeq 0.385$, $\eta
=\lambda ^{2}\simeq 0.148$ and $\varpi =3$. Blue, green, orange, and red
lines (plots from left to right) correspond to the populations of
the 1st, the 2nd, the 3rd and the 4th atomic energy levels, respectively.
They are computed from the master equation. The dashed dotted lines are the same
objects computed from the pre--master equation.}}
\label{level}
\end{figure*}

As we can observe in figures \ref{inversion} and \ref{level}, the
qualitative difference in the behavior of the solutions of the master,
pre--master and Pauli master equations for population densities are clear:
The pre--master equation removes the small, but fast oscillations due to the
cosine in the master equation (figure \ref{level}), and the Pauli master
equation additionally cancels the Rabi (moderate) oscillations still present
in the pre--master equation (compare the left and the right plot of figure %
\ref{inversion}).

Note that Rabi oscillations, which depends on the coupling constant $\eta$,
are progressively suppressed by dissipative processes. Removing the
atom--reservoir interaction by setting $\lambda=0$ while keeping $%
\eta\simeq0.148$ we observe non--suppressed oscillations, see figure \ref%
{rabi}. In this case the stable inversion of population also disappears, as
explained in the beginning of Section \ref{Section def model}.
%TCIMACRO{%
%\TeXButton{level}{\begin{figure*}[hbtp]
%\begin{center}
%\mbox{
%\leavevmode
%\subfigure
%{ \includegraphics[angle=0,scale=1,clip=true,width=4.1cm]{level1.eps} }
%\leavevmode
%\subfigure
%{ \includegraphics[angle=0,scale=1,clip=true,width=4.1cm]{level2.eps} }
%\leavevmode
%\subfigure
%{ \includegraphics[angle=0,scale=1,clip=true,width=4.1cm]{level3.eps} }
%\leavevmode
%\subfigure
%{ \includegraphics[angle=0,scale=1,clip=true,width=4.1cm]{level4.eps} }
%}
%\end{center}
%\caption{\emph{Illustration of the populations as functions of $\alpha \equiv
%t\in \left[ 0,2\right] $ at $\beta =0.5$, $\lambda \simeq 0.385$, $\eta
%=\lambda ^{2}\simeq 0.148$ and $\varpi =3$. Blue, green, orange, and red
%lines (plots from left to right) correspond to the populations of
%the 1st, the 2nd, the 3rd and the 4th atomic energy levels, respectively.
%They are computed from the master equation. The dashed dotted lines are the same
%objects computed from the pre--master equation.}}
%\label{level}
%\end{figure*}}}%
%BeginExpansion
%EndExpansion
%TCIMACRO{%
%\TeXButton{rabi}{\begin{figure*}[hbtp]
%\begin{center}
%\mbox{
%\leavevmode
%\subfigure
%{ \includegraphics[angle=0,scale=1,clip=true,width=8cm]{rabi1.eps} }
%}
%\end{center}
%\caption{\emph{Illustration of the populations as functions of $\alpha \equiv
%t\in \left[ 0,300\right] $ at $\beta =0.5$, $\lambda =0$, $\eta \simeq 0.148$
%and $\varpi =3$. Blue, green, orange, and red lines correspond to the
%populations of the 1st, the 2nd, the 3rd and the 4th atomic
%energy levels, respectively. The dashed dotted line marks the time $\alpha =180$ when the pump is turned off, i.e., $\eta =0$ for all times
%beyond this line. The four dashed line mark the populations of the
%Gibbs state $\rho _{\mathfrak{g}}$ at inverse temperature $\beta =0.5$ of
%the thermal reservoir. The figure is computed from the master equation. Note
%that, for $\alpha \geq 180$, the system do not relax to the Gibbs state $\rho _{\mathfrak{g}}$ at inverse temperature $\beta =0.5$ of the thermal
%reservoir because there is no atom--reservoir interaction as $\lambda =0$.}}
%\label{rabi}
%\end{figure*}}}%
%BeginExpansion
\begin{figure*}[hbtp]
\begin{center}
\mbox{
\leavevmode
\subfigure
{ \includegraphics[angle=0,scale=1,clip=true,width=8cm]{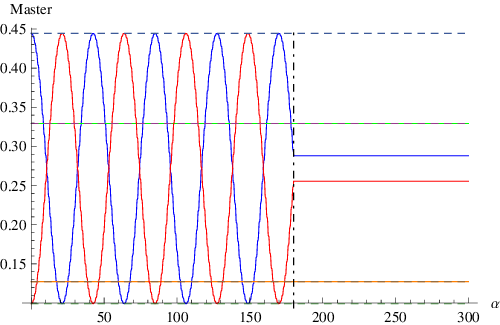} }
}
\end{center}
\caption{\emph{Illustration of the populations as functions of $\alpha \equiv
t\in \left[ 0,300\right] $ at $\beta =0.5$, $\lambda =0$, $\eta \simeq 0.148$
and $\varpi =3$. Blue, green, orange, and red lines correspond to the
populations of the 1st, the 2nd, the 3rd and the 4th atomic
energy levels, respectively. The dashed dotted line marks the time $\alpha =180$ when the pump is turned off, i.e., $\eta =0$ for all times
beyond this line. The four dashed line mark the populations of the
Gibbs state $\rho _{\mathfrak{g}}$ at inverse temperature $\beta =0.5$ of
the thermal reservoir. The figure is computed from the master equation. Note
that, for $\alpha \geq 180$, the system do not relax to the Gibbs state $\rho _{\mathfrak{g}}$ at inverse temperature $\beta =0.5$ of the thermal
reservoir because there is no atom--reservoir interaction as $\lambda =0$.}}
\label{rabi}
\end{figure*}%
%EndExpansion

\begin{figure*}[hbtp]
\begin{center}
\mbox{
\leavevmode
\subfigure
{ \includegraphics[angle=0,scale=1,clip=true,width=7cm]{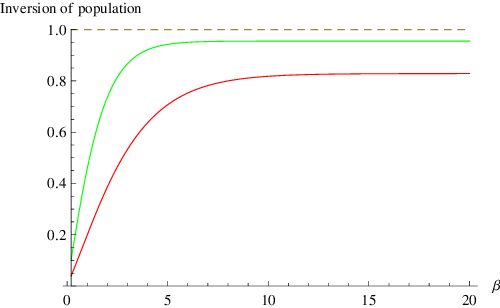} }
\leavevmode
\subfigure
{ \includegraphics[angle=0,scale=1,clip=true,width=7cm]{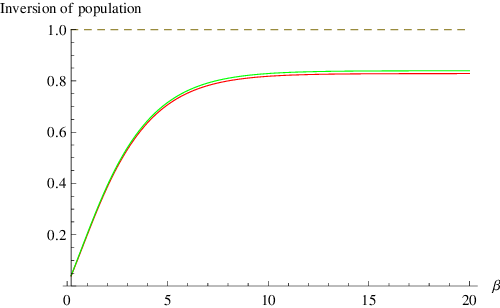} }
}
\end{center}
\caption{\emph{Illustration of the inversion of population $\mathrm{d}_{3}(\infty )-\mathrm{d}_{1}(\infty )$ for large times, computed from the pre--master equation, as
a\ function of the inverse temperature $\beta \in \left[ 0.2,20\right] $ at $\lambda \simeq 0.385$.
In the left and right plots, the red lines
correspond to $\eta =\lambda ^{2}\simeq 0.148$, $\varpi =3$, whereas the
green line represents the choices $\eta =\lambda ^{2}$, $\varpi =6$ (left plot) and
$\eta =10\lambda ^{2}$ , $\varpi =3$ (right plot). Note
that the atomic Hamiltonian $H_{\mathrm{at}}$ depends on the frequency $\varpi $,
see (\ref{atomic Hamiltonian example}).}}
\label{inversion-temp}
\end{figure*}

To conclude, our results permit purely quantum mechanical detailed studies
of many--level optically active impurities used to produce lasing materials.
For instance, they can be used to analyze the influence of temperature and
other relevant physical parameters on inversion of population: At least in
the example considered here, the inversion of population is decreasing with
the temperature, i.e., is an increasing function of $\beta $ as shown figure %
\ref{inversion-temp}, and the most efficient parameter to strengthen it is
curiously the frequency $\varpi $ and not the parameter $\eta $, provided
the last is neither too small nor too large. The effect of the degeneracy
and the dynamics for very low frequencies can also be analyzed. The latter
corresponds to a situation close to the adiabatic limit and shows an unusual
behavior. We postpone this kind of studies to a further paper and start now
the rigorous proofs related to these numerical observations.
%TCIMACRO{%
%\TeXButton{inversion-temp}{\begin{figure*}[hbtp]
%\begin{center}
%\mbox{
%\leavevmode
%\subfigure
%{ \includegraphics[angle=0,scale=1,clip=true,width=7cm]{inversion-temp1.eps} }
%\leavevmode
%\subfigure
%{ \includegraphics[angle=0,scale=1,clip=true,width=7cm]{inversion-temp2.eps} }
%}
%\end{center}
%\caption{\emph{Illustration of the inversion of population $\mathrm{d}_{3}(\infty )-\mathrm{d}_{1}(\infty )$ for large times, computed from the pre--master equation, as
%a\ function of the inverse temperature $\beta \in \left[ 0.2,20\right] $ at $\lambda \simeq 0.385$.
%In the left and right plots, the red lines
%correspond to $\eta =\lambda ^{2}\simeq 0.148$, $\varpi =3$, whereas the
%green line represents the choices $\eta =\lambda ^{2}$, $\varpi =6$ (left plot) and
%$\eta =10\lambda ^{2}$ , $\varpi =3$ (right plot). Note
%that the atomic Hamiltonian $H_{\mathrm{at}}$ depends on the frequency $\varpi $,
%see (\ref{atomic Hamiltonian example}).}}
%\label{inversion-temp}
%\end{figure*}}}%
%BeginExpansion
%EndExpansion

\subsection{From the non--autonomous master equation to an autonomous
effective dynamics, evolution semigroups\label{Section effective howland}}

Consider the Hilbert space
\begin{equation}
\mathfrak{H}_{\mathrm{evo}}:=L^{2}\left( \left[ 0,2\pi \varpi ^{-1}\right) ,%
\mathfrak{H}_{\mathrm{at}}\right)  \label{H evo}
\end{equation}%
of time--dependent $2\pi \varpi ^{-1}$--periodic $\mathfrak{H}_{\mathrm{at}}$%
--valued functions. The scalar product on $\mathfrak{H}_{\mathrm{evo}}$ is
naturally defined by%
\begin{equation*}
\left\langle f,g\right\rangle _{\mathrm{evo}}:=\frac{\varpi }{2\pi }%
\int_{0}^{\frac{2\pi }{\varpi }}\left\langle f\left( t\right) ,g\left(
t\right) \right\rangle _{\mathrm{at}}\mathrm{d}t=\frac{\varpi }{2\pi }%
\int_{0}^{\frac{2\pi }{\varpi }}\mathrm{Tr}_{\mathbb{C}^{d}}\left( f\left(
t\right) ^{\ast }g\left( t\right) \right) \mathrm{d}t
\end{equation*}%
for all $f,g\in \mathfrak{H}_{\mathrm{evo}}$, see (\ref{trace h at}). $%
\mathfrak{H}_{\mathrm{at}}\varsubsetneq \mathfrak{H}_{\mathrm{evo}}$ is seen
as the subspace of constant functions of $\mathfrak{H}_{\mathrm{evo}}$.

From the continuous two--parameter family $\{\hat{\tau}_{t,s}^{(\lambda
,\eta )}\}_{t\geq s}$ corresponding to the non--autonomous master equation (%
\ref{effective atomic master equation}) we can uniquely define a strongly
continuous one--parameter semigroup $\{\mathcal{T}_{\alpha }\}_{\alpha \geq
0}$ by the condition%
\begin{equation}
\forall t\in \left[ 0,2\pi \varpi ^{-1}\right) \ \mathrm{a.e.}:\qquad
\mathcal{T}_{\alpha }\left( f\right) \left( t\right) =\hat{\tau}_{t,t-\alpha
}^{(\lambda ,\eta )}f\left( t-\alpha \right) \ .  \label{definition howland}
\end{equation}%
for all $\alpha \geq 0$ and $f\in \mathfrak{H}_{\mathrm{evo}}$. Because of (%
\ref{periodic t effective}), $\mathcal{T}_{\alpha }$ is an operator acting
on $\mathfrak{H}_{\mathrm{evo}}$ for any $\alpha \geq 0$. The strong
continuity of $\alpha \mapsto \mathcal{T}_{\alpha }$ follows from the
continuity of $(s,t)\mapsto \hat{\tau}_{t,s}^{(\lambda ,\eta )}$, and the
semigroup property of $\{\mathcal{T}_{\alpha }\}_{\alpha \geq 0}$ from the
cocycle property of the two--parameter family $\{\hat{\tau}_{t,s}^{(\lambda
,\eta )}\}_{t\geq s}$. Moreover, by the norm boundedness (\ref{gnackidiot})
of the evolution family $\{\hat{\tau}_{t,s}^{(\lambda ,\eta )}\}_{t\geq s}$,
the semigroup $\{\mathcal{T}_{\alpha }\}_{\alpha \geq 0}\subset \mathcal{B}(%
\mathfrak{H}_{\mathrm{evo}})$ is uniformly norm bounded:
\begin{equation}
\forall \lambda ,\eta \in \mathbb{R}{,\ }\alpha \geq 0:\qquad \parallel
\mathcal{T}_{\alpha }\parallel \leq C\ ,  \label{ninja bound semi}
\end{equation}%
for some finite constant $C\in \left( 0,\infty \right) $ not depending on $%
\lambda $, $\eta $, and $\alpha $.

The generator of the strongly continuous semigroup $\{\mathcal{T}_{\alpha
}\}_{\alpha \geq 0}$ is the closed unbounded operator%
\begin{equation}
G^{\left( \lambda ,\eta \right) }:=-\frac{d}{dt}+\mathfrak{L}_{\mathrm{evo}%
}^{\left( \lambda ,\eta \right) }\ ,  \label{ninja bound semibis}
\end{equation}%
the so--called \emph{Howland} operator of the non--autonomous atomic
dynamics, with dense domain
\begin{equation}
\mathcal{D}(G^{\left( \lambda ,\mu \right) }):=\left\{ \sum_{k=-\infty
}^{\infty }a_{k}e^{ik\varpi t}\;:\;a_{k}\in \mathfrak{H}_{\mathrm{at}},\
\sum_{k=-\infty }^{\infty }\left\Vert k\,a_{k}\right\Vert _{\mathrm{at}%
}^{2}<\infty \right\} \subset \mathfrak{H}_{\mathrm{evo}}\ .
\label{ninja bound semibisbis}
\end{equation}%
Here,%
\begin{equation*}
\frac{d}{dt}f(t)=\sum_{k=-\infty }^{\infty }ika_{k}e^{ik\varpi t}
\end{equation*}%
in the $L^{2}\left( \left[ 0,2\pi \varpi ^{-1}\right) ,\mathfrak{H}_{\mathrm{%
at}}\right) $ sense for all
\begin{equation*}
f=\sum_{k=-\infty }^{\infty }a_{k}e^{ik\varpi t}\in \mathcal{D}(G^{\left(
\lambda ,\mu \right) })\ ,
\end{equation*}%
and $\mathfrak{L}_{\mathrm{evo}}^{\left( \lambda ,\eta \right) }\in \mathcal{%
B}(\mathfrak{H}_{\mathrm{evo}})$ is the bounded operator defined, for all $%
f\in \mathfrak{H}_{\mathrm{evo}}$, by%
\begin{equation}
\forall t\in \left[ 0,2\pi \varpi ^{-1}\right) \ \mathrm{a.e.}:\qquad
\mathfrak{L}_{\mathrm{evo}}^{\left( \lambda ,\eta \right) }(f)\left(
t\right) :=\mathfrak{L}_{t}^{\left( \lambda ,\eta \right) }(f(t))\ .
\label{ninja bound semibisbisbis}
\end{equation}

\begin{remark}[The uncoupled atom--reservoir case]
\mbox{ }\newline
The spectrum of $G^{\left( 0,0\right) }$ is purely discrete. Therefore,
Kato's perturbation theory \cite{Kato} of discrete eigenvalues can be used
to study the spectral properties of the generator $G^{\left( \lambda ,\eta
\right) }$ for small $\lambda $ and $\eta $.
\end{remark}

The time--behavior of the solution $\rho (t)\in \mathfrak{H}_{\mathrm{at}}$
of the non--autonomous master equation (\ref{effective atomic master
equation}) can be studied on the subspace $\mathfrak{D}\subset \mathcal{B}(%
\mathbb{C}^{d})\equiv \mathfrak{H}_{\mathrm{at}}$ of block--diagonal
matrices (cf. (\ref{ninja0})) by using the $C_{0}$--semigroup $\{\mathcal{T}%
_{\alpha }\}_{\alpha \geq 0}$. More precisely, we prove in the following
lemma that, for any (block--diagonal) $A\in \mathfrak{D}$, the scalar
products of the form
\begin{equation*}
\left\langle \mathcal{T}_{\alpha }\left( \rho \right) ,A\right\rangle _{%
\mathrm{evo}}
\end{equation*}%
properly describe (i.e. up to small oscillations uniformly bounded in time)
the time evolution of
\begin{equation*}
\left\langle \rho (\alpha ),A\right\rangle _{\mathrm{at}}:=\mathrm{Tr}\left(
\rho (\alpha )A\right)
\end{equation*}%
whenever the pump frequency $\varpi $ is sufficiently large or the couplings
$\lambda $, $\eta $ are sufficiently small.

\begin{lemma}[Average dynamics over a moving period of length $2\protect\pi
\protect\varpi ^{-1}$]
\label{lemmalongtime}\mbox{ }\newline
Assume that $\rho _{\mathrm{at}}\in \mathfrak{D}$. The unique solution $%
\{\rho (t)\}_{t\geq 0}$ of the effective atomic master equation (\ref%
{effective atomic master equation}) satisfies the bound%
\begin{equation}
\left\vert \left\langle \rho (\alpha ),A\right\rangle _{\mathrm{at}%
}-\left\langle \mathcal{T}_{\alpha }\left( \rho _{\mathrm{at}}\right)
,A\right\rangle _{\mathrm{evo}}\right\vert \leq C\lambda ^{2}\varpi
^{-1}\left\Vert A\right\Vert  \label{longtimebehavior}
\end{equation}%
for all (block--diagonal matrices) $A\in \mathfrak{D}$, all $\left\vert
\lambda \right\vert \leq 1$ and all $\alpha \in \mathbb{R}_{0}^{+}$. Here, $%
C\in (0,\infty )$ is a finite constant not depending on $\rho _{\mathrm{at}}$%
, $A$, $\lambda $, $\eta $, $\varpi $, and $\alpha $.
\end{lemma}

\noindent \textit{Proof.} By (\ref{definition howland}),
\begin{equation}
\left\langle \rho (\alpha ),A\right\rangle _{\mathrm{at}}-\left\langle
\mathcal{T}_{\alpha }\left( \rho _{\mathrm{at}}\right) ,A\right\rangle _{%
\mathrm{evo}}=\frac{\varpi }{2\pi }\int_{0}^{\frac{2\pi }{\varpi }%
}\left\langle (\hat{\tau}_{\alpha ,0}^{(\lambda ,\eta )}-\hat{\tau}%
_{t,t-\alpha }^{(\lambda ,\eta )})(\rho _{\mathrm{at}}),A\right\rangle _{%
\mathrm{at}}\mathrm{d}t\ .  \label{super gnack2}
\end{equation}%
Therefore, we need to estimate the integrand in this last equality for any $%
\alpha \in \mathbb{R}_{0}^{+}$ and $t\in \lbrack 0,2\pi \varpi ^{-1})$. To
this end, we choose, for any $\alpha \in \mathbb{R}_{0}^{+}$, $r\left(
\alpha \right) \in 2\pi \varpi ^{-1}\mathbb{N}_{0}$ such that%
\begin{equation*}
0\leq r\left( \alpha \right) -\alpha \leq 2\pi \varpi ^{-1}\ .
\end{equation*}%
Using the $2\pi \varpi ^{-1}$--periodicity (\ref{periodic t effective}) of
the evolution family $\{\hat{\tau}_{t,s}^{(\lambda ,\eta )}\}_{t\geq s}$ we
obtain that%
\begin{equation}
\hat{\tau}_{\alpha ,0}^{(\lambda ,\eta )}-\hat{\tau}_{t,t-\alpha }^{(\lambda
,\eta )}=\hat{\tau}_{\alpha ,0}^{(\lambda ,\eta )}-\hat{\tau}_{\delta
+\alpha ,\delta }^{(\lambda ,\eta )}  \label{super gnack}
\end{equation}%
for any $t\in \lbrack 0,2\pi \varpi ^{-1}]$ with%
\begin{equation*}
\delta :=t+r\left( \alpha \right) -\alpha \in \lbrack 0,4\pi \varpi ^{-1}]\ .
\end{equation*}%
Using the cocycle property of the two--parameter family $\{\hat{\tau}%
_{t,s}^{(\lambda ,\eta )}\}_{t\geq s}$ together with (\ref{super gnack}) we
have%
\begin{equation}
\hat{\tau}_{\alpha ,0}^{(\lambda ,\eta )}-\hat{\tau}_{t,t-\alpha }^{(\lambda
,\eta )}=\hat{\tau}_{\alpha +\delta ,\delta }^{(\lambda ,\eta )}(\hat{\tau}%
_{\delta ,0}^{(\lambda ,\eta )}-\mathbf{1})+(\mathbf{1-}\hat{\tau}_{\alpha
+\delta ,\alpha }^{(\lambda ,\eta )})\hat{\tau}_{\alpha ,0}^{(\lambda ,\eta
)}  \label{ninja triangle}
\end{equation}%
with $\delta \in \lbrack 0,4\pi \varpi ^{-1}]$. Note that $\left\Vert [H_{%
\mathrm{at}},\ \cdot \ ]\right\Vert =\mathcal{O}(\varpi )$ ($\varpi
=E_{N}-E_{1}\in \sigma ([H_{\mathrm{at}},\ \cdot \ ])$) and thus, we cannot
expect the norms
\begin{equation*}
\Vert \hat{\tau}_{\delta ,0}^{(\lambda ,\eta )}-\mathbf{1}\Vert \quad
\mathrm{and}\quad \Vert \mathbf{1}-\hat{\tau}_{\alpha +\delta ,\alpha
}^{(\lambda ,\eta )}\Vert
\end{equation*}%
to be small for large pump frequencies $\varpi >>1$ when $\delta =\mathcal{O}%
(\varpi ^{-1})$.

However, the integral equation (\ref{ninja intergal eq}) implies%
\begin{equation}
\hat{\tau}_{s+\delta ,s}^{(\lambda ,\eta )}-\mathbf{1}=\hat{\tau}_{\delta
}^{(0,0)}-\mathbf{1}+\int_{s}^{s+\delta }\hat{\tau}_{s+\delta -v}^{(0,0)}%
\mathfrak{W}_{v}^{\left( \lambda ,\eta \right) }\hat{\tau}_{v,s}^{(\lambda
,\eta )}\mathrm{d}v\ ,  \label{ninja intergal eqbisbis}
\end{equation}%
for all $s\in \mathbb{R}$ and $\delta \in \lbrack 0,4\pi \varpi ^{-1}]$.
Meanwhile, for all $A\in \mathfrak{D}$,%
\begin{equation}
(\mathbf{1}-\hat{\tau}_{\delta }^{(0,0)})^{\ast }(A)=0\ ,
\label{ninja intergal eqbisbisbis}
\end{equation}%
as $(\hat{\tau}_{\delta }^{(0,0)})^{\ast }=\mathrm{e}^{-\delta \mathfrak{L}_{%
\mathrm{at}}}$ and $\mathfrak{L}_{\mathrm{at}}(\mathfrak{D})=\{0\}$. Recall
that the pump is moderate with respect to the atom--reservoir interaction,
i.e., $|\eta |\leq C\lambda ^{2}$ for some fixed constant $C\in (0,\infty )$%
, see Assumption \ref{assumption important}. Hence, since the evolution
family $\{\hat{\tau}_{t,s}^{(\lambda ,\eta )}\}_{t\geq s}$ is uniformly norm
bounded (cf. (\ref{gnackidiot})), we deduce from (\ref{ninja intergal eqbis}%
), (\ref{ninja intergal eqbisbis}) and (\ref{ninja intergal eqbisbisbis})
that%
\begin{equation}
\Vert (\mathbf{1}-\hat{\tau}_{\alpha +\delta ,\alpha }^{(\lambda ,\eta
)})^{\ast }(A)\Vert \leq C\lambda ^{2}\varpi ^{-1}\left\Vert A\right\Vert
\label{ninja intergal eqbisbisbisbis}
\end{equation}%
for some constant $C\in (0,\infty )$ not depending on $A\in \mathfrak{D}$, $%
\lambda $, $\eta $, $\varpi $, $\alpha $ and $\delta \in \lbrack 0,4\pi
\varpi ^{-1}]$. By (\ref{gnackidiot}),
\begin{equation}
\left\vert \left\langle (\mathbf{1-}\hat{\tau}_{\alpha +\delta ,\alpha
}^{(\lambda ,\eta )})\hat{\tau}_{\alpha ,0}^{(\lambda ,\eta )}(\rho _{%
\mathrm{at}}),A\right\rangle _{\mathrm{at}}\right\vert \leq C\lambda
^{2}\varpi ^{-1}\left\Vert A\right\Vert  \label{gnackidiotbis}
\end{equation}%
for some constant $C\in (0,\infty )$ not depending on $\rho _{\mathrm{at}}$,
$A$, $\lambda \in \lbrack -1,1]$, $\eta $, $\varpi $, $\alpha $ and $\delta
\in \lbrack 0,4\pi \varpi ^{-1}]$.

Similarly, as $\rho _{\mathrm{at}}\in \mathfrak{D}$ and $\left\Vert \rho _{%
\mathrm{at}}\right\Vert =1$,%
\begin{equation*}
\Vert (\hat{\tau}_{\delta ,0}^{(\lambda ,\eta )}-\mathbf{1})(\rho _{\mathrm{%
at}})\Vert \leq C\lambda ^{2}\varpi ^{-1}
\end{equation*}%
for some constant $C\in (0,\infty )$ not depending on $\rho _{\mathrm{at}}$,
$\lambda \in \lbrack -1,1]$, $\eta $, $\varpi $, and $\delta \in \lbrack
0,4\pi \varpi ^{-1}]$. Using this together with (\ref{gnackidiot}) we
conclude that
\begin{equation}
\left\vert \left\langle \hat{\tau}_{\alpha +\delta ,\delta }^{(\lambda ,\eta
)}(\hat{\tau}_{\delta ,0}^{(\lambda ,\eta )}-\mathbf{1})(\rho _{\mathrm{at}%
}),A\right\rangle _{\mathrm{at}}\right\vert \leq C\lambda ^{2}\varpi
^{-1}\left\Vert A\right\Vert  \label{gnackidiotbisbis}
\end{equation}%
for some constant $C\in (0,\infty )$ not depending on $\rho _{\mathrm{at}%
}\in \mathfrak{D}$, $A\in \mathfrak{D}$, $\lambda \in \lbrack -1,1]$, $\eta $%
, $\varpi $, $\alpha $ and $\delta \in \lbrack 0,4\pi \varpi ^{-1}]$.

From (\ref{ninja triangle}), (\ref{gnackidiotbis}) and (\ref%
{gnackidiotbisbis}) we obtain%
\begin{equation*}
\forall t\in \left[ 0,2\pi \varpi ^{-1}\right] :\qquad \left\vert
\left\langle (\hat{\tau}_{\alpha ,0}^{(\lambda ,\eta )}-\hat{\tau}%
_{t,t-\alpha }^{(\lambda ,\eta )})(\rho _{\mathrm{at}}),A\right\rangle _{%
\mathrm{at}}\right\vert \leq C\lambda ^{2}\varpi ^{-1}\left\Vert A\right\Vert
\end{equation*}%
with some constant $C\in (0,\infty )$ not depending on $\rho _{\mathrm{at}%
}\in \mathfrak{D}$, $A\in \mathfrak{D}$, $\lambda \in \lbrack -1,1]$, $\eta $%
, $\varpi $, $t$ and $\alpha $. Combining this with (\ref{super gnack2}),
estimate (\ref{longtimebehavior}) follows. \hfill {}{}$\Box $

\begin{remark}[General atomic initial states]
\mbox{ }\newline
If the density matrix $\rho _{\mathrm{at}}\in \mathfrak{H}_{\mathrm{at}}$ of
the initial state $\omega _{\mathrm{at}}$ is not block--diagonal, i.e., $%
\rho _{\mathrm{at}}\in \mathfrak{H}_{\mathrm{at}}\backslash \mathfrak{D}$,
then the assertion of Lemma \ref{lemmalongtime} holds at large times.
Indeed, the transient behavior of $\{\rho _{\mathrm{at}}\left( t\right)
\}_{t\in \mathbb{R}_{0}^{+}}$ strongly depends on the quantum correlations
of the initial atomic state, whereas its long time behavior does not depend
on the initial conditions. The following bound can be shown for arbitrary
density matrices $\rho _{\mathrm{at}}\in \mathfrak{H}_{\mathrm{at}}$ (i.e., $%
\rho _{\mathrm{at}}\in \mathfrak{D}$ is not assumed):%
\begin{equation*}
\left\vert \left\langle \rho (\alpha ),A\right\rangle _{\mathrm{at}%
}-\left\langle \mathcal{T}_{\alpha }\left( \rho _{\mathrm{at}}\right)
,A\right\rangle _{\mathrm{evo}}\right\vert \leq C(\lambda ^{2}\varpi ^{-1}+%
\mathrm{e}^{-c\alpha })\left\Vert A\right\Vert
\end{equation*}%
for some constants $c,C\in (0,\infty )$ not depending on $\rho _{\mathrm{at}%
} $, $A$, $\lambda $, $\eta $, $\varpi $, and $\alpha $. We omit the details
and illustrate this fact with a numerical example: see figure \ref%
{inversion-new}. Indeed, we focus on the study of the dynamics of
populations and the initial state is in most cases of interest the Gibbs
state $\mathfrak{g}_{\mathrm{at}}\in \mathfrak{D}$. For sake of technical
simplicity we assume that $\rho _{\mathrm{at}}\in \mathfrak{D}$ and only
observe at this point that all results below on the dynamics $\{\rho _{%
\mathrm{at}}\left( t\right) \}_{t\in \mathbb{R}_{0}^{+}}$ stay correct for
all $\rho _{\mathrm{at}}\in \mathfrak{H}_{\mathrm{at}}\backslash \mathfrak{D}
$ up to a transient factor decaying as $\mathrm{e}^{-c\alpha }$ for some $%
c>0 $.
\end{remark}

\subsection{Dimensional restriction of $\{\mathcal{T}_{\protect\alpha}\}_{%
\protect\alpha\geq0}$\label{Sectino restriction finite}}

%TCIMACRO{%
%\TeXButton{inversion-new}{\begin{figure*}[hbtp]
%\begin{center}
%\mbox{
%\leavevmode
%\subfigure
%{ \includegraphics[angle=0,scale=1,clip=true,width=12cm]{premaster-mastertestoffdiag.eps} }
%}
%\end{center}
%\caption{\emph{Illustration of the populations computed from the master equation as functions of $\alpha \equiv
%t\in \mathbb{R}_{0}^{+}$ at $\beta =0.5$, $\lambda \simeq 0.385$, $\eta
%=\lambda ^{2}\simeq 0.148$ and $\varpi =3$. Blue, green, orange, and red
%lines correspond to the populations of the 1st, the 2nd, the 3rd
%and the 4th atomic energy levels, respectively. The dashed dotted line
%marks the time $\alpha=100$ when the pump is turned off, i.e., $\eta =100$ for all times
%beyond this line. The four dashed lines are the populations computed from the pre--master equation. The figure is computed with initial matrix coefficients $(\rho _{\mathrm{at}})_{j,k}=0.5$ for $(j,k)\in \{(1,1),(4,4),(1,4),(4,1)\}$ and  $(\rho _{\mathrm{at}})_{j,k}=0$ otherwise. I.e., $\rho _{\mathrm{at}}\notin
%\mathfrak{D}$.}}
%\label{inversion-new}
%\end{figure*}}}%
%BeginExpansion
\begin{figure*}[hbtp]
\begin{center}
\mbox{
\leavevmode
\subfigure
{ \includegraphics[angle=0,scale=1,clip=true,width=12cm]{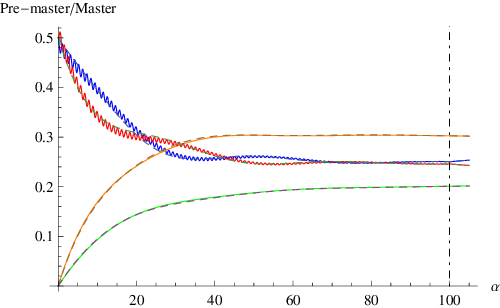} }
}
\end{center}
\caption{\emph{Illustration of the populations computed from the master equation as functions of $\alpha \equiv
t\in \mathbb{R}_{0}^{+}$ at $\beta =0.5$, $\lambda \simeq 0.385$, $\eta
=\lambda ^{2}\simeq 0.148$ and $\varpi =3$. Blue, green, orange, and red
lines correspond to the populations of the 1st, the 2nd, the 3rd
and the 4th atomic energy levels, respectively. The dashed dotted line
marks the time $\alpha=100$ when the pump is turned off, i.e., $\eta =100$ for all times
beyond this line. The four dashed lines are the populations computed from the pre--master equation. The figure is computed with initial matrix coefficients $(\rho _{\mathrm{at}})_{j,k}=0.5$ for $(j,k)\in \{(1,1),(4,4),(1,4),(4,1)\}$ and  $(\rho _{\mathrm{at}})_{j,k}=0$ otherwise. I.e., $\rho _{\mathrm{at}}\notin
\mathfrak{D}$.}}
\label{inversion-new}
\end{figure*}%
%EndExpansion
As explained at the beginning of Section \ref{sec:The-effective-atomic}, the
semigroup $\{\mathcal{T}_{\alpha }\}_{\alpha \geq 0}$ acts on an infinite
dimensional Hilbert space $\mathfrak{H}_{\mathrm{evo}}$, but the initial
conditions we are interested in are constant functions, i.e., elements of
the finite dimensional subspace $\mathfrak{H}_{\mathrm{at}}\subset \mathfrak{%
H}_{\mathrm{evo}}$. It turns out that $\mathfrak{H}_{\mathrm{at}}$ is
contained in some finite dimensional subspace $\mathfrak{H}_{\mathrm{inv}%
}^{\left( 0,0\right) }$ which is almost parallel to a finite dimensional
subspace $\mathfrak{H}_{\mathrm{inv}}^{\left( \lambda ,\eta \right) }\subset
\mathfrak{H}_{\mathrm{evo}}$. The latter is invariant with respect to $\{%
\mathcal{T}_{\alpha }\}_{\alpha \geq 0}$, see (\ref{gnack info}) below. As
this semigroup is bounded (cf. (\ref{ninja bound semi})), the restriction of
the autonomous dynamics to this invariant subspace describes -- up to small
errors -- the evolution of the solution $\{\rho (t)\}_{t\geq 0}$ of the
effective atomic master equation (\ref{effective atomic master equation}).

To define $\mathfrak{H}_{\mathrm{inv}}^{\left( \lambda ,\eta \right) }$
precisely we need some preliminary definitions. We denote by
\begin{equation}
P_{\epsilon }^{\left( \lambda ,\eta \right) }:=\frac{1}{2\pi i}%
\oint\limits_{\left\vert z+i\epsilon \right\vert =\frac{R}{4}}\big(%
z-G^{\left( \lambda ,\eta \right) }\big)^{-1}\mathrm{d}z
\label{Kato projection}
\end{equation}%
the Riesz projection \cite[Chapter II]{Kato} associated with the generator $%
G^{\left( \lambda ,\eta \right) }$ defined by (\ref{ninja bound semibis}).
Here,
\begin{equation}
R:=\min \left\{ \left\vert \epsilon -\epsilon ^{\prime }\right\vert :\
\epsilon ,\epsilon ^{\prime }\in \sigma ([H_{\mathrm{at}},\ \cdot \ ]),\
\epsilon \neq \epsilon ^{\prime }\right\} >0  \label{kato R}
\end{equation}%
and we assume that the atom--reservoir coupling $\lambda $ -- and thus $\eta
$, by Assumption \ref{assumption important} -- is sufficiently small to
ensure that the Kato projection $P_{\epsilon }^{\left( \lambda ,\eta \right)
}$ is well--defined and has the same dimension as $P_{\epsilon }^{\left(
0,0\right) }$. Then, for each $\epsilon \in \sigma ([H_{\mathrm{at}},\ \cdot
\ ])$,%
\begin{equation}
\mathfrak{H}_{\epsilon }^{\left( \lambda ,\eta \right) }:=P_{\epsilon
}^{\left( \lambda ,\eta \right) }\mathfrak{H}_{\mathrm{evo}}
\label{invariant subspace}
\end{equation}%
is an invariant, finite dimensional subspace of the evolution semigroup $\{%
\mathcal{T}_{\alpha }\}_{\alpha \geq 0}$. Consequently,%
\begin{equation}
\mathfrak{H}_{\mathrm{inv}}^{\left( \lambda ,\eta \right) }:=\mathrm{span}%
\left\{ \underset{\epsilon \in \sigma ([H_{\mathrm{at}},\cdot ])}{\bigcup }%
\mathfrak{H}_{\epsilon }^{\left( \lambda ,\eta \right) }\right\} .
\label{H inv}
\end{equation}%
is a finite dimensional invariant subspace of $\{\mathcal{T}_{\alpha
}\}_{\alpha \geq 0}$ and $\mathfrak{H}_{\mathrm{at}}\subset \mathfrak{H}_{%
\mathrm{inv}}^{\left( 0,0\right) }$ and $P_{\epsilon }^{\left( \lambda ,\eta
\right) }-P_{\epsilon }^{\left( 0,0\right) }=\mathcal{O}(\lambda ^{2})$.

As a consequence, if $\lambda $ is sufficiently small and $\varpi $ is large
enough then the restriction of the semigroup $\{\mathcal{T}_{\alpha
}\}_{\alpha \geq 0}$ to its invariant space $\mathfrak{H}_{\mathrm{inv}%
}^{\left( \lambda ,\eta \right) }$ accurately describes the time evolution
of $\{\left\langle \rho (\alpha ),A\right\rangle _{\mathrm{at}}\}_{\alpha
\geq 0}$:

\begin{lemma}[Finite dimensional effective autonomous dynamics]
\label{gnack lemma1}\mbox{ }\newline
Assume that $\rho _{\mathrm{at}}\in \mathfrak{D}$. The unique solution $%
\{\rho (t)\}_{t\geq 0}$ of the effective atomic master equation (\ref%
{effective atomic master equation}) satisfies the bound%
\begin{equation*}
\left\vert \left\langle \rho (\alpha ),A\right\rangle _{\mathrm{at}%
}-\sum\limits_{\epsilon \in \sigma ([H_{\mathrm{at}},\cdot ])}\left\langle
\exp \left( \alpha \,G^{\left( \lambda ,\eta \right) }P_{\epsilon }^{\left(
\lambda ,\eta \right) }\right) P_{\epsilon }^{\left( \lambda ,\eta \right)
}\rho _{\mathrm{at}},A\right\rangle _{\mathrm{evo}}\right\vert \leq C\lambda
^{2}\left( 1+\varpi ^{-1}\right) \left\Vert A\right\Vert
\end{equation*}%
for any (block--diagonal) $A\in \mathfrak{D}$, $\left\vert \lambda
\right\vert <<1$ and $\alpha \in \mathbb{R}_{0}^{+}$. Here, $C\in (0,\infty
) $ is a finite constant not depending on $\rho _{\mathrm{at}}$, $A$, $%
\lambda $, $\eta $, $\varpi $, and $\alpha $.
\end{lemma}

\noindent \textit{Proof.} Consider the projection
\begin{equation}
P_{\mathrm{inv}}^{\left( \lambda ,\eta \right) }:=\sum\limits_{\epsilon \in
\sigma ([H_{\mathrm{at}},\cdot ])}P_{\epsilon }^{\left( \lambda ,\eta
\right) }  \label{gnack info00}
\end{equation}%
onto the invariant subspace $\mathfrak{H}_{\mathrm{inv}}^{\left( \lambda
,\eta \right) }$. Note, that $\mathfrak{H}_{\mathrm{at}}\subset \mathfrak{H}%
_{\mathrm{inv}}^{\left( 0,0\right) }$ and
\begin{equation}
\forall \rho \in \mathfrak{H}_{\mathrm{at}}:\qquad P_{\mathrm{inv}}^{\left(
0,0\right) }\rho =\rho \ .  \label{gnack info0}
\end{equation}%
By Assumption \ref{assumption important} and Kato's perturbation theory \cite%
{Kato} for discrete eigenvalues, there is a constant $C\in (0,\infty )$ not
depending on coupling constants $\lambda $ and $\eta $ such that%
\begin{equation}
\left\Vert P_{\mathrm{inv}}^{\left( \lambda ,\eta \right) }-P_{\mathrm{inv}%
}^{\left( 0,0\right) }\right\Vert \leq C\lambda ^{2}  \label{gnack info}
\end{equation}%
at small $\lambda $. By (\ref{ninja bound semi}), we also observe that the
operator family
\begin{equation*}
\left\{ \exp \left( \alpha \,G^{\left( \lambda ,\eta \right) }P_{\epsilon
}^{\left( \lambda ,\eta \right) }\right) \right\} _{\alpha \geq 0}
\end{equation*}%
is a bounded semigroup for any eigenvalue $\epsilon \in \sigma ([H_{\mathrm{%
at}},\ \cdot \ ])$ as it is a restriction of the bounded semigroup $\{%
\mathcal{T}_{\alpha }\}_{\alpha \geq 0}$ onto the invariant subspace $%
\mathfrak{H}_{\epsilon }^{\left( \lambda ,\eta \right) }$ (\ref{invariant
subspace}). Indeed,
\begin{equation}
\forall \alpha \geq 0,\ \epsilon \in \sigma ([H_{\mathrm{at}},\ \cdot \
]):\qquad \exp \left( \alpha \,G^{\left( \lambda ,\eta \right) }P_{\epsilon
}^{\left( \lambda ,\eta \right) }\right) =\mathcal{T}_{\alpha }P_{\epsilon
}^{\left( \lambda ,\eta \right) }+(1-P_{\epsilon }^{\left( \lambda ,\eta
\right) }).  \label{gnack info1}
\end{equation}%
Hence, by (\ref{gnack info00}), (\ref{gnack info0}) and (\ref{gnack info1}),
for any $\rho \in \mathfrak{H}_{\mathrm{at}}$,
\begin{eqnarray*}
\mathcal{T}_{\alpha }\left( \rho \right) &=&\mathcal{T}_{\alpha }\left(
\left( \mathbf{1}-P_{\mathrm{inv}}^{\left( \lambda ,\eta \right) }\right)
\rho _{\mathrm{at}}\right) +\mathcal{T}_{\alpha }\left( P_{\mathrm{inv}%
}^{\left( \lambda ,\eta \right) }\rho _{\mathrm{at}}\right) \\
&=&\mathcal{T}_{\alpha }\left( \left( P_{\mathrm{inv}}^{\left( 0,0\right)
}-P_{\mathrm{inv}}^{\left( \lambda ,\eta \right) }\right) \rho \right)
+\sum\limits_{\epsilon \in \sigma ([H_{\mathrm{at}},\cdot ])}\exp \left(
\alpha \,G^{\left( \lambda ,\eta \right) }P_{\epsilon }^{\left( \lambda
,\eta \right) }\right) P_{\epsilon }^{\left( \lambda ,\eta \right) }\rho
\end{eqnarray*}%
which, combined with (\ref{ninja bound semi}) and (\ref{gnack info}), in
turn implies%
\begin{equation*}
\left\vert \left\langle \mathcal{T}_{\alpha }\left( \rho _{\mathrm{at}%
}\right) ,A\right\rangle _{\mathrm{evo}}-\sum\limits_{\epsilon \in \sigma
([H_{\mathrm{at}},\cdot ])}\left\langle \exp \left( \alpha \,G^{\left(
\lambda ,\eta \right) }P_{\epsilon }^{\left( \lambda ,\eta \right) }\right)
P_{\epsilon }^{\left( \lambda ,\eta \right) }\rho _{\mathrm{at}%
},A\right\rangle _{\mathrm{evo}}\right\vert \leq C\left\Vert A\right\Vert
\lambda ^{2}
\end{equation*}%
with $C\in (0,\infty )$ not depending on $\rho _{\mathrm{at}}$, $A$, $%
\lambda $, $\eta $, $\varpi $, and $\alpha $. The assertion follows now from
Lemma \ref{lemmalongtime}.\hfill {}$\Box $

\noindent As a consequence, we can restrict the autonomous dynamics
described by the evolution semigroup $\{\mathcal{T}_{\alpha }\}_{\alpha \geq
0}$ to the finite dimensional subspace $\mathfrak{H}_{\mathrm{inv}}^{\left(
\lambda ,\eta \right) }\subset \mathfrak{H}_{\mathrm{evo}}$.

\subsection{Effective block--diagonal dynamics\label{Section average
block--diagonal dynamics}}

Since Theorem \ref{ninja thm cool} only compares the orthogonal projections $%
P_{\mathfrak{D}}\left(\rho_{\mathrm{at}}\left(t\right)\right)$ (\ref%
{ninja0bis}) and $P_{\mathfrak{D}}\left(\rho\left(t\right)\right)$ of the
atomic density matrix $\rho_{\mathrm{at}}\left(t\right)$ and $\rho(t)$,
respectively, we are only interested in the effective block--diagonal
dynamics defined by $\{P_{\mathfrak{D}}\left(\rho\left(t\right)\right)\}_{t%
\geq0}$. As shown in the following lemma, this quantity is related to the
finite dimensional, invariant subspace $\mathfrak{H}_{0}^{\left(\lambda,\eta%
\right)}$ defined by (\ref{invariant subspace}) for $\epsilon=0\in\sigma([H_{%
\mathrm{at}},\ \cdot\ ])$, see also (\ref{definition spectre atomique}).

\begin{lemma}[Effective block--diagonal dynamics -- I]
\label{lemmalongtime copy(1)}\mbox{
}\newline
Assume that $\rho _{\mathrm{at}}\in \mathfrak{D}$. The effective
block--diagonal density matrix $\{P_{\mathfrak{D}}\left( \rho \left(
t\right) \right) \}_{t\geq 0}$ satisfies the bound%
\begin{equation*}
\left\vert \left\langle P_{\mathfrak{D}}\left( \rho (\alpha )\right)
,A\right\rangle _{\mathrm{at}}-\left\langle \exp \left( \alpha
\,P_{0}^{\left( \lambda ,\eta \right) }G^{\left( \lambda ,\eta \right)
}P_{0}^{\left( \lambda ,\eta \right) }\right) P_{0}^{\left( 0,0\right) }\rho
_{\mathrm{at}},A\right\rangle _{\mathrm{evo}}\right\vert \leq C\lambda
^{2}\left( 1+\varpi ^{-1}\right) \left\Vert A\right\Vert
\end{equation*}%
for any $A\in \mathfrak{H}_{\mathrm{at}}$, $\lambda $ sufficiently small and
$\alpha \in \mathbb{R}_{0}^{+}$. Here, $C\in (0,\infty )$ is a finite
constant not depending on $\rho _{\mathrm{at}}$, $A$, $\lambda $, $\eta $, $%
\varpi $, and $\alpha $.
\end{lemma}

\noindent \textit{Proof.} The orthogonal projection $P_{\mathfrak{D}}$ acts
in $\mathfrak{H}_{\mathrm{at}}$ and naturally induces an orthogonal
projection, again denoted by $P_{\mathfrak{D}}$, in the Hilbert space $%
\mathfrak{H}_{\mathrm{evo}}$ by:%
\begin{equation*}
\forall f\in\mathfrak{H}_{\mathrm{evo}},\
t\in\lbrack0,2\pi\varpi^{-1}):\qquad P_{\mathfrak{D}}\left(f\right)\left(t%
\right):=P_{\mathfrak{D}}\left(f(t)\right).
\end{equation*}
In particular, for any $A\in\mathfrak{H}_{\mathrm{at}}$, $%
\epsilon\in\sigma([H_{\mathrm{at}},\ \cdot\ ])$ and $\alpha\in\mathbb{R}%
_{0}^{+}$,%
\begin{eqnarray*}
& & \left\langle \exp\left(\alpha\,
G^{\left(\lambda,\eta\right)}P_{\epsilon}^{\left(\lambda,\eta\right)}%
\right)P_{\epsilon}^{\left(\lambda,\eta\right)}\rho_{\mathrm{at}},P_{%
\mathfrak{D}}\left(A\right)\right\rangle _{\mathrm{evo}} \\
& = & \left\langle P_{\mathfrak{D}}\left(P_{\epsilon}^{\left(\lambda,\eta%
\right)}\exp\left(\alpha\,
G^{\left(\lambda,\eta\right)}P_{\epsilon}^{\left(\lambda,\eta\right)}\right)%
\rho_{\mathrm{at}}\right),A\right\rangle _{\mathrm{evo}}
\end{eqnarray*}
and similarly,
\begin{equation*}
\left\langle \rho(\alpha),P_{\mathfrak{D}}\left(A\right)\right\rangle _{%
\mathrm{at}}=\left\langle P_{\mathfrak{D}}\left(\rho(\alpha)\right),A\right%
\rangle _{\mathrm{at}}\ .
\end{equation*}
With the last two equalities we use now Lemma \ref{gnack lemma1} to obtain
the bound%
\begin{equation}
\left\vert \left\langle P_{\mathfrak{D}}\left(\rho(\alpha)\right),A\right%
\rangle _{\mathrm{at}}-\left\langle P_{\mathfrak{D}}\left(P_{\epsilon}^{%
\left(\lambda,\eta\right)}\exp\left(\alpha\,
G^{\left(\lambda,\eta\right)}P_{\epsilon}^{\left(\lambda,\eta\right)}\right)%
\rho_{\mathrm{at}}\right),A\right\rangle _{\mathrm{evo}}\right\vert \leq
C\lambda^{2}\left(1+\varpi^{-1}\right)\left\Vert A\right\Vert ,
\label{lemmalongtime copy(1)-1}
\end{equation}
where $C\in(0,\infty)$ is some constant not depending on $\rho_{\mathrm{at}}$%
, $A$, $\lambda$, $\eta$, $\varpi$, and $\alpha$. For any $%
\epsilon\in\sigma([H_{\mathrm{at}},\ \cdot\ ])$, note that
\begin{equation}
P_{\mathfrak{D}}P_{\epsilon}^{\left(\lambda,\eta\right)}=P_{\mathfrak{D}%
}P_{\epsilon}^{\left(0,0\right)}+P_{\mathfrak{D}}(P_{\epsilon}^{\left(%
\lambda,\eta\right)}-P_{\epsilon}^{\left(0,0\right)})=\delta_{\epsilon,0}P_{%
\mathfrak{D}}+P_{\mathfrak{D}}(P_{\epsilon}^{\left(\lambda,\eta\right)}-P_{%
\epsilon}^{\left(0,0\right)})  \label{lemmalongtime copy(1)-2}
\end{equation}
with $\delta_{\epsilon,\epsilon^{\prime}}$ being the Kronecker delta.
Similar to (\ref{gnack info}), there is a constant $C\in(0,\infty)$ not
depending on the coupling constants $\lambda$ and $\eta$ such that%
\begin{equation}
\underset{\epsilon\in\sigma([H_{\mathrm{at}},\cdot])}{\max}\left\Vert
P_{\epsilon}^{\left(\lambda,\eta\right)}-P_{\epsilon}^{\left(0,0\right)}%
\right\Vert \leq C\lambda^{2},  \label{lemmalongtime copy(1)-3}
\end{equation}
for $\lambda$ sufficiently small. From (\ref{ninja bound semi}) and (\ref%
{lemmalongtime copy(1)-1})--(\ref{lemmalongtime copy(1)-3})
\begin{equation}
\left\vert \left\langle P_{\mathfrak{D}}\left(\rho(\alpha)\right),A\right%
\rangle _{\mathrm{at}}-\left\langle P_{\mathfrak{D}}\left(\exp\left(\alpha\,
G^{\left(\lambda,\eta\right)}P_{0}^{\left(\lambda,\eta\right)}\right)\rho_{%
\mathrm{at}}\right),A\right\rangle _{\mathrm{evo}}\right\vert \leq
C\lambda^{2}\left(1+\varpi^{-1}\right)\left\Vert A\right\Vert ,
\label{lemmalongtime copy(1)-4}
\end{equation}
with $C\in(0,\infty)$ not depending on $\rho_{\mathrm{at}}$, $A$, $\lambda$,
$\eta$, $\varpi$, and $\alpha$. Finally, for all $A\in\mathfrak{H}_{\mathrm{%
at}}$, observe that%
\begin{eqnarray*}
& & \left\langle P_{\mathfrak{D}}\left(\exp\left(\alpha\,
G^{\left(\lambda,\eta\right)}P_{0}^{\left(\lambda,\eta\right)}\right)\rho_{%
\mathrm{at}}\right),A\right\rangle _{\mathrm{evo}} \\
& = & \left\langle \exp\left(\alpha\,
G^{\left(\lambda,\eta\right)}P_{0}^{\left(\lambda,\eta\right)}\right)\rho_{%
\mathrm{at}},P_{\mathfrak{D}}\left(A\right)\right\rangle _{\mathrm{evo}} \\
& = & \left\langle \exp\left(\alpha\,
G^{\left(\lambda,\eta\right)}P_{0}^{\left(\lambda,\eta\right)}\right)\rho_{%
\mathrm{at}},P_{0}^{\left(0,0\right)}\left(A\right)\right\rangle _{\mathrm{%
evo}} \\
& = & \left\langle \exp\left(\alpha\,
P_{0}^{\left(0,0\right)}G^{\left(\lambda,\eta\right)}P_{0}^{\left(\lambda,%
\eta\right)}\right)P_{0}^{\left(0,0\right)}\rho_{\mathrm{at}},A\right\rangle
_{\mathrm{evo}}.
\end{eqnarray*}
Using (\ref{ninja bound semi}), (\ref{lemmalongtime copy(1)-3}) and (\ref%
{lemmalongtime copy(1)-4}) the assertion follows.\hfill{}{}$\Box$

The invariant spaces $\mathfrak{H}_{0}^{\left( \lambda ,\eta \right) }$ (\ref%
{invariant subspace}) associated with the projectors $P_{0}^{\left( \lambda
,\eta \right) }$ are, however, not explicit enough for practical purposes.
Therefore, the next step is to represent the effective block--diagonal
dynamics onto the explicitly known eigenspace $\mathfrak{H}_{0}^{\left(
0,0\right) }$. To this end, we denote the restriction of $G^{\left( \lambda
,\eta \right) }$ onto the space $\mathfrak{H}_{0}^{\left( 0,0\right) }$ by
\begin{equation}
\Lambda ^{\left( \lambda ,\eta \right) }:=P_{0}^{\left( 0,0\right)
}G^{\left( \lambda ,\eta \right) }P_{0}^{\left( 0,0\right) }\ .
\label{lambda}
\end{equation}%
Observe that the semigroups generated by $\Lambda ^{\left( \lambda ,\eta
\right) }$ and $P_{0}^{\left( \lambda ,\eta \right) }G^{\left( \lambda ,\eta
\right) }P_{0}^{\left( \lambda ,\eta \right) }$ are very close to each other
for small coupling constants $\lambda $ and $\eta $:

\begin{satz}[Uniform norm estimates on semigroups]
\label{lemmalongtime copy(2)}\mbox{
}\newline
For any $\varpi \in \mathbb{R}$, any $\varepsilon >0$, and any $\alpha \in
\mathbb{R}_{0}^{+}$, there is a constant $C_{\varpi ,\varepsilon }\in
(0,\infty )$ not depending on $\lambda $, $\eta $, and $\alpha $ such that%
\begin{equation*}
\left\Vert \exp \left( \alpha \Lambda ^{\left( \lambda ,\eta \right)
}\right) -\exp \left( \alpha P_{0}^{\left( \lambda ,\eta \right) }G^{\left(
\lambda ,\eta \right) }P_{0}^{\left( \lambda ,\eta \right) }\right)
\right\Vert \leq C_{\varpi ,\varepsilon }\left\vert \lambda \right\vert
^{2\left( 1-\varepsilon \right) }.
\end{equation*}
\end{satz}

\noindent The proof of Theorem \ref{lemmalongtime copy(2)} needs some
technical preparations. For the sake of clarity we defer it to Section \ref%
{Theorem blabla}. We stress that $C_{\varpi ,\varepsilon }$ can vary
considerably with the pump frequency $\varpi $.

Thus, Lemma \ref{lemmalongtime copy(1)} and Theorem \ref{lemmalongtime
copy(2)} imply that the evolution of $P_{\mathfrak{D}}\left(\rho\left(t%
\right)\right)$ can be approximated by the semigroup generated by $%
\Lambda^{\left(\lambda,\eta\right)}$:

\begin{koro}[Effective block--diagonal dynamics -- II]
\label{lemmalongtime copy(3)}\mbox{
}\newline
Assume that $\rho _{\mathrm{at}}\in \mathfrak{D}$. The effective
block--diagonal density matrix $\{P_{\mathfrak{D}}\left( \rho \left(
t\right) \right) \}_{t\geq 0}$ satisfies the bound%
\begin{equation*}
\left\vert \left\langle P_{\mathfrak{D}}\left( \rho (\alpha )\right)
,A\right\rangle _{\mathrm{at}}-\left\langle P_{0}^{\left( 0,0\right) }\exp
\left( \alpha \,\Lambda ^{\left( \lambda ,\eta \right) }\right)
P_{0}^{\left( 0,0\right) }\rho _{\mathrm{at}},A\right\rangle _{\mathrm{evo}%
}\right\vert \leq C_{\varpi ,\varepsilon }\left\vert \lambda \right\vert
^{2\left( 1-\varepsilon \right) }\left\Vert A\right\Vert ,
\end{equation*}%
for any $A\in \mathfrak{H}_{\mathrm{at}}$, $\lambda $ sufficiently small,
any $\varepsilon >0$ and $\alpha \in \mathbb{R}_{0}^{+}$. Here, $C_{\varpi
,\varepsilon }\in (0,\infty )$ is a finite constant depending on the pump
frequency $\varpi $ and $\varepsilon $ but not on $\rho _{\mathrm{at}}$, $A$%
, $\lambda $, $\eta $, and $\alpha $.
\end{koro}

In Corollary \ref{lemmalongtime copy(3)} we approximate the effective
block--diagonal dynamics by some time--evolution on the finite dimensional
subspace $\mathfrak{H}_{0}^{\left( 0,0\right) }\varsubsetneq \mathfrak{H}_{%
\mathrm{evo}}$ (\ref{invariant subspace}). This Hilbert space is not a
subspace of $\mathfrak{H}_{\mathrm{at}}\varsubsetneq \mathfrak{H}_{\mathrm{%
evo}}$ because of oscillating terms present in it, but it can be explicitly
defined as follows.

Recall that the eigenspaces of the atomic Hamiltonian $H_{\mathrm{at}}\in%
\mathcal{B}(\mathbb{C}^{d})$, associated with the eigenvalues $E_{k}$ for $%
k\in\{1,\ldots,N\}$, and their dimensions are denoted by ${\mathcal{H}}%
_{k}\subset\mathbb{C}^{d}$ and $n_{k}\in\mathbb{N}$, respectively. See
Section \ref{part.syst}. By taking any arbitrary orthonormal basis $%
\{e_{n}^{\left(k\right)}\}_{n=1}^{n_{k}}$ of ${\mathcal{H}}_{k}$ for each $%
k\in\{1,\ldots,N\}$ we define the elements
\begin{equation*}
W_{\left(k,n\right)}^{\left(k^{\prime},n^{\prime}\right)}\in\mathfrak{H}_{%
\mathrm{at}}\equiv\mathcal{B}(\mathbb{C}^{d})
\end{equation*}
for any $k,k^{\prime}\in\{1,\ldots,N\}$, $n\in\{1,\ldots,n_{k}\}$ and $%
n^{\prime}\in\{1,\ldots,n_{k^{\prime}}\}$ by%
\begin{equation}
\forall k^{\prime\prime}\in\{1,\ldots,N\},\
n^{\prime\prime}\in\{1,\ldots,n_{k^{\prime\prime}}\}:\qquad
W_{\left(k,n\right)}^{\left(k^{\prime},n^{\prime}\right)}e_{n^{\prime%
\prime}}^{\left(k^{\prime\prime}\right)}:=\delta_{n,n^{\prime\prime}}%
\delta_{k,k^{\prime\prime}}e_{n^{\prime}}^{\left(k^{\prime}\right)}\ .
\label{W1}
\end{equation}
Then, straightforward computations show that the Hilbert space $\mathfrak{H}%
_{0}^{\left(0,0\right)}$ equals%
\begin{equation*}
\mathfrak{H}_{0}^{\left(0,0\right)}=\mathrm{span}\Big\{\mathrm{e}%
^{-it\left(E_{k^{\prime}}-E_{k}\right)}W_{\left(k,n\right)}^{\left(k^{%
\prime},n^{\prime}\right)}\,|\,(k,k^{\prime})\in\mathfrak{t}_{-\varpi}\cup%
\mathfrak{t}_{0}\cup\mathfrak{t}_{\varpi},\; n\in\{1,\ldots,n_{k}\},\;
n^{\prime}\in\{1,\ldots,n_{k^{\prime}}\}\Big\}.
\end{equation*}
Recall that $\mathfrak{t}_{m\varpi}$ is the set defined by (\ref{t eps}) for
$\epsilon=m\varpi\in\sigma([H_{\mathrm{at}},\ \cdot\ ])$ and $m\in\left\{
-1,0,1\right\} $, that is explicitly,%
\begin{equation}
\mathfrak{t}_{-\varpi}=\left\{ (1,N)\right\} ,\quad\mathfrak{t}%
_{0}=\{(j,j):j\in\{1,2,\ldots,N\}\},\quad\mathfrak{t}_{\varpi}=\left\{
(N,1)\right\} ,  \label{W1bis}
\end{equation}
because $\varpi:=E_{N}-E_{1}>0$, see (\ref{pump ninja1}).

Obviously,
\begin{equation*}
\mathfrak{H}_{0}^{\left(0,0\right)}\cap\mathfrak{H}_{\mathrm{at}}=P_{%
\mathfrak{D}}\left(\mathfrak{H}_{\mathrm{at}}\right)
\end{equation*}
and $\mathfrak{H}_{0}^{\left(0,0\right)}\nsubseteq\mathfrak{H}_{\mathrm{at}}$%
. Nevertheless, we can remove the oscillating terms by defining a unitary
map $\mathrm{U}_{0}$ from $\mathfrak{H}_{0}^{\left(0,0\right)}$ to the
atomic subspace%
\begin{equation}
\mathfrak{\tilde{H}}_{0}^{\left(0,0\right)}:=\mathrm{span}\Big\{ %
W_{\left(k,n\right)}^{\left(k^{\prime},n^{\prime}\right)}\,|\,(k,k^{\prime})%
\in\mathfrak{t}_{-\varpi}\cup\mathfrak{t}_{0}\cup\mathfrak{t}_{\varpi},\;
n\in\{1,\ldots,n_{k}\},\; n^{\prime}\in\{1,\ldots,n_{k^{\prime}}\}\Big\}%
\subset\mathfrak{H}_{\mathrm{at}}  \label{H tilde}
\end{equation}
as follows:
\begin{equation}
\mathrm{U}_{0}\Big(\mathrm{e}^{-it(E_{k}-E_{k^{\prime}})}W_{\left(k,n%
\right)}^{\left(k^{\prime},n^{\prime}\right)}\Big):=W_{\left(k,n\right)}^{%
\left(k^{\prime},n^{\prime}\right)}\in\mathfrak{\tilde{H}}%
_{0}^{\left(0,0\right)}  \label{U}
\end{equation}
for any $k,k^{\prime}\in\{1,\ldots,N\}$, $n\in\{1,\ldots,n_{k}\}$ and $%
n^{\prime}\in\{1,\ldots,n_{k^{\prime}}\}$. Clearly, $P_{\mathfrak{D}}\left(%
\mathfrak{H}_{\mathrm{at}}\right)\subset\mathfrak{\tilde{H}}%
_{0}^{\left(0,0\right)}$ as $\mathrm{U}_{0}P_{\mathfrak{D}}\left(\mathfrak{H}%
_{\mathrm{at}}\right)=P_{\mathfrak{D}}\left(\mathfrak{H}_{\mathrm{at}%
}\right) $.

Hence, by Corollary \ref{lemmalongtime copy(3)}, the behavior of $P_{%
\mathfrak{D}}\left( \rho \left( \alpha \right) \right) $ can be studied
through the semigroup acting on $\mathfrak{\tilde{H}}_{0}^{\left( 0,0\right)
}\subset \mathfrak{H}_{\mathrm{at}}$ and generated by the operator%
\begin{equation}
\tilde{\Lambda}^{\left( \lambda ,\eta \right) }:=\mathrm{U}_{0}\Lambda
^{\left( \lambda ,\eta \right) }\mathrm{U}_{0}^{\ast }\ .
\label{lambda tilde}
\end{equation}%
This follows, for any initial density matrix $\rho _{\mathrm{at}}\in
\mathfrak{H}_{\mathrm{at}}$, from the equality%
\begin{eqnarray}
P_{\mathfrak{D}}\left( \exp \left( \alpha \Lambda ^{\left( \lambda ,\eta
\right) }\right) P_{0}^{\left( 0,0\right) }\left( \rho _{\mathrm{at}}\right)
\right) &=&P_{\mathfrak{D}}\left( \mathrm{U}_{0}^{\ast }\exp \left( \alpha
\tilde{\Lambda}^{\left( \lambda ,\eta \right) }\right) \mathrm{U}_{0}P_{%
\mathfrak{D}}\left( \rho _{\mathrm{at}}\right) \right)  \notag \\
&=&P_{\mathfrak{D}}\left( \exp \left( \alpha \tilde{\Lambda}^{\left( \lambda
,\eta \right) }\right) P_{\mathfrak{D}}\left( \rho _{\mathrm{at}}\right)
\right) .  \label{important equation}
\end{eqnarray}

As a first application of the above results we are now in position to study
the large time behavior of the effective block--diagonal density matrix $P_{%
\mathfrak{D}}\left( \rho \left( t\right) \right) $:

\begin{satz}[Large time behavior of $P_{\mathfrak{D}}\left( \protect\rho %
\left( t\right) \right) $]
\label{lemmalongtime copy(4)}\mbox{ }\newline
\emph{(i)} For all $\rho \in \mathfrak{\tilde{H}}_{0}^{\left( 0,0\right) }$,
\begin{equation*}
\tilde{\Lambda}^{\left( \lambda ,\eta \right) }\left( \rho \right) =\frac{%
\eta }{2}\mathfrak{L}_{\mathrm{p}}(\rho )+\lambda ^{2}\mathfrak{L}_{\mathcal{%
R}}(\rho )\in \mathfrak{\tilde{H}}_{0}^{\left( 0,0\right) }
\end{equation*}%
with $\mathfrak{L}_{\mathrm{p}}$ and $\mathfrak{L}_{\mathcal{R}}$ defined by
(\ref{L pump}) and (\ref{L R}), respectively. \newline
\emph{(ii)} Assume that $\rho _{\mathrm{at}}\in \mathfrak{D}$. There is a
unique density matrix $\tilde{\rho}_{\infty }\in \mathfrak{\tilde{H}}%
_{0}^{\left( 0,0\right) }$ such that $\tilde{\Lambda}^{\left( \lambda ,\eta
\right) }\left( \tilde{\rho}_{\infty }\right) =0$ and
\begin{equation}
\underset{\alpha \rightarrow \infty }{\lim \sup }\left\Vert P_{\mathfrak{D}%
}\left( \rho \left( \alpha \right) \right) -P_{\mathfrak{D}}\left( \tilde{%
\rho}_{\infty }\right) \right\Vert \leq C_{\varpi ,\varepsilon }\left\vert
\lambda \right\vert ^{2\left( 1-\varepsilon \right) }
\label{inequality cool}
\end{equation}%
for all $\lambda $ sufficiently small and any $\varepsilon >0$. Here, $%
C_{\varpi ,\varepsilon }\in (0,\infty )$ is a finite constant depending on $%
\varpi $ and $\varepsilon $ but not on $\rho _{\mathrm{at}}$, $\lambda $, $%
\eta $.
\end{satz}

\noindent \textit{Proof.} (i) By construction,
\begin{eqnarray*}
\left[ G^{\left( \lambda ,\eta \right) }P_{0}^{\left( 0,0\right) }\mathrm{U}%
_{0}^{\ast }\rho \right] \left( t\right) &=&-\frac{d}{dt}\left[ \mathrm{U}%
_{0}^{\ast }\rho \right] \left( t\right) +\left[ \mathrm{U}_{0}^{\ast }%
\mathfrak{L}_{\mathrm{at}}(\rho )\right] (t)+\mathfrak{L}_{t}^{\left(
\lambda ,\eta \right) }\left[ \mathrm{U}_{0}^{\ast }\rho \right] \left(
t\right) \\
&=&\frac{\eta }{2}\left( \mathrm{e}^{i\varpi t}+\mathrm{e}^{-i\varpi
t}\right) \mathfrak{L}_{\mathrm{p}}\left[ \mathrm{U}_{0}^{\ast }\rho \right]
\left( t\right) +\lambda ^{2}\mathfrak{L}_{\mathcal{R}}\left[ \mathrm{U}%
_{0}^{\ast }\rho \right] \left( t\right) \ ,
\end{eqnarray*}%
for any $\rho \in \mathfrak{\tilde{H}}_{0}^{\left( 0,0\right) }$ and $t\in %
\left[ 0,2\pi \varpi ^{-1}\right) $ a.e., see (\ref{lindblad ninja1}), (\ref%
{ninja bound semibis})--(\ref{ninja bound semibisbisbis}) and (\ref{Kato
projection})--(\ref{invariant subspace}). Using the explicit expressions (%
\ref{Effective atomic dissipation}) and (\ref{eq:Effective atomic
dissipation 2}) one easily checks%
\begin{equation*}
\left[ \mathfrak{L}_{\mathrm{at}},\mathrm{U}_{0}\right] =0\quad \mathrm{and}%
\quad \left[ \mathfrak{L}_{\mathcal{R}},\mathrm{U}_{0}\right] =0\ .
\end{equation*}%
Note that%
\begin{equation*}
-\frac{d}{dt}\left[ \mathrm{U}_{0}^{\ast }\rho \right] \left( t\right) +%
\left[ \mathrm{U}_{0}^{\ast }\mathfrak{L}_{\mathrm{at}}\rho \right] (t)=0
\end{equation*}%
for any $\rho \in \mathfrak{\tilde{H}}_{0}^{\left( 0,0\right) }$. Then, we
have%
\begin{equation*}
\left( \mathbf{1-1}\left[ \mathfrak{L}_{\mathrm{p}}=0\right] \right) \left[
\mathrm{U}_{0}^{\ast }\rho \right] \left( t\right) =\left(
\begin{array}{ccc}
P_{1,1}\left( \rho \right) & 0 & \mathrm{e}^{it\varpi }P_{1,N}\left( \rho
\right) \\
0 & 0 & 0 \\
\mathrm{e}^{-it\varpi }P_{N,1}\left( \rho \right) & 0 & P_{N,N}\left( \rho
\right)%
\end{array}%
\right) \ ,
\end{equation*}%
where $\mathbf{1}\left[ \mathfrak{L}_{\mathrm{p}}=0\right] $ is the
projection onto the kernel of $\mathfrak{L}_{\mathrm{p}}$ and $P_{j,k}$ are
the orthogonal projections%
\begin{equation}
P_{j,k}:=\underrightarrow{\mathbf{1}\left[ H_{\mathrm{at}}=E_{j}\right] }\
\underleftarrow{\mathbf{1}\left[ H_{\mathrm{at}}=E_{k}\right] }\in \mathcal{B%
}(\mathfrak{H}_{\mathrm{at}})  \label{eq:Pjk}
\end{equation}%
for all $j,k\in \{1,\dots ,N\}$. Hence, we obtain%
\begin{eqnarray*}
&&P_{0}^{\left( 0,0\right) }\left( \mathrm{e}^{it\varpi }+\mathrm{e}%
^{-it\varpi }\right) \mathfrak{L}_{\mathrm{p}}\left[ \mathrm{U}_{0}^{\ast
}\rho \right] \left( t\right) \\
&=&i\left(
\begin{array}{ccc}
P_{1,N}\left( \rho \right) h_{\mathrm{p}}-h_{\mathrm{p}}^{\ast
}P_{N,1}\left( \rho \right) & 0 & \mathrm{e}^{it\varpi }\left( P_{1,1}\left(
\rho \right) h_{\mathrm{p}}^{\ast }-h_{\mathrm{p}}^{\ast }P_{N,N}\left( \rho
\right) \right) \\
0 & 0 & 0 \\
\mathrm{e}^{-it\varpi }\left( P_{N,N}\left( \rho \right) h_{\mathrm{p}}-h_{%
\mathrm{p}}P_{1,1}\left( \rho \right) \right) & 0 & P_{N,1}\left( \rho
\right) h_{\mathrm{p}}^{\ast }-h_{\mathrm{p}}P_{1,N}\left( \rho \right)%
\end{array}%
\right) \ .
\end{eqnarray*}%
Multiplying now $\mathrm{U}_{0}$ from the left we arrive at%
\begin{equation*}
\mathrm{U}_{0}P_{0}^{\left( 0,0\right) }G^{\left( \lambda ,\eta \right)
}P_{0}^{\left( 0,0\right) }\mathrm{U}_{0}^{\ast }\rho =\frac{\eta }{2}%
\mathfrak{L}_{\mathrm{p}}\left( \rho \right) +\lambda ^{2}\mathfrak{L}_{%
\mathcal{R}}\left( \rho \right)
\end{equation*}%
for any $\rho \in \mathfrak{\tilde{H}}_{0}^{\left( 0,0\right) }$.\newline
(ii) Note that the bounded operator $\tilde{\Lambda}^{\left( \lambda ,\eta
\right) }$ given by the equality of the first assertion (i) makes sense for
all atomic density matrices $\rho \in \mathfrak{H}_{\mathrm{at}}$, whereas $%
\mathfrak{\tilde{H}}_{0}^{\left( 0,0\right) }\subset \mathfrak{H}_{\mathrm{at%
}}$ is an invariant space of $\tilde{\Lambda}^{\left( \lambda ,\eta \right)
}\in \mathcal{B}(\mathfrak{H}_{\mathrm{at}})$. By Remark \ref{remark
projection}, Assumption \ref{assumption3}, Theorem \ref{Spohn}, and
assertion (i), $\tilde{\Lambda}^{\left( \lambda ,\eta \right) }\in \mathcal{B%
}(\mathfrak{H}_{\mathrm{at}})$ is the generator of a relaxing Markov CP
semigroup for all $\lambda ,\eta \in \mathbb{R}^{2}$, $\lambda \neq 0$. See
Definition \ref{relaxing} and Remark \ref{markov CP semigroup}. I.e., there
is a unique density matrix $\tilde{\rho}_{\infty }\in \mathfrak{H}_{\mathrm{%
at}}$ such that, for any $\rho \in \mathfrak{H}_{\mathrm{at}}$,%
\begin{equation}
\underset{\alpha \rightarrow \infty }{\lim }\left( \exp \left( \alpha \tilde{%
\Lambda}^{\left( \lambda ,\eta \right) }\right) \rho \right) =\tilde{\rho}%
_{\infty }\ .  \label{limit alpha}
\end{equation}%
It follows that $\tilde{\rho}_{\infty }\in \mathfrak{H}_{\mathrm{at}}$ is
the unique density matrix satisfying $\tilde{\Lambda}^{\left( \lambda ,\eta
\right) }\left( \tilde{\rho}_{\infty }\right) =0$, see Theorem \ref{Spohn}.
As $\mathfrak{\tilde{H}}_{0}^{\left( 0,0\right) }$ is an invariant space of $%
\tilde{\Lambda}^{\left( \lambda ,\eta \right) }\in \mathcal{B}(\mathfrak{H}_{%
\mathrm{at}})$ containing density matrices, one must have $\tilde{\rho}%
_{\infty }\in \mathfrak{\tilde{H}}_{0}^{\left( 0,0\right) }$. Using (\ref%
{important equation}) and the fact that $P_{\mathfrak{D}}\left( \rho _{%
\mathrm{at}}\right) \in \mathfrak{H}_{\mathrm{at}}$ is also a density
matrix, we obtain%
\begin{equation*}
\underset{\alpha \rightarrow \infty }{\lim }P_{0}^{\left( 0,0\right) }\exp
\left( \alpha \Lambda ^{\left( \lambda ,\eta \right) }\right) P_{0}^{\left(
0,0\right) }\left( \rho _{\mathrm{at}}\right) =P_{\mathfrak{D}}\left( \tilde{%
\rho}_{\infty }\right) .
\end{equation*}%
The inequality (\ref{inequality cool}) then results from Corollary \ref%
{lemmalongtime copy(3)} and the finite dimensionality of the Hilbert space $%
\mathfrak{H}_{\mathrm{at}}$.\hfill {}{}$\Box $

\noindent It now remains to characterize more precisely the block diagonal
projection $\rho_{\infty}:=P_{\mathfrak{D}}\left(\tilde{\rho}%
_{\infty}\right)\in\mathfrak{D}$ of the density matrix $\tilde{\rho}%
_{\infty} $ of Theorem \ref{lemmalongtime copy(4)} (ii). This is done via a
\emph{balance condition for populations} in Theorem \ref{corollary
uniqueness limit density matrix} below.

We first define the orthogonal projection
\begin{equation*}
P_{\mathfrak{D}}^{\bot}:=P_{\mathfrak{\tilde{H}}_{0}^{\left(0,0\right)}}-P_{%
\mathfrak{D}}\in\mathcal{B}(\mathfrak{H}_{\mathrm{at}})
\end{equation*}
and the operator
\begin{equation*}
\mathfrak{C}:=P_{\mathfrak{D}}^{\bot}\left(\frac{\eta}{2}\mathfrak{L}_{%
\mathrm{p}}+\lambda^{2}\mathfrak{L}_{\mathcal{R}}\right)P_{\mathfrak{D}%
}^{\bot}\in\mathcal{B}(\mathfrak{H}_{\mathrm{at}})\ ,
\end{equation*}
i.e., $\mathfrak{C}=P_{\mathfrak{D}}^{\bot}\tilde{\Lambda}%
^{\left(\lambda,\eta\right)}P_{\mathfrak{D}}^{\bot}$.

The operator $\mathfrak{C}\equiv \mathfrak{C}_{\lambda }$ only depends on $%
\lambda \in \mathbb{R}$: Observe that $P_{\mathfrak{D}}$ and $P_{\mathfrak{D}%
}^{\bot }$ are orthogonal projections onto the Hilbert space $\mathfrak{%
\tilde{H}}_{0}^{\left( 0,0\right) }\subset $ $\mathfrak{H}_{\mathrm{at}}$
for which $\mathrm{ran}(P_{\mathfrak{D}})$ and $\mathrm{ran}(P_{\mathfrak{D}%
}^{\bot })$ are invariant spaces of $\mathfrak{L}_{\mathcal{R}}$. The
operator $\mathfrak{L}_{\mathrm{p}}$ however maps the subspace $\mathrm{ran}%
(P_{\mathfrak{D}})$ of block diagonal matrices to the subspace $\mathrm{ran}%
(P_{\mathfrak{D}}^{\bot })$ of off--diagonal matrices in $\mathfrak{\tilde{H}%
}_{0}^{\left( 0,0\right) }$ and vice versa. In particular,
\begin{equation}
\mathfrak{L}_{\mathrm{p}}P_{\mathfrak{D}}=P_{\mathfrak{D}}^{\bot }\mathfrak{L%
}_{\mathrm{p}}P_{\mathfrak{D}}\quad \mathrm{and}\quad P_{\mathfrak{D}}%
\mathfrak{L}_{\mathrm{p}}P_{\mathfrak{D}}=0,  \label{ine facile}
\end{equation}%
and hence,
\begin{equation*}
\mathfrak{C}\equiv \mathfrak{C}_{\lambda }=\lambda ^{2}P_{\mathfrak{D}%
}^{\bot }\mathfrak{L}_{\mathcal{R}}P_{\mathfrak{D}}^{\bot }\quad \mathrm{and}%
\quad \mathfrak{C}^{\ast }\equiv \mathfrak{C}_{\lambda }^{\ast }=\lambda
^{2}P_{\mathfrak{D}}^{\bot }\mathfrak{L}_{\mathcal{R}}^{\ast }P_{\mathfrak{D}%
}^{\bot }
\end{equation*}%
for all $\lambda ,\eta \in \mathbb{R}$.

Recall that, by Remarks \ref{remark projection}--\ref{markov CP semigroup},
Assumption \ref{assumption3} and Theorem \ref{Spohn}, the atom--reservoir
Lindbladian $\mathfrak{L}_{\mathcal{R}}$ is the generator of a relaxing
Markov CP semigroup on $\mathfrak{H}_{\mathrm{at}}$ and all non--zero
elements $p\in \sigma (\mathfrak{L}_{\mathcal{R}})\backslash \left\{
0\right\} $ have a strictly negative real part $\mathop{\rm Re}\left(
p\right) <0$. On the other hand, explicit computations show that the density
matrix $\rho _{\mathfrak{g}}=P_{\mathfrak{D}}\left( \rho _{\mathfrak{g}%
}\right) $ of the Gibbs state $\mathfrak{g}_{\mathrm{at}}$ belongs to the
kernel of $\mathfrak{L}_{\mathcal{R}}$, i.e., $\mathfrak{L}_{\mathcal{R}%
}\left( \rho _{\mathfrak{g}}\right) =0$, provided that the parameter $\beta
\in (0,\infty )$ in (\ref{Gibbs.init}) is the inverse temperature of the
reservoir. Thus,
\begin{equation*}
\ker \mathfrak{L}_{\mathcal{R}}=\mathbb{C}\cdot \mathfrak{g}_{\mathrm{at}%
}\subset \mathfrak{D}\quad \mathrm{and}\quad \ker \mathfrak{L}_{\mathcal{R}%
}^{\ast }=\mathbb{C}\cdot \mathbf{1}_{\mathbb{C}^{d}}\subset \mathfrak{D}\ .
\end{equation*}%
Note that the second equality is an obvious consequence of Theorem \ref%
{Theorem generator CP}.

It follows that the operator $\mathfrak{C}$ and its adjoint $\mathfrak{C}%
^{\ast}$ are both invertible on the subspace $\mathrm{ran}(P_{\mathfrak{D}%
}^{\bot})$. Therefore, using (\ref{ine facile}) and standard results on
Feshbach maps \cite[Theorem 2.1, Remark 2.2]{Bach Fr Sigal Chen}, we deduce
that $\rho\mapsto P_{\mathfrak{D}}(\rho)$ defines a one--to--one map from%
\begin{equation*}
\ker\tilde{\Lambda}^{\left(\lambda,\eta\right)}\cap\mathfrak{\tilde{H}}%
_{0}^{\left(0,0\right)}=\ker\left(\lambda^{2}\mathfrak{L}_{\mathcal{R}}+%
\frac{\eta}{2}\mathfrak{L}_{\mathrm{p}}\right)\cap\mathfrak{\tilde{H}}%
_{0}^{\left(0,0\right)}
\end{equation*}
to the space
\begin{equation*}
\ker\left(\lambda^{2}\mathfrak{L}_{\mathcal{R}}+\frac{\eta^{2}}{4\lambda^{2}}%
\mathfrak{B}_{\mathrm{p},\mathcal{R}}\right)\cap\mathfrak{D}\ ,
\end{equation*}
with
\begin{equation}
\mathfrak{B}_{\mathrm{p},\mathcal{R}}:=-\lambda^{2}P_{\mathfrak{D}}\mathfrak{%
L}_{\mathrm{p}}\left(\mathfrak{C}|_{\mathrm{ran}(P_{\mathfrak{D}%
}^{\bot})}\right)^{-1}\mathfrak{L}_{\mathrm{p}}P_{\mathfrak{D}}\in\mathcal{B}%
(\mathfrak{H}_{\mathrm{at}})\ .  \label{balanced pump operator0}
\end{equation}
By uniqueness of the density matrix $\tilde{\rho}_{\infty}\in\mathfrak{%
\tilde{H}}_{0}^{\left(0,0\right)}$ satisfying $\tilde{\Lambda}%
^{\left(\lambda,\eta\right)}(\tilde{\rho}_{\infty})=0$ (Theorem \ref%
{lemmalongtime copy(4)} (ii)), the following unique characterization of the
populations $\rho_{\infty}=P_{\mathfrak{D}}\left(\tilde{\rho}%
_{\infty}\right) $ holds:

\begin{satz}[Characterization of $\protect\rho_{\infty}$ via a balance
condition]
\label{corollary uniqueness limit density matrix}\mbox{
}\newline
$\rho_{\infty}=P_{\mathfrak{D}}\left(\tilde{\rho}_{\infty}\right)$ is the
unique (block diagonal) density matrix $\rho_{\infty}\in\mathfrak{D}$
satisfying the balance condition%
\begin{equation}
\mathfrak{L}_{\mathcal{R}}\left(\rho_{\infty}\right)+\frac{\eta^{2}}{%
4\lambda^{4}}\mathfrak{B}_{\mathrm{p},\mathcal{R}}\left(\rho_{\infty}%
\right)=0\ .  \label{balance conditoin}
\end{equation}
\end{satz}

As explained above, all non--zero elements $p\in \sigma (\mathfrak{L}_{%
\mathcal{R}})\backslash \left\{ 0\right\} $ have a strictly negative real
part $\mathop{\rm Re}\left( p\right) <0$. In particular, there are constants
$C,c\in \left( 0,\infty \right) $ such that%
\begin{equation}
\left\Vert \mathrm{e}^{s\mathfrak{L}_{\mathcal{R}}}P_{\mathfrak{D}}^{\bot
}\right\Vert \leq C\mathrm{e}^{-sc}\ .  \label{eq:mathfrakBconvergence}
\end{equation}%
As a consequence, by using (\ref{ine facile}) and expressing resolvents of
generator of semigroups through Laplace transform, we can rewrite the
operator $\mathfrak{B}_{\mathrm{p},\mathcal{R}}$, which describes the pump
contribution to the (quasi--) steady populations, as%
\begin{equation}
\mathfrak{B}_{\mathrm{p},\mathcal{R}}=\int_{0}^{\infty }\mathfrak{L}_{%
\mathrm{p}}\mathrm{e}^{s\mathfrak{L}_{\mathcal{R}}}\mathfrak{L}_{\mathrm{p}%
}P_{\mathfrak{D}}\ \mathrm{d}s\in \mathcal{B}(\mathfrak{H}_{\mathrm{at}})\ .
\label{balanced pump operator}
\end{equation}%
This formulation is important to get the balance condition from a dynamical
principle.

\subsection{The pre--master equation and the balance condition\label{Section
The pre--master equation}}

We analyze in the present section the time--evolution of density matrices
\begin{equation}
\rho _{\mathfrak{D}}\left( \alpha \right) :=P_{\mathfrak{D}}\exp \left(
\alpha \tilde{\Lambda}^{\left( \lambda ,\eta \right) }\right) P_{\mathfrak{D}%
}\left( \rho _{\mathrm{at}}\right) \in P_{\mathfrak{D}}(\mathfrak{\tilde{H}}%
_{0}^{\left( 0,0\right) })\ .  \label{important equationbis}
\end{equation}%
Since
\begin{equation*}
\left\Vert P_{\mathfrak{D}}\left( \rho \left( \alpha \right) \right) -\rho _{%
\mathfrak{D}}\left( \alpha \right) \right\Vert \leq C_{\varpi ,\varepsilon
}\left\vert \lambda \right\vert ^{2\left( 1-\varepsilon \right) }
\end{equation*}%
with $C_{\varpi ,\varepsilon }\in (0,\infty )$ not depending on $\rho _{%
\mathrm{at}}$, $A$, $\lambda $, $\eta $, and $\alpha $ (cf. Corollary \ref%
{lemmalongtime copy(3)} and (\ref{important equation})), the density matrix $%
\rho _{\mathfrak{D}}\left( \alpha \right) $ accurately describes the real
evolution of atomic populations at small couplings. One important
consequence of this analysis is the derivation, after Theorem \ref{The
pre--master equation}, of the balance condition (\ref{balance conditoin})
from a dynamical principle. Indeed, $\rho _{\mathfrak{D}}\left( \alpha
\right) $ satisfies an integro--differential equation, called \emph{%
pre--master equation}, and Theorem \ref{corollary uniqueness limit density
matrix} is equivalent to the fact that $\rho _{\infty }$ is the unique
stationary state of the Markov approximation of this integro--differential
equation.

\begin{satz}[The pre--master equation]
\label{The pre--master equation}\mbox{
}\newline
The family $\{\rho _{\mathfrak{D}}\left( \alpha \right) \}_{\alpha \geq
0}\subset \mathfrak{D}$ of block--diagonal density matrices obeys the
integro--differential equation%
\begin{equation}
\forall \alpha \geq 0:\hspace{0.3cm}\frac{d}{d\alpha }\rho _{\mathfrak{D}%
}\left( \alpha \right) =\lambda ^{2}\mathfrak{L}_{\mathcal{R}}\left( \rho _{%
\mathfrak{D}}\left( \alpha \right) \right) +\frac{\eta ^{2}}{4\lambda ^{2}}%
\int\limits_{0}^{\lambda ^{2}\alpha }\mathfrak{L}_{\mathrm{p}}\mathrm{e}^{s%
\mathfrak{L}_{\mathcal{R}}}\mathfrak{L}_{\mathrm{p}}\left( \rho _{\mathfrak{D%
}}\left( \alpha -s\lambda ^{-2}\right) \right) \mathrm{d}s
\label{pre master eq}
\end{equation}%
with $\rho _{\mathfrak{D}}\left( 0\right) =\rho _{\mathrm{at}}\in \mathfrak{D%
}$.
\end{satz}

\noindent \textit{Proof.} The proof of this assertion is standard (see for
instance \cite[Chapter 7]{Joosetal2003}) and is given here for completeness.
The two--fold iteration of {}``variation of constants formula%
\textquotedblright{}\ yields the equality
\begin{eqnarray}
\mathrm{e}^{\alpha\tilde{\Lambda}^{\left(\lambda,\eta\right)}} & = & \mathrm{%
e}^{\alpha\lambda^{2}\mathfrak{L}_{\mathcal{R}}}+\frac{\eta}{2}%
\int_{0}^{\alpha}\mathrm{e}^{\left(\alpha-s\right)\lambda^{2}\mathfrak{L}_{%
\mathcal{R}}}\mathfrak{L}_{\mathrm{p}}\mathrm{e}^{s\lambda^{2}\mathfrak{L}_{%
\mathcal{R}}}\mathrm{d}s  \label{ptt eq trivial} \\
& & +\frac{\eta^{2}}{4}\int_{0}^{\alpha}\mathrm{d}s_{1}\int_{0}^{s_{1}}%
\mathrm{d}s_{2}\ \mathrm{e}^{\left(\alpha-s_{1}\right)\lambda^{2}\mathfrak{L}%
_{\mathcal{R}}}\mathfrak{L}_{\mathrm{p}}\mathrm{e}^{\left(s_{1}-s_{2}\right)%
\lambda^{2}\mathfrak{L}_{\mathcal{R}}}\mathfrak{L}_{\mathrm{p}}\mathrm{e}%
^{s_{2}\tilde{\Lambda}^{\left(\lambda,\eta\right)}}.  \notag
\end{eqnarray}
Using that
\begin{equation*}
\mathfrak{L}_{\mathcal{R}}(P_{\mathfrak{D}}(\mathfrak{\tilde{H}}%
_{0}^{\left(0,0\right)}))\subset P_{\mathfrak{D}}(\mathfrak{\tilde{H}}%
_{0}^{\left(0,0\right)}),\ \mathfrak{L}_{\mathcal{R}}(P_{\mathfrak{D}%
}^{\bot}(\mathfrak{\tilde{H}}_{0}^{\left(0,0\right)}))\subset P_{\mathfrak{D}%
}^{\bot}(\mathfrak{\tilde{H}}_{0}^{\left(0,0\right)})
\end{equation*}
and%
\begin{equation*}
\mathfrak{L}_{\mathrm{p}}(P_{\mathfrak{D}}(\mathfrak{\tilde{H}}%
_{0}^{\left(0,0\right)}))\subset P_{\mathfrak{D}}^{\bot}(\mathfrak{\tilde{H}}%
_{0}^{\left(0,0\right)}),\ \mathfrak{L}_{\mathrm{p}}(P_{\mathfrak{D}}^{\bot}(%
\mathfrak{\tilde{H}}_{0}^{\left(0,0\right)}))\subset P_{\mathfrak{D}}(%
\mathfrak{\tilde{H}}_{0}^{\left(0,0\right)})
\end{equation*}
we readily deduce from (\ref{ptt eq trivial}) that%
\begin{eqnarray*}
& & P_{\mathfrak{D}}\mathrm{e}^{\alpha\tilde{\Lambda}^{\left(\lambda,\eta%
\right)}}P_{\mathfrak{D}} \\
& = & \mathrm{e}^{\alpha\lambda^{2}\mathfrak{L}_{\mathcal{R}}}P_{\mathfrak{D}%
}+\frac{\eta^{2}}{4}\int_{0}^{\alpha}\mathrm{d}s_{1}\int_{0}^{s_{1}}\mathrm{d%
}s_{2}\ \mathrm{e}^{\left(\alpha-s_{1}\right)\lambda^{2}\mathfrak{L}_{%
\mathcal{R}}}\mathfrak{L}_{\mathrm{p}}\mathrm{e}^{\left(s_{1}-s_{2}\right)%
\lambda^{2}\mathfrak{L}_{\mathcal{R}}}\mathfrak{L}_{\mathrm{p}}P_{\mathfrak{D%
}}\mathrm{e}^{s_{2}\tilde{\Lambda}^{\left(\lambda,\eta\right)}}P_{\mathfrak{D%
}}.
\end{eqnarray*}
Deriving this last equation we get
\begin{equation*}
\frac{d}{d\alpha}\rho_{\mathfrak{D}}\left(\alpha\right)=\lambda^{2}\mathfrak{%
L}_{\mathcal{R}}\left(\rho_{\mathfrak{D}}\left(\alpha\right)\right)+\frac{%
\eta^{2}}{4}\int_{0}^{\alpha}\mathfrak{L}_{\mathrm{p}}\mathrm{e}%
^{s\lambda^{2}\mathfrak{L}_{\mathcal{R}}}\mathfrak{L}_{\mathrm{p}%
}\left(\rho_{\mathfrak{D}}\left(\alpha-s\right)\right)\mathrm{d}s
\end{equation*}
from which we deduce the theorem by a\ trivial change of variable.\hfill{}$%
\Box$

By combining (\ref{limit alpha}) with the equality $\tilde{\Lambda}%
^{\left(\lambda,\eta\right)}\left(\tilde{\rho}_{\infty}\right)=0$ (Theorem %
\ref{lemmalongtime copy(4)} (ii)), the density matrix $\rho_{\mathfrak{D}%
}\left(\alpha\right)$ must converge to $P_{\mathfrak{D}}\left(\tilde{\rho}%
_{\infty}\right)$ and its derivative must vanish in the limit $%
\alpha\rightarrow\infty$. By (\ref{balanced pump operator}) and Lebesgue's
dominated convergence theorem, the limit $\varrho_{\infty}=P_{\mathfrak{D}%
}\left(\tilde{\rho}_{\infty}\right)$ must solve the balance condition\emph{\
}(\ref{balance conditoin}), as already proven in Theorem \ref{corollary
uniqueness limit density matrix}.

\begin{remark}[Moderate optical pump]
\label{remark Moderate optical pump}\mbox{
}\newline
The balance condition shows that the contribution of the pump to the final
atomic state is of order $\eta ^{2}/\lambda ^{4}$ whereas the contribution
of the atom--reservoir interaction is of order one (in the parameter $\eta
^{2}/\lambda ^{4}$). As explained in Section \ref{moderate section}, this
justifies Assumption \ref{assumption important}, that is, $|\eta |\leq
C\lambda ^{2}$. In particular, this regime follows Observation (b) given at
the beginning of Section \ref{Section def model}.
\end{remark}

\section{Generalized Einstein Coefficients and Pauli equations \label%
{sec:The-Generalized-Einstein}}

Because of Theorem \ref{corollary uniqueness limit density matrix}, we can
interpret $\lambda ^{2}\mathfrak{A}_{\mathcal{R}}$, with
\begin{equation*}
\mathfrak{A}_{\mathcal{R}}:=P_{\mathfrak{D}}\mathfrak{L}_{\mathcal{R}}P_{%
\mathfrak{D}}\ ,
\end{equation*}%
as spontaneous transitions rates and $\frac{\eta ^{2}}{4\lambda ^{2}}%
\mathfrak{B}_{\mathrm{p},\mathcal{R}}$ as effective stimulated rates between
atomic energy levels. In order to make this precise it is natural to impose
that $\mathfrak{A}_{\mathcal{R}}$ and $\mathfrak{B}_{\mathrm{p},\mathcal{R}}$
generate Markov semigroups on $\mathfrak{D}$ which preserves positivity.

The operator $\mathfrak{A}_{\mathcal{R}}$ has this property for any choice
of parameters because $\mathfrak{L}_{\mathcal{R}}$ generates a Markov CP
semigroup, which preserves the subspace $\mathfrak{D}$. Note however that
this feature is in general \emph{not} satisfied by the operator $\mathfrak{B}%
_{\mathrm{p},\mathcal{R}}$. A simple counter--example with $N=2$ and $d=3$,
where one energy level is two--fold degenerated, is given in Section \ref%
{contra exemplo}. This fact is not very surprising. Indeed, as discussed in
Section \ref{Section The pre--master equation}, the balance condition comes
from a Markov approximation of the restriction of a CP dynamics. It is
well--known that this kind of construction can destroy positivity \cite[%
Section III.1]{AlickiLendi2007}.

Hence, we will assume in this section the following:

\begin{assumption}[$\mathfrak{B}_{\mathrm{p},\mathcal{R}}$ as transition
rates]
\label{assumption Markov CP B effect}\mbox{
}\newline
The operator $\mathfrak{B}_{\mathrm{p},\mathcal{R}}$ defined by (\ref%
{balanced pump operator0}) is the restriction on $\mathfrak{D}$ of the
generator of a Markov CP semigroup on $\mathcal{B}(\mathbb{C}^{d})$ with an
invariant space $\mathfrak{D\subset}\mathcal{B}(\mathbb{C}^{d})$. In
particular, $\mathfrak{B}_{\mathrm{p},\mathcal{R}}$ generates a Markov
semigroup on $\mathfrak{D}$ which preserves positivity.
\end{assumption}

\noindent A sufficient condition on the Lindbladian $\mathfrak{L}_{\mathcal{R%
}}$ to satisfy Assumption \ref{assumption Markov CP B effect} is given by
Theorem \ref{suff cond CP B} in Section \ref{contra exemplo}. It is always
satisfied if the $1$st and the $N$th atomic energy levels are
non--degenerate. Indeed, the condition stated in Theorem \ref{suff cond CP B}
physically corresponds to the following:

\begin{itemize}
\item The pump is uniformly resonant, i.e., the reservoir--impurity
interaction does not split the spectral line corresponding to the $N$--$1$
atomic transition. This atomic spectral line may however move as a whole
under the influence of the reservoir (uniform Lamb shift).

\item The decoherence time is uniform for the $N$--$1$ correlations, i.e.,
the reservoir does not induce a splitting of $\mathrm{ran}(P_{N,1})$ in
smaller independent coherence subspaces.
\end{itemize}

Assuming from now Assumption \ref{assumption Markov CP B effect}, we are in
position to define in Section \ref{Generalized Einstein} what we call \emph{%
generalized Einstein coefficients}. These coefficients yield the Pauli
master equation and Einstein's relations, respectively described in Sections %
\ref{section Pauli master equation} and \ref{subsec:The-Generalized-Einstein}%
.

\subsection{Generalized Einstein coefficients\label{Generalized Einstein}}

Recall that the left and right multiplications are defined by (\ref{left
multi}), whereas the orthogonal projections $P_{j,k}$ are defined, for all $%
j,k\in \{1,\dots ,N\}$, by (\ref{eq:Pjk}). Clearly,
\begin{equation}
\mathbf{1}_{\mathfrak{H}_{\mathrm{at}}}=\sum_{j,k=1}^{N}P_{j,k}\qquad
\mathrm{and}\qquad \mathfrak{A}_{\mathcal{R}}=\sum_{j,k\in \{1,\dots ,N\}}%
\mathbf{A}_{j,k}\ ,  \label{eq:Pjkbis}
\end{equation}%
where, for all $j,k\in \{1,\dots ,N\}$, $\mathbf{A}_{j,k}:=P_{j,j}\mathfrak{L%
}_{\mathcal{R}}P_{k,k}$. Since $\mathfrak{A}_{\mathcal{R}}$ generates a
semigroup which always preserves positivity, for any $j,k\in \{1,\dots ,N\}$
such that $j\neq k$, the operator $A_{j,k}:=\lambda ^{2}\mathbf{A}_{j,k}$
defines a map from $\mathcal{B}^{+}(\mathcal{H}_{k})$ to $\mathcal{B}^{+}(%
\mathcal{H}_{j})$ and can hence be interpreted as the spontaneous transition
rate from the $k$th to the $j$th atomic energy level. Here, $\mathcal{B}^{+}(%
\mathfrak{h})$ denotes the set of positive operators on the Hilbert space $%
\mathfrak{h}$. The operator $A_{j,j}:=\lambda ^{2}\mathbf{A}_{j,j}$ for any $%
j\in \{1,\dots ,N\}$ is then responsible for the trace preservation of the
total dynamics generated by the operator $\lambda ^{2}\mathfrak{A}_{\mathcal{%
R}}$.

The spontaneous transition rates $\mathbf{A}_{j,k}$ can explicitly be
computed from the quantities defining the microscopic model. By (\ref{Lat}),
(\ref{L R}), (\ref{Atomic Lamb shift}), (\ref{Effective atomic dissipation})
and (\ref{eq:Effective atomic dissipation 2}), we have
\begin{eqnarray}
\mathbf{A}_{j,k} &=&2\left( 1-\delta _{j,k}\right) \sum_{\ell
=1}^{m}c_{j,k}^{(\ell )}\underrightarrow{V_{j,k}^{(\ell )}}\ \underleftarrow{%
V_{j,k}^{(\ell )\ast }}  \notag \\
&&+\delta _{j,k}\sum_{\ell =1}^{m}\left[ 2c_{k,k}^{(\ell )}\underrightarrow{%
V_{k,k}^{(\ell )}}\ \underleftarrow{V_{k,k}^{(\ell )\ast }}%
-\sum_{l=1}^{N}c_{l,k}^{(\ell )}\left( \underrightarrow{V_{l,k}^{(\ell )\ast
}V_{l,k}^{(\ell )}}+\underleftarrow{V_{l,k}^{(\ell )\ast }V_{l,k}^{(\ell )}}%
\right) \right]  \label{effective transition ratebisbis}
\end{eqnarray}%
for all $j,k\in \left\{ 1,\ldots ,N\right\} $ and with $\delta _{j,k}$ being
the Kronecker delta. With this expression, and the corresponding ones (\ref%
{effective stimulated transition rate}) for $\mathbf{B}_{j,k}$ below, we see
what role is played by the dissipative effects due to the atom--reservoir
interaction for the behavior of (quasi--) steady populations of optically
pumped atomic energy levels. This is already mentioned in Observation (a) at
the beginning of Section \ref{Section def model}.

Analogously, define%
\begin{equation}
\begin{array}{l}
\mathbf{B}_{N,1}:=-P_{N,N}\underleftarrow{h_{\mathrm{p}}^{\ast}}%
\left(P_{N,1}\left(\mathfrak{C}|_{\mathrm{ran}(P_{\mathfrak{D}%
}^{\bot})}\right)^{-1}P_{N,1}\right)\underrightarrow{h_{\mathrm{p}}}P_{1,1}
\\
\quad\quad\quad\,\,-P_{N,N}\underrightarrow{h_{\mathrm{p}}}%
\left(P_{1,N}\left(\mathfrak{C}|_{\mathrm{ran}(P_{\mathfrak{D}%
}^{\bot})}\right)^{-1}P_{1,N}\right)\underleftarrow{h_{\mathrm{p}}^{\ast}}%
P_{1,1}\ , \\
\mathbf{B}_{1,N}:=-P_{1,1}\underleftarrow{h_{\mathrm{p}}}\left(P_{1,N}\left(%
\mathfrak{C}|_{\mathrm{ran}(P_{\mathfrak{D}}^{\bot})}\right)^{-1}P_{1,N}%
\right)\underrightarrow{h_{\mathrm{p}}^{\ast}}P_{N,N} \\
\quad\quad\quad\,\,-P_{1,1}\underrightarrow{h_{\mathrm{p}}^{\ast}}%
\left(P_{N,1}\left(\mathfrak{C}|_{\mathrm{ran}(P_{\mathfrak{D}%
}^{\bot})}\right)^{-1}P_{N,1}\right)\underleftarrow{h_{\mathrm{p}}}P_{N,N}\ ,
\\
\mathbf{B}_{N,N}:=P_{N,N}\underrightarrow{h_{\mathrm{p}}}\left(P_{1,N}\left(%
\mathfrak{C}|_{\mathrm{ran}(P_{\mathfrak{D}}^{\bot})}\right)^{-1}P_{1,N}%
\right)\underrightarrow{h_{\mathrm{p}}^{\ast}}P_{N,N} \\
\quad\quad\quad\,\,+P_{N,N}\underleftarrow{h_{\mathrm{p}}^{\ast}}%
\left(P_{N,1}\left(\mathfrak{C}|_{\mathrm{ran}(P_{\mathfrak{D}%
}^{\bot})}\right)^{-1}P_{N,1}\right)\underleftarrow{h_{\mathrm{p}}}P_{N,N}\ ,
\\
\mathbf{B}_{1,1}:=P_{1,1}\underrightarrow{h_{\mathrm{p}}^{\ast}}%
\left(P_{N,1}\left(\mathfrak{C}|_{\mathrm{ran}(P_{\mathfrak{D}%
}^{\bot})}\right)^{-1}P_{N,1}\right)\underrightarrow{h_{\mathrm{p}}}P_{1,1}
\\
\quad\quad\quad\,\,+P_{1,1}\underleftarrow{h_{\mathrm{p}}}\left(P_{1,N}\left(%
\mathfrak{C}|_{\mathrm{ran}(P_{\mathfrak{D}}^{\bot})}\right)^{-1}P_{1,N}%
\right)\underleftarrow{h_{\mathrm{p}}^{\ast}}P_{1,1}\ .%
\end{array}
\label{effective stimulated transition rate}
\end{equation}
Recall that $h_{\mathrm{p}}\in\mathcal{B}(\mathbb{C}^{d})$ maps ${\mathcal{H}%
}_{1}$ to ${\mathcal{H}}_{N}$ and its kernel equals $\ker\left(h_{\mathrm{p}%
}\right)={\mathcal{H}}_{1}^{\perp}$, whereas $h_{\mathrm{p}}^{\ast}$ maps ${%
\mathcal{H}}_{N}$ to ${\mathcal{H}}_{1}$ with $\ker\left(h_{\mathrm{p}%
}^{\ast}\right)={\mathcal{H}}_{N}^{\perp}$, see Section \ref{section pump
nija}. Observe meanwhile that $\mathrm{ran}\left(P_{1,N}\right),\mathrm{ran}%
\left(P_{N,1}\right)\subset\mathrm{ran}(P_{\mathfrak{D}}^{\bot})$ are
invariant spaces of the Lindbladian $\mathfrak{L}_{\mathcal{R}}$. As a
consequence, from (\ref{pump ninja1bis}), (\ref{L pump}) and (\ref{balanced
pump operator0}),%
\begin{equation}
\mathfrak{B}_{\mathrm{p},\mathcal{R}}=\mathbf{B}_{N,1}+\mathbf{B}_{1,N}+%
\mathbf{B}_{N,N}+\mathbf{B}_{1,1}\ .  \label{operatorB}
\end{equation}

For any $j,k\in \{1,N\}$ such that $j\neq k$, the operator%
\begin{equation*}
B_{j,k}:=\frac{\eta ^{2}}{4\lambda ^{2}}\mathbf{B}_{j,k}
\end{equation*}%
maps $\mathcal{B}^{+}(\mathcal{H}_{k})$ to $\mathcal{B}^{+}(\mathcal{H}_{j})$
and is interpreted as the stimulated transition rate from the $k$th to the $%
j $th atomic energy level. Similar to the spontaneous transition rates, the
operators $B_{j,j}:=\frac{\eta ^{2}}{4\lambda ^{2}}\mathbf{B}_{j,j}$ are
such that the full dynamics generated by $\frac{\eta ^{2}}{4\lambda ^{2}}%
\mathfrak{B}_{\mathrm{p},\mathcal{R}}$ preserves traces.

The objects $A_{j,k}$, $B_{j,k}$ seen as maps $K_{k}\rightarrow K_{j}$
between cones $K_{k}\subset \mathcal{B}^{+}(\mathcal{H}_{k})$, $K_{j}\subset
\mathcal{B}^{+}(\mathcal{H}_{j})$ are called here \emph{generalized Einstein
coefficients}. They satisfy strong constraints, named here \emph{generalized
Einstein relations}, which have consequences for the structure of the
(quasi--) steady populations $\rho _{\infty }$ through the corresponding
balance condition satisfied by $\rho _{\infty }$. These relations are
discussed in Section \ref{subsec:The-Generalized-Einstein} below, after
introducing the Pauli master equation.

\subsection{The Pauli master equation\label{section Pauli master equation}}

By Theorem \ref{corollary uniqueness limit density matrix}, we can see the
unique (quasi--) steady populations $\rho _{\infty }$ as the stationary
state of the \emph{phenomenological (quantum) Pauli master equation}%
\begin{equation}
\forall t\geq 0:\hspace{0.3cm}\frac{\mathrm{d}}{\mathrm{d}t}\varrho
(t)=\lambda ^{2}\mathfrak{A}_{\mathcal{R}}\left( \varrho (t)\right) +\frac{%
\eta ^{2}}{4\lambda ^{2}}\mathfrak{B}_{\mathrm{p},\mathcal{R}}\left( \varrho
(t)\right) ,\qquad \varrho (0)=P_{\mathfrak{D}}(\rho _{\mathrm{at}})\ ,
\label{phen Pauli I}
\end{equation}%
which generalizes the classical Pauli equation for populations found in
standard textbooks on laser physics.

From Assumption \ref{assumption Markov CP B effect} and Theorem \ref{Spohn}
the unique solution $\varrho(t)=P_{\mathfrak{D}}\left(\varrho(t)\right)$ of
the Pauli master equation converges to the unique density matrix $%
\rho_{\infty}$ solution of the balance condition (\ref{balance conditoin})
as $t\rightarrow\infty$. By Theorems \ref{ninja thm cool} and \ref%
{lemmalongtime copy(4)} (ii), one thus extracts from the Pauli master
equation the correct asymptotic behavior of the atomic dynamics at small
couplings and large times:
\begin{equation*}
\underset{t\rightarrow\infty}{\lim\sup}\left\Vert P_{\mathfrak{D}%
}\left(\rho_{\mathrm{at}}\left(t\right)\right)-\varrho(t)\right\Vert \leq
C_{\varpi}\lambda^{2}
\end{equation*}
for some finite constant $C_{\varpi}\in(0,\infty)$ depending on $\varpi$ but
not on $\rho_{\mathrm{at}}$, $\lambda$, and $\eta$.

Observe also that the Pauli master equation gives the true effective
dynamics when $\eta=0$, i.e., when there is no optical pump. In this case,
the master, pre--master and Pauli equations are the same. However, for $%
\eta\neq0$, the dynamics governed by (\ref{phen Pauli I})\ is \emph{%
generically quite different} from the time evolution of the density matrix $%
\{\rho_{\mathfrak{D}}\left(\alpha\right)\}_{\alpha\geq0}$, which corresponds
-- up to small errors -- to the real atomic dynamics $\{P_{\mathfrak{D}%
}(\rho_{\mathrm{at}}(t))\}_{t\geq0}$, cf. Theorem \ref{ninja thm cool} and %
\ref{lemmalongtime copy(4)} (ii). See, for instance, the numerical examples
in Section \ref{section inv of pop}.

The reason for such a discrepancy are the memory effects related to the
stimulated processes: $\{\rho _{\mathfrak{D}}\left( \alpha \right)
\}_{\alpha \geq 0}$ is governed by a (pre--master) integro--differential
equation (\ref{pre master eq}), the Pauli master equation being its Markov
approximation. Thus, the smaller the decoherence rates and Lamb shifts of
the $N$--$1$ correlations as compared to the spontaneous transition rates $%
A_{j,k}$ of the atom, the less accurate is the dynamics given by the Pauli
master equation compared to the microscopic atomic dynamics. (The
decoherence rate and Lamb shift of the $N$--$1$ correlations correspond in
Theorem \ref{suff cond CP B} to $|\mathop{\rm Re}\{\xi _{N,1}\}|$ and $%
\mathop{\rm Im}\{\xi _{N,1}\}$, respectively.)

By Assumption \ref{assumption Markov CP B effect}, the solution of (\ref%
{phen Pauli I}) evolves in the positive cone
\begin{equation*}
\mathfrak{D}^{+}:=\mathcal{B}^{+}(\mathbb{C}^{d})\cap \mathfrak{D}=\mathrm{co%
}\left\langle \bigcup\limits_{k=1}^{N}\mathcal{B}^{+}(\mathcal{H}%
_{k})\right\rangle \ .
\end{equation*}%
Here, $\mathrm{co}\left\langle \mathfrak{m}\right\rangle $ stands for the
convex hull of the set $\mathfrak{m}$. By using the spontaneous and
stimulated atomic transition rates $A_{j,k},B_{j,k}:\mathcal{B}^{+}(\mathcal{%
H}_{k})\rightarrow \mathcal{B}^{+}(\mathcal{H}_{j})$ defined in Section \ref%
{Generalized Einstein}, the Pauli master equation (\ref{phen Pauli I})
reads, for all $j\in \left\{ 1,\ldots ,N\right\} $,%
\begin{equation}
\forall t\geq 0:\hspace{0.3cm}\dfrac{\mathrm{d}}{\mathrm{d}t}\varrho _{j}(t)=%
\underset{k=1}{\overset{N}{\sum }}\left( A_{j,k}+B_{j,k}\right) \varrho
_{k}(t),\quad \varrho _{j}(0)=\varrho _{j}\in \mathcal{B}^{+}(\mathcal{H}%
_{j})  \label{Pauli master equation}
\end{equation}%
with $\varrho _{j}(t)=P_{j,j}\left( \varrho (t)\right) $ and where $%
B_{j,k}:=0$ if $\{j,k\}\varsubsetneq \{1,N\}$.

In many situations, for instance in presence of symmetries, $A_{j,k}$ and $%
B_{j,k}$ define maps from subcones $K_{k}\subseteq \mathcal{B}^{+}(\mathcal{H%
}_{k})$ to subcones $K_{j}\subseteq \mathcal{B}^{+}(\mathcal{H}_{j})$ and we
can write the evolution equation (\ref{Pauli master equation}) with initial
conditions $\varrho _{j}(0)=\varrho _{j}\in K_{j}$ for $j\in \left\{
1,\ldots ,N\right\} $. The unique solution of this initial value problem
satisfies in this case $\{\varrho _{j}(t)\}_{t\geq 0}\subset K_{j}$ for all $%
j\in \left\{ 1,\ldots ,N\right\} $. If the initial density matrix $\varrho _{%
\mathrm{at}}\in \mathfrak{H}_{\mathrm{at}}$ is chosen such that $%
P_{j,j}\left( \varrho _{\mathrm{at}}\right) \in K_{j}$ for all $j\in \left\{
1,\ldots ,N\right\} $, the $j$th population $\varrho _{j}(t)$ then converges
to $P_{j,j}\left( \varrho _{\infty }\right) \subset K_{j}$. This system of $%
N $ differential equations is then the \emph{Pauli master equation} of the
invariant family $\left\{ K_{k}\right\} _{k=1}^{N}$ of subcones $%
K_{k}\subseteq \mathcal{B}^{+}(\mathcal{H}_{k})$.

A simple, sufficient and necessary condition on subcones $\left\{
K_{k}\right\} _{k=1}^{N}$ to ensure that $A_{j,k}$ and $B_{j,k}$ both map $%
K_{k}$ to $K_{j}$ is given in the following definition:

\begin{definition}[Invariant family of cones]
\label{Invariant family of cones}\mbox{
}\newline
A family $\left\{ K_{k}\right\} _{k=1}^{N}$ of subcones $K_{k}\subseteq%
\mathcal{B}^{+}(\mathcal{H}_{k})$ is an invariant family whenever%
\begin{equation*}
K:=\mathrm{co}\left\langle \bigcup\limits _{k=1}^{N}K_{k}\right\rangle
\end{equation*}
is invariant under the action of the semigroups $\left\{ \mathrm{e}^{t%
\mathfrak{A}_{\mathcal{R}}}\right\} _{t\geq0}$ and $\left\{ \mathrm{e}^{t%
\mathfrak{B}_{\mathrm{p},\mathcal{R}}}\right\} _{t\geq0}$.
\end{definition}

\noindent By the Trotter product formula, observe that the subset $K$
defined in this definition is also invariant under the action of the
semigroup
\begin{equation*}
\left\{ \mathrm{\exp}\left(t(\lambda^{2}\mathfrak{A}_{\mathcal{R}}+\frac{%
\eta^{2}}{4\lambda^{2}}\mathfrak{B}_{\mathrm{p},\mathcal{R}})\right)\right\}
_{t\geq0}.
\end{equation*}
Furthermore, the invariance of the family $\left\{ K_{k}\right\} _{k=1}^{N}$
yields
\begin{equation*}
\forall j,k\in\left\{ 1,\ldots,N\right\} :\quad
A_{j,k}(K_{k}),B_{j,k}(K_{k})\subset K_{j}\ .
\end{equation*}
Conversely, if a family $\left\{ K_{k}\right\} _{k=1}^{N}$ of subcones $%
K_{k}\subset\mathcal{B}^{+}(\mathcal{H}_{k})$ is such that $A_{j,k}$ and $%
B_{j,k}$ map $K_{k}$ to $K_{j}$ then $\left\{ K_{k}\right\} _{k=1}^{N}$ is
clearly an invariant family.

One trivial example of an invariant family of cones is given by taking $%
K_{k}=\mathcal{B}^{+}(\mathcal{H}_{k})$ for all $k\in \left\{ 1,\ldots
,N\right\} $. But the smaller the dimension of such cones is, the more
classical is the description of the final state via the transition rates $%
\left\{ A_{j,k}\right\} _{j,k=1}^{N}$ and $\left\{ B_{j,k}\right\}
_{j,k=1}^{N}$. It can even happen that the dimension of all subcones $%
\left\{ K_{k}\right\} _{k=1}^{N}$ can be chosen to be one. The latter
trivially happens with $K_{k}=\mathcal{B}^{+}(\mathcal{H}_{k})$ when all
atomic levels are non--degenerated, i.e., the dimension $n_{k}$ of the
eigenspace ${\mathcal{H}}_{k}$ is one for all $k\in \left\{ 1,\ldots
,N\right\} $. Note however that the non--degeneracy of \emph{all} atomic
energy level is not a necessary condition for the existence of such
one--dimensional cones and under certain circumstances the fully classical
picture of the (quasi--) steady populations $\rho _{\infty }$ is valid. In
such a case the results in this paper directly relate the coefficients
corresponding to the classical Pauli equation to microscopic quantities.

\subsection{Einstein's relations\label{subsec:The-Generalized-Einstein}}

By (\ref{effective stimulated transition rate})--(\ref{operatorB}),
\begin{equation*}
B_{j,k}=\eta^{2}F_{j,k}(\tilde{A}_{N,1},\tilde{A}_{1,N}),\quad\mathrm{with}%
\quad\tilde{A}_{j,k}:=\lambda^{2}P_{j,k}\mathfrak{L}_{\mathcal{R}}P_{j,k}\ .
\end{equation*}
In other words, $B_{j,k}$ is proportional to the intensity $\eta^{2}$ of the
pumping (monochromatic) light and proportional to a fixed function $F_{j,k}(%
\tilde{A}_{N,1},\tilde{A}_{1,N})$ of the spontaneous {}``\emph{off--diagonal}%
\textquotedblright{}\ transition rates $\tilde{A}_{N,1}$ and $\tilde{A}%
_{1,N} $. Einstein derived similar relations, called here \emph{Einstein
AB--relations}, for an atom interacting with a (broad--band, i.e.,
non--monochromatic) black--body radiation field in his seminal paper \cite%
{Einstein1916}. This was performed by using phenomenological considerations
about the expected final state of the atomic populations and the asymptotics
of the light intensity at large wave--numbers (Maxwell distribution).

Note that $F_{j,k}$ strongly depends on the specific setting. The function $%
F_{j,k}$ appearing in the present paper cannot be compared to the one
appearing in Einstein's work. However, the fact that the stimulated,
operator--value coefficients $B_{j,k}$ only depend on light intensity and
spontaneous (operator--valued) coefficients $\tilde{A}_{j,k}$ seems to be
universal. We stress that this property is rigorously derived here from a
microscopic quantum mechanical description of the system under consideration
and \emph{not} from phenomenological assumptions.

Einstein also gives in his works a relation between the stimulated
transition rates $B_{j,k}$ and $B_{k,j}$: Denoting the degeneracy of the $k$%
th atomic level by $n_{k}$, he obtained the equations%
\begin{equation*}
\forall j,k\in \left\{ 1,\ldots ,N\right\} :\qquad
n_{k}B_{j,k}=n_{j}B_{k,j}\ ,
\end{equation*}%
named here \emph{Einstein BB--relations}.

Let $p_{k}$ denote the population in the $k$th atomic level and define the
\emph{stimulated flux} from the $k$th to $j$th atomic level by $%
f_{j,k}:=B_{j,k}p_{k}$. Then the Einstein BB--relations for fluxes reads%
\begin{equation}
\forall j,k\in\left\{ 1,\ldots,N\right\} :\qquad
p_{j}n_{k}f_{j,k}-p_{k}n_{j}f_{k,j}=0\ .  \label{Einstein.flux}
\end{equation}

In our setting, a variant of BB--relations holds for fluxes, at least for
density matrices in the subcone
\begin{equation*}
K^{0}:=\mathrm{co}\left\langle
\bigcup\limits_{k=1}^{N}K_{k}^{0}\right\rangle \ ,
\end{equation*}%
where, for all $k\in \left\{ 1,\ldots ,N\right\} $,
\begin{equation*}
K_{k}^{0}:=\mathbb{R}_{0}^{+}\cdot \mathbf{1}\left[ H_{\mathrm{at}}=E_{k}%
\right] \subset \mathcal{B}^{+}(\mathcal{H}_{k})\ .
\end{equation*}

In this context and for any density matrix $\rho \in \mathfrak{H}_{\mathrm{at%
}}$, the population in the $k$th atomic level is naturally defined to be
\begin{equation*}
p_{k}(\rho )=\mathrm{Tr}_{\mathbb{C}^{d}}\left( P_{k,k}\left( \rho \right)
\right) =\mathrm{Tr}_{\mathbb{C}^{d}}(\mathbf{1}\left[ H_{\mathrm{at}}=E_{k}%
\right] \ \rho \ \mathbf{1}\left[ H_{\mathrm{at}}=E_{k}\right] )\geq 0
\end{equation*}%
for $k\in \left\{ 1,\ldots ,N\right\} $, see (\ref{population}). Similarly,
for all $j,k\in \left\{ 1,\ldots ,N\right\} $,
\begin{equation*}
f_{j,k}(\rho ):=\mathrm{Tr}_{\mathbb{C}^{d}}\left( B_{j,k}(\rho )\right) =%
\mathrm{Tr}_{\mathbb{C}^{d}}\left( B_{j,k}P_{k,k}\left( \rho \right) \right)
\end{equation*}%
represents the stimulated flux from the $k$th to the $j$th atomic level with
respect to the density matrix $\rho \in \mathfrak{H}_{\mathrm{at}}$. Then
one proves the following:

\begin{proposition}[Einstein BB--relations for states in $K^{0}$]
\label{BB--Relations}\mbox{ }\newline
For any $\rho\in K^{0}$,
\begin{equation*}
\forall j,k\in\left\{ 1,\ldots,N\right\} :\qquad
p_{j}(\rho)n_{k}f_{j,k}(\rho)-p_{k}(\rho)n_{j}f_{k,j}(\rho)=0\ .
\end{equation*}
\end{proposition}

\noindent \textit{Proof. }Clearly, for any $k\in\{1,\ldots,N\}$ and $\rho\in
K^{0}$,%
\begin{equation*}
n_{k}P_{k,k}\left(\rho\right)=p_{k}(\rho)\mathbf{1}\left[H_{\mathrm{at}%
}=E_{k}\right]\ .
\end{equation*}
As a consequence, it suffices to prove the equality%
\begin{equation}
\mathrm{Tr}_{\mathbb{C}^{d}}\left(B_{j,k}(\mathbf{1}\left[H_{\mathrm{at}%
}=E_{k}\right])\right)=\mathrm{Tr}_{\mathbb{C}^{d}}\left(B_{k,j}(\mathbf{1}%
\left[H_{\mathrm{at}}=E_{j}\right])\right)\ .  \label{equation idiote}
\end{equation}
Recall that $B_{j,k}$ equals $B_{j,k}=0$ for all $j\in\{2,\ldots,N-1\}$ or $%
k\in\{2,\ldots,N-1\}$. So, we only need to prove (\ref{equation idiote}) for
$j,k\in\{1,N\}$ and $j\neq k$. By (\ref{effective stimulated transition rate}%
),
\begin{eqnarray*}
B_{1,N}\left(\mathbf{1}\left[H_{\mathrm{at}}=E_{N}\right]\right) & = & -%
\frac{\eta^{2}}{4\lambda^{2}}\left(h_{\mathrm{p}}^{\ast}(\mathfrak{C}|_{%
\mathrm{ran}(P_{\mathfrak{D}}^{\bot})})^{-1}(h_{\mathrm{p}})+(\mathfrak{C}|_{%
\mathrm{ran}(P_{\mathfrak{D}}^{\bot})})^{-1}(h_{\mathrm{p}}^{\ast})h_{%
\mathrm{p}}\right)\ , \\
B_{N,1}\left(\mathbf{1}\left[H_{\mathrm{at}}=E_{1}\right]\right) & = & -%
\frac{\eta^{2}}{4\lambda^{2}}\left(h_{\mathrm{p}}(\mathfrak{C}|_{\mathrm{ran}%
(P_{\mathfrak{D}}^{\bot})})^{-1}(h_{\mathrm{p}}^{\ast})+(\mathfrak{C}|_{%
\mathrm{ran}(P_{\mathfrak{D}}^{\bot})})^{-1}(h_{\mathrm{p}})h_{\mathrm{p}%
}^{\ast}\right)\ .
\end{eqnarray*}
Therefore, using the cyclicity of the trace and the identity%
\begin{equation*}
(\mathfrak{C}|_{\mathrm{ran}(P_{\mathfrak{D}}^{\bot})})^{-1}(A^{\ast})=\left[%
(\mathfrak{C}|_{\mathrm{ran}(P_{\mathfrak{D}}^{\bot})})^{-1}(A)\right]%
^{\ast},\qquad A\in\mathrm{ran}(P_{\mathfrak{D}}^{\bot})\subset\mathfrak{H}_{%
\mathrm{at}},
\end{equation*}
the assertion (\ref{equation idiote}) follows for $j,k\in\{1,N\}$ and $j\neq
k$.\hfill{}$\Box$

Now, without any optical pump, i.e., for $\eta =0$, the final density matrix
$\tilde{\rho}_{\infty }$ belongs to the subcone $K^{0}$. More precisely, in
this case $\rho _{\infty }=\rho _{\mathfrak{g}}\in K^{0}$ is the atomic
Gibbs state (\ref{Gibbs.init}) with same inverse temperature $\beta $ as the
one of the reservoir. As a consequence, we infer from Proposition \ref%
{BB--Relations}\ together with Kato's perturbation theory \cite{Kato} for
non--degenerated eigenvectors that approximated Einstein BB--relations hold
for the (quasi--)steady populations $\rho _{\infty }$ for weak pumps.

\begin{koro}[Einstein BB--relations at weak optical pump]
\mbox{
}\newline
For all $(\eta ,\lambda )\in \mathbb{R}^{2}$,
\begin{equation*}
\forall j,k\in \left\{ 1,\ldots ,N\right\} :\qquad \left\vert p_{j}(\rho
_{\infty })n_{k}f_{j,k}(\rho _{\infty })-p_{k}(\rho _{\infty
})n_{j}f_{k,j}(\rho _{\infty })\right\vert \leq C_{\varpi }\frac{\eta ^{2}}{%
\lambda ^{4}}\ .
\end{equation*}%
Here, $C_{\varpi }\in (0,\infty )$ is a constant depending on $\varpi $ but
not on $j$, $k$, $\rho _{\mathrm{at}}$, $\lambda $ and $\eta $.
\end{koro}

\section{Appendix\label{sectino appendix}}

We give the proof of Theorem \ref{lemmalongtime copy(2)} in Section \ref%
{Theorem blabla}. Section \ref{section compl positive} is a short review on
completely positive (CP) semigroups by focussing on results which are
relevant for our analysis in order to facilitate the reading of the paper.
Finally, in Section \ref{contra exemplo} we discuss positivity questions
related to the interpretation as transitions rates of the coefficients of
the balance condition (\ref{balance conditoin}) characterizing uniquely the
(quasi--) steady populations.

\subsection{Proof of Theorem \protect\ref{lemmalongtime copy(2)}\label%
{Theorem blabla}}

\noindent Before starting the proof, we first extend the definitions of $%
\Lambda ^{\left( \lambda ,\eta \right) }$ (\ref{lambda}), $\mathfrak{\tilde{H%
}}_{0}^{\left( 0,0\right) }$ (\ref{H tilde}), $\mathrm{U}_{0}$ (\ref{U}),
and $\tilde{\Lambda}^{\left( \lambda ,\eta \right) }$ (\ref{lambda tilde})
to all eigenvalues $\epsilon \in \sigma ([H_{\mathrm{at}},\ \cdot \ ])$\ as
follows:%
\begin{equation*}
\mathfrak{\tilde{H}}_{\epsilon }^{\left( 0,0\right) }:=\mathrm{span}\Big\{%
W_{\left( k,n\right) }^{\left( k^{\prime },n^{\prime }\right)
}\,|\,(k,k^{\prime })\in \underset{m\in \left\{ -2,-1,0,1,2\right\} }{%
\bigcup }\mathfrak{t}_{\epsilon +m\varpi },\;n\in \{1,\ldots
,n_{k}\},\;n^{\prime }\in \{1,\ldots ,n_{k^{\prime }}\}\Big\}\ .
\end{equation*}%
The unitary operator $\mathrm{U}_{\epsilon }$ is defined by
\begin{equation*}
\mathrm{U}_{\epsilon }\Big(\mathrm{e}^{it\left( \epsilon +E_{k}-E_{k^{\prime
}}\right) }W_{\left( k,n\right) }^{\left( k^{\prime },n^{\prime }\right) }%
\Big):=W_{\left( k,n\right) }^{\left( k^{\prime },n^{\prime }\right) }\in
\mathfrak{\tilde{H}}_{\epsilon }^{\left( 0,0\right) },
\end{equation*}%
whereas%
\begin{equation*}
\Lambda _{\epsilon }^{\left( \lambda ,\eta \right) }:=P_{\epsilon }^{\left(
0,0\right) }G^{\left( \lambda ,\eta \right) }P_{\epsilon }^{\left(
0,0\right) }\qquad \mathrm{and}\qquad \tilde{\Lambda}_{\epsilon }^{\left(
\lambda ,\eta \right) }:=\mathrm{U}_{\epsilon }\Lambda _{\epsilon }^{\left(
\lambda ,\eta \right) }\mathrm{U}_{\epsilon }^{\ast }\ .
\end{equation*}%
See, e.g., (\ref{t eps}), (\ref{definition spectre atomique}) and (\ref{W1}%
). Observe that%
\begin{equation*}
\tilde{\mathfrak{H}}_{\epsilon }^{\left( 0,0\right) }\subset \mathfrak{H}_{%
\mathrm{at}},\quad \mathfrak{\tilde{H}}_{-\varpi }^{\left( 0,0\right) }=%
\mathfrak{\tilde{H}}_{0}^{\left( 0,0\right) }=\mathfrak{\tilde{H}}_{\varpi
}^{\left( 0,0\right) },
\end{equation*}%
whereas
\begin{equation}
\forall \epsilon \in \sigma ([H_{\mathrm{at}},\ \cdot \ ])\backslash
\{-\varpi ,0,\varpi \}:\qquad \mathfrak{\tilde{H}}_{0}^{\left( 0,0\right)
}\perp \mathfrak{\tilde{H}}_{\epsilon }^{\left( 0,0\right) }.
\label{assertion useful}
\end{equation}%
Using these observations we can deduce the spectral structure of the
operators $\Lambda _{\epsilon }^{\left( \lambda ,\eta \right) }$:

\begin{lemma}[Spectral properties of operators $\Lambda _{\protect\epsilon %
}^{\left( \protect\lambda ,\protect\eta \right) }$]
\label{specLambda}\mbox{ }\newline
For all $\epsilon \in \sigma ([H_{\mathrm{at}},\ \cdot \ ])$,
\begin{equation*}
\sigma (\Lambda _{\epsilon }^{\left( \lambda ,\eta \right) })\backslash
\{-i\epsilon \}\subset i\mathbb{R}{-}\mathbb{R}^{+}
\end{equation*}%
and $-i\varpi $, $0$, $i\varpi $ are simple eigenvalues of $\Lambda _{\varpi
}^{\left( \lambda ,\eta \right) }$, $\Lambda _{0}^{\left( \lambda ,\eta
\right) }$ and $\Lambda _{-\varpi }^{\left( \lambda ,\eta \right) }$,
respectively. Moreover, for all $\epsilon \in \sigma ([H_{\mathrm{at}},\
\cdot \ ])\backslash \{-\varpi ,0,\varpi \}$,
\begin{equation*}
\sigma (\Lambda _{\epsilon }^{\left( \lambda ,\eta \right) })\subset i%
\mathbb{R}{-}\mathbb{R}^{+}.
\end{equation*}%
I.e., any eigenvalue of $\Lambda _{\epsilon }^{\left( \lambda ,\eta \right)
} $ has a strictly negative real part for every $\epsilon \in \sigma ([H_{%
\mathrm{at}},\ \cdot \ ])\backslash \{-\varpi ,0,\varpi \}$.
\end{lemma}

\noindent \textit{Proof.} Similar to the proof of Theorem \ref{lemmalongtime
copy(4)} (i), we compute that%
\begin{equation}
\tilde{\Lambda}_{\epsilon}^{\left(\lambda,\eta\right)}=\left.\left(-i%
\epsilon+\frac{\eta}{2}\mathfrak{L}_{\mathrm{p}}+\lambda^{2}\mathfrak{L}_{%
\mathcal{R}}\right)\right\vert _{\mathfrak{\tilde{H}}_{\epsilon}^{\left(0,0%
\right)}}.  \label{eq}
\end{equation}
Indeed, if $\epsilon\in\{-\varpi,0,\varpi\}$ then $\mathfrak{\tilde{H}}%
_{\pm\varpi}^{\left(0,0\right)}=\mathfrak{\tilde{H}}_{0}^{\left(0,0\right)}$
and the computations for this case are exactly those given in the proof of
Theorem \ref{lemmalongtime copy(4)}. For $\epsilon\in\sigma([H_{\mathrm{at}%
},\ \cdot\ ])\backslash\{-\varpi,0,\varpi\}$ the properties (\ref{pump
ninja1})--(\ref{pump ninja1bisbis}) of the optical pump yield $\mathfrak{L}_{%
\mathrm{p}}|_{\mathfrak{\tilde{H}}_{\epsilon}^{\left(0,0\right)}}=0$ whereas
one can readily check that the Lindbladian $\mathfrak{L}_{\mathcal{R}}$
conserves the atomic subspace $\mathfrak{\tilde{H}}_{\epsilon}^{\left(0,0%
\right)}$. These properties lead to (\ref{eq}) for $\epsilon\in\sigma([H_{%
\mathrm{at}},\ \cdot\ ])\backslash\{-\varpi,0,\varpi\}$.

As a consequence,
\begin{equation}
\sigma (\Lambda _{\epsilon }^{\left( \lambda ,\eta \right) })=\sigma (\tilde{%
\Lambda}_{\epsilon }^{\left( \lambda ,\eta \right) })=-i\epsilon +\sigma %
\Big(\left. \tilde{\Lambda}^{\left( \lambda ,\eta \right) }\right\vert _{%
\mathfrak{\tilde{H}}_{\epsilon }^{\left( 0,0\right) }}\Big)\ ,  \label{idiot}
\end{equation}%
where $\tilde{\Lambda}^{\left( \lambda ,\eta \right) }$ is seen as an
operator acting on $\mathfrak{H}_{\mathrm{at}}$, see again proof of Theorem %
\ref{lemmalongtime copy(4)}. Moreover, $\mathfrak{\tilde{H}}_{\epsilon
}^{\left( 0,0\right) }$ is an invariant space of $\tilde{\Lambda}^{\left(
\lambda ,\eta \right) }\in \mathcal{B}(\mathfrak{H}_{\mathrm{at}})$. As
explained in the proof of Theorem \ref{lemmalongtime copy(4)}, the operator $%
\tilde{\Lambda}^{\left( \lambda ,\eta \right) }\in \mathcal{B}(\mathfrak{H}_{%
\mathrm{at}})$ is the generator of a relaxing, Markov CP semigroup, see
Definition \ref{relaxing}. In particular, $0$ is a non--degenerate
eigenvalue of $\tilde{\Lambda}^{\left( \lambda ,\eta \right) }$ and the
corresponding eigenvector is an element of $\mathfrak{\tilde{H}}_{\pm \varpi
}^{\left( 0,0\right) }=\mathfrak{\tilde{H}}_{0}^{\left( 0,0\right) }$. As a
consequence, from (\ref{idiot}), $-i\varpi $, $0$ and $i\varpi $ must be
non--degenerated eigenvalues of $\Lambda _{\varpi }^{\left( \lambda ,\eta
\right) }$, $\Lambda _{0}^{\left( \lambda ,\eta \right) }$ and $\Lambda
_{-\varpi }^{\left( \lambda ,\eta \right) }$, respectively.

Finally, by Theorem \ref{Spohn},
\begin{equation*}
\mathop{\rm Re}\left\{ \sigma\Big(\tilde{\Lambda}^{\left(\lambda,\eta\right)}%
\Big)\right\} \backslash\left\{ 0\right\} \subset\left(-\infty,0\right)
\end{equation*}
which, for any $\epsilon\in\sigma([H_{\mathrm{at}},\ \cdot\
])\backslash\{-\varpi,0,\varpi\}$, implies that
\begin{equation*}
\mathop{\rm Re}\left\{ \sigma\Big(\left.\tilde{\Lambda}^{\left(\lambda,\eta%
\right)}\right\vert _{\mathfrak{\tilde{H}}_{\epsilon}^{\left(0,0\right)}}%
\Big)\right\} =\mathop{\rm Re}\left\{ \sigma\Big(\left.\tilde{\Lambda}%
^{\left(\lambda,\eta\right)}\right\vert _{\mathfrak{\tilde{H}}%
_{\epsilon}^{\left(0,0\right)}}\Big)\right\} \backslash\left\{ 0\right\}
\subset\left(-\infty,0\right)
\end{equation*}
because of (\ref{assertion useful}).\hfill{}$\Box$

The proof of Theorem \ref{lemmalongtime copy(2)} needs further technical
results. The next one concerns the stability of the irreducibility of
quantum Markov chains (Assumption \ref{assumption3}) under block
localization:

\begin{lemma}[Stability of Assumption \protect\ref{assumption3} under block
localization]
\label{super Spohn}\mbox{ }\newline
Let $H\in \mathcal{B}(\mathbb{C}^{d})$ be any self--adjoint operator and
denote by $\mathbf{1}_{\varepsilon }$ the orthogonal projection onto the
eigenspace of $[H,\ \cdot \ ]$ associated with the eigenvalue $\varepsilon
\in \sigma ([H,\ \cdot \ ])$. Then, there are $\tilde{m}\in \mathbb{N}$,
non--negative real numbers $\{\tilde{c}_{j,k}^{(\tilde{\ell})}\}_{j,k,\tilde{%
\ell}}\subset \mathbb{R}^{+}$, real numbers $\{\tilde{d}_{j,k}^{(\tilde{\ell}%
)}\}_{j,k,\tilde{\ell}}\subset \mathbb{R}$, and operators $\{\tilde{V}%
_{j,k}^{(\tilde{\ell})}\}_{j,k,\tilde{\ell}}\subset \mathcal{B}(\mathbb{C}%
^{d})$ such that
\begin{equation*}
\mathfrak{\tilde{L}}_{\mathcal{R}}(\rho ):=\underset{\varepsilon \in \sigma
\left( \left[ H,\cdot \right] \right) }{\sum }\mathbf{1}_{\varepsilon }%
\mathfrak{L}_{\mathcal{R}}\mathbf{1}_{\varepsilon }=-i[\tilde{H}_{\mathrm{%
Lamb}},\ \cdot \ ]+\mathfrak{\tilde{L}}_{d}\ ,
\end{equation*}%
where%
\begin{equation*}
\tilde{H}_{\mathrm{Lamb}}:=-\frac{1}{2}\sum_{\epsilon \in \sigma ([H_{%
\mathrm{at}},\cdot ])\backslash \{0\}}\ \sum_{(j,k)\in \mathfrak{t}%
_{\epsilon }}\sum_{\tilde{\ell}=1}^{\tilde{m}}\tilde{d}_{j,k}^{(\tilde{\ell}%
)}\tilde{V}_{j,k}^{(\tilde{\ell})\ast }\tilde{V}_{j,k}^{(\tilde{\ell})}
\end{equation*}%
and, for any $\rho \in \mathfrak{H}_{\mathrm{at}}$,
\begin{equation*}
\mathfrak{\tilde{L}}_{d}\left( \rho \right) :=\frac{1}{2}\sum_{\epsilon \in
\sigma ([H_{\mathrm{at}},\cdot ])\backslash \{0\}}\sum_{(j,k)\in \mathfrak{t}%
_{\epsilon }}\sum_{\tilde{\ell}=1}^{\tilde{m}}\tilde{c}_{j,k}^{(\tilde{\ell}%
)}\left( 2\tilde{V}_{j,k}^{(\tilde{\ell})}\rho \tilde{V}_{j,k}^{(\tilde{\ell}%
)\ast }-\tilde{V}_{j,k}^{(\tilde{\ell})\ast }\tilde{V}_{j,k}^{(\tilde{\ell}%
)}\rho -\rho \tilde{V}_{j,k}^{(\tilde{\ell})\ast }\tilde{V}_{j,k}^{(\tilde{%
\ell})}\right) .
\end{equation*}%
Moreover,%
\begin{equation*}
\Big(\bigcup\limits_{\{(j,k,\tilde{\ell})\ :\ \tilde{c}_{j,k}^{(\tilde{\ell}%
)}\neq 0\}}\left\{ \tilde{V}_{j,k}^{(\tilde{\ell})}\right\} \Big)^{\prime
\prime }=\mathcal{B}(\mathbb{C}^{d})\ .
\end{equation*}%
In particular, $\mathfrak{\tilde{L}}_{\mathcal{R}}$ is the generator of a
relaxing, Markov, CP semigroup satisfying Assumption \ref{assumption3}.
\end{lemma}

\noindent \textit{Proof.} Let $\{\tilde{E}_{j}\}_{j=1}^{M}$ be the
eigenvalues of the self--adjoint operator $H$ ($M\leq d$) and%
\begin{equation*}
\mathfrak{\tilde{t}}_{\varepsilon }:=\{(j,k):\,\tilde{E}_{j}-\tilde{E}%
_{k}=\varepsilon \}\subset \{1,2,\ldots M\}\times \{1,2,\ldots M\}\ .
\end{equation*}%
For any $V\in \mathcal{B}(\mathbb{C}^{d})$ and each eigenvalue%
\begin{equation*}
\varepsilon \in \sigma ([H,\ \cdot \ ])=\{\tilde{E}_{j}-\tilde{E}_{k}:j,k\in
\{1,2,\ldots M\}\}\ ,
\end{equation*}%
we define
\begin{equation*}
V_{\varepsilon }:=\mathbf{1}_{\varepsilon }\left( V\right) =\sum_{(j,k)\in
\mathfrak{\tilde{t}}_{\varepsilon }}\mathbf{1}[H=\tilde{E}_{j}]\ V\ \mathbf{1%
}[H=\tilde{E}_{k}]\in \mathcal{B}(\mathbb{C}^{d})\ .
\end{equation*}%
By construction, note that
\begin{equation}
\underset{\varepsilon \in \sigma \left( \left[ H,\cdot \right] \right) }{%
\sum }V_{\varepsilon }=V\ .  \label{trivial}
\end{equation}%
In particular, one has
\begin{equation}
V\in \mathrm{span\ }\Big\{\bigcup\limits_{\varepsilon \in \sigma \left( %
\left[ H,\cdot \right] \right) }V_{\varepsilon }\Big\}.  \label{trivial2}
\end{equation}%
For $(j,k)\in \mathfrak{t}_{\epsilon }$, $\epsilon \in \sigma ([H_{\mathrm{at%
}},\cdot ])\backslash \{0\}$, and $\varepsilon \in \sigma ([H,\ \cdot \ ])$
let%
\begin{equation}
\tilde{V}_{j,k}^{(\ell ,\varepsilon )}:=\mathbf{1}_{\varepsilon
}(V_{j,k}^{(\ell )})\ ,\qquad \tilde{c}_{j,k}^{(\ell ,\varepsilon
)}:=c_{j,k}^{(\ell )}\ ,\qquad \tilde{d}_{j,k}^{(\ell ,\varepsilon
)}:=d_{j,k}^{(\ell )}\ .  \label{trivial2bis}
\end{equation}%
By identifying the finite sets $\{1,2,\ldots m\}\times \sigma ([H,\ \cdot \
] $ and $\{1,2,\ldots \tilde{m}\}$, we infer from (\ref{trivial2}) and
Assumption \ref{assumption3} that the family $\{\tilde{V}_{j,k}^{(\tilde{\ell%
})}\}_{j,k,\tilde{\ell}}\subset \mathcal{B}(\mathbb{C}^{d})$ of operators
satisfies
\begin{equation}
\Big(\bigcup\limits_{\{(j,k,\tilde{\ell})\ :\ \tilde{c}_{j,k}^{(\tilde{\ell}%
)}\neq 0\}}\left\{ \tilde{V}_{j,k}^{(\tilde{\ell})}\right\} \Big)^{\prime
\prime }=\mathcal{B}(\mathbb{C}^{d})\ .  \label{assumption B}
\end{equation}

For all eigenvectors $A_{\varepsilon _{1}},B_{\varepsilon _{2}}\in \mathcal{B%
}(\mathbb{C}^{d})$ associated with eigenvalues $\varepsilon _{1},\varepsilon
_{2}\in \sigma ([H,\ \cdot \ ])$ respectively, i.e., $\mathbf{1}%
_{\varepsilon _{1}}\left( A\right) =A_{\varepsilon _{1}}$ and $\mathbf{1}%
_{\varepsilon _{2}}\left( B\right) =B_{\varepsilon _{2}}$, let
\begin{equation*}
D:=A_{\varepsilon _{1}}B_{\varepsilon _{2}}\in \mathcal{B}(\mathbb{C}^{d})\ .
\end{equation*}%
If $D\neq 0$ then we observe that $(\varepsilon _{1}+\varepsilon _{2})\in
\sigma ([H,\ \cdot \ ])$ and $\mathbf{1}_{\varepsilon _{1}+\varepsilon
_{2}}\left( D\right) =D$, i.e., $D=D_{\varepsilon _{1}+\varepsilon _{2}}$
must be an eigenvector associated with the eigenvalue $\varepsilon
_{1}+\varepsilon _{2}$. Therefore, by using (\ref{trivial}) and the equality
$\mathbf{1}_{-\varepsilon }\left( V^{\ast }\right) =V_{\varepsilon }^{\ast }$%
, we deduce that%
\begin{eqnarray}
\underset{\varepsilon \in \sigma \left( \left[ H,\cdot \right] \right) }{%
\sum }\mathbf{1}_{\varepsilon }\left( V\mathbf{1}_{\varepsilon }\left( \rho
\right) V^{\ast }\right) &=&\underset{\varepsilon \in \sigma \left( \left[
H,\cdot \right] \right) }{\sum }\mathbf{1}_{\varepsilon }\left( \underset{%
\tilde{\varepsilon}\in \sigma \left( \left[ H,\cdot \right] \right) }{\sum }%
V_{\tilde{\varepsilon}}\mathbf{1}_{\varepsilon }\left( \rho \right) \underset%
{\hat{\varepsilon}\in \sigma \left( \left[ H,\cdot \right] \right) }{\sum }%
\mathbf{1}_{\hat{\varepsilon}}\left( V^{\ast }\right) \right)  \notag \\
&=&\underset{\varepsilon \in \sigma \left( \left[ H,\cdot \right] \right) }{%
\sum }\text{ }\underset{\tilde{\varepsilon}\in \sigma \left( \left[ H,\cdot %
\right] \right) }{\sum }V_{\tilde{\varepsilon}}\mathbf{1}_{\varepsilon
}\left( \rho \right) \mathbf{1}_{-\tilde{\varepsilon}}\left( V^{\ast }\right)
\notag \\
&=&\underset{\tilde{\varepsilon}\in \sigma \left( \left[ H,\cdot \right]
\right) }{\sum }V_{\tilde{\varepsilon}}\rho V_{\tilde{\varepsilon}}^{\ast }
\label{trivial3}
\end{eqnarray}%
for any $\rho \in \mathfrak{H}_{\mathrm{at}}\equiv \mathcal{B}(\mathbb{C}%
^{d})$. Similarly,
\begin{eqnarray}
\underset{\varepsilon \in \sigma \left( \left[ H,\cdot \right] \right) }{%
\sum }\mathbf{1}_{\varepsilon }\left( V^{\ast }V\mathbf{1}_{\varepsilon
}\left( \rho \right) \right) &=&\underset{\tilde{\varepsilon}\in \sigma
\left( \left[ H,\cdot \right] \right) }{\sum }V_{\tilde{\varepsilon}}^{\ast
}V_{\tilde{\varepsilon}}\rho \ , \\
\underset{\varepsilon \in \sigma \left( \left[ H,\cdot \right] \right) }{%
\sum }\mathbf{1}_{\varepsilon }\left( \mathbf{1}_{\varepsilon }\left( \rho
\right) V^{\ast }V\right) &=&\underset{\tilde{\varepsilon}\in \sigma \left( %
\left[ H,\cdot \right] \right) }{\sum }\rho V_{\tilde{\varepsilon}}^{\ast
}V_{\tilde{\varepsilon}}  \label{trivial4}
\end{eqnarray}%
for any $\rho \in \mathfrak{H}_{\mathrm{at}}$. Therefore, by using (\ref{L R}%
), (\ref{Atomic Lamb shift})--(\ref{eq:Effective atomic dissipation 2}), (%
\ref{trivial2bis}) and (\ref{trivial3})--(\ref{trivial4}) with the
identification of the finite sets $\{1,2,\ldots m\}\times \sigma ([H,\ \cdot
\ ]$ and $\{1,2,\ldots \tilde{m}\}$ we obtain the explicit decomposition of $%
\mathfrak{\tilde{L}}_{\mathcal{R}}$ stated in the lemma. By Theorems \ref%
{Theorem generator CP} and \ref{Spohn} together with (\ref{assumption B}),
it is then straightforward to verify that $\mathfrak{\tilde{L}}_{\mathcal{R}%
} $ is the generator of a relaxing, Markov CP semigroup.\hfill {}$\Box $

We need more precise information about the behavior of the spectral gap
given in Lemma \ref{specLambda} with respect to the coupling constants $\eta$
and $\lambda$. This is achieved by using the last lemma:

\begin{lemma}[Behavior of the spectral gap of $\Lambda ^{\left( \protect%
\lambda ,\protect\eta \right) }$]
\label{g(0,eta)}\mbox{ }\newline
For all $(\lambda ,\eta )\in \mathbb{R}\times \mathbb{R}$,
\begin{equation*}
\min \left\{ \ \left\vert \mathrm{\mathop{\rm Re}}\left\{ p\right\}
\right\vert :p\in \sigma (\Lambda ^{\left( \lambda ,\eta \right)
})\backslash \{0\}\right\} \geq C_{\varpi }\lambda ^{2}
\end{equation*}%
with $C_{\varpi }\in (0,\infty )$ being a constant depending on $\varpi $
but not on $\lambda $, $\eta $.
\end{lemma}

\noindent \textit{Proof.} We define the function
\begin{equation*}
g(\lambda ,\eta ):=\lambda ^{-2}\min \left\{ \ \left\vert \mathrm{%
\mathop{\rm Re}}\left\{ p\right\} \right\vert :p\in \sigma (\Lambda ^{\left(
\lambda ,\eta \right) })\backslash \{0\}\right\}
\end{equation*}%
on the set $\mathbb{R}{\backslash \{0\}}\times \mathbb{R}$. Observe that $%
g(\lambda ,\eta )$ only depends on the ratio $\varkappa :=\eta /\lambda ^{2}$
and is strictly positive, by Theorems \ref{Theorem generator CP} and \ref%
{Spohn}. Indeed, by the proof of Lemma \ref{specLambda},%
\begin{equation*}
g(\lambda ,\eta )=\min \left\{ \ \left\vert \mathrm{\mathop{\rm Re}}\left\{
p\right\} \right\vert :p\in \sigma \left( \frac{\eta }{2\lambda ^{2}}%
\mathfrak{L}_{\mathrm{p}}+\mathfrak{L}_{\mathcal{R}}\right) \backslash
\{0\}\right\} .
\end{equation*}%
Furthermore, by Remark \ref{markov CP semigroup}, $\varkappa \mathfrak{L}_{%
\mathrm{p}}/2+\mathfrak{L}_{\mathcal{R}}$ is the generator of a Markov CP
semigroup satisfying Assumption \ref{assumption3} for any $\varkappa \in
\mathbb{R}$ (see also Remark \ref{remark projection}) and Theorem \ref{Spohn}
yields that this semigroup must be relaxing.

By Kato's perturbation theory \cite{Kato}, for some constants $C,c\in
(0,\infty )$, $g(\lambda ,\eta )\geq C$ whenever $\eta \leq c\lambda ^{2}$,
i.e., when $\varkappa \leq c$. Using again Kato's perturbation theory \cite%
{Kato} and Theorem \ref{Spohn}, $\varkappa \mapsto g(\varkappa ^{-\frac{1}{2}%
},1)$ is a strictly positive continuous function on the interval $%
[c,c^{\prime }]$ for any finite constant $c^{\prime }>c$. By compactness of
the interval $[c,c^{\prime }]$, it follows that
\begin{equation*}
\min \left\{ g(\varkappa ^{-\frac{1}{2}},1)\,|\,\varkappa \in \lbrack
c,c^{\prime }]\right\} >0\ .
\end{equation*}%
So, it remains to prove that $g(\lambda ,\eta )\geq C$ whenever $\eta
>c^{\prime }\lambda ^{2}$, i.e., when $\varkappa >c^{\prime }$, for some
constant $C\in \left( 0,\infty \right) $ and sufficiently large $c^{\prime
}<\infty $. By (\ref{idiot}) for $\lambda ,\epsilon =0$, note that
\begin{equation*}
\sigma (\Lambda ^{\left( 0,\eta \right) })\subset i\mathbb{R}\ .
\end{equation*}%
Thus, by Kato's perturbation theory \cite{Kato} for the spectrum of $%
\varkappa \mathfrak{L}_{\mathrm{p}}/2+\mathfrak{L}_{\mathcal{R}}$, the limit%
\begin{equation}
\lim_{\varkappa \rightarrow \infty }g(\varkappa ^{-\frac{1}{2}},1)\in
\lbrack 0,\infty )  \label{definition de gnack g}
\end{equation}%
exists and satisfies
\begin{equation}
\lim_{\varkappa \rightarrow \infty }g(\varkappa ^{-\frac{1}{2}},1)\geq \min
\left\{ \left\vert \mathrm{\mathop{\rm Re}}\left\{ p\right\} \right\vert
:p\in \sigma (\mathfrak{\tilde{L}}_{\mathcal{R}})\backslash \{0\}\right\} \ ,
\label{gnack inequality}
\end{equation}%
where
\begin{equation*}
\mathfrak{\tilde{L}}_{\mathcal{R}}:=\sum\limits_{\varepsilon \in \frac{\eta
}{2}\sigma \left( \left[ H_{\mathrm{p}},\cdot \right] \right) }\mathbf{1}%
_{\varepsilon }\mathfrak{L}_{\mathcal{R}}\mathbf{1}_{\varepsilon }\ .
\end{equation*}%
Here, $\mathbf{1}_{\varepsilon }$ denotes the spectral projection of $[H_{%
\mathrm{p}},\ \cdot \ ]$ onto the eigenspace associated with an eigenvalue $%
\varepsilon \in \sigma ([H_{\mathrm{p}},\ \cdot \ ])$. Using Lemma \ref%
{super Spohn}, $\mathfrak{\tilde{L}}_{\mathcal{R}}$ is the generator of a
relaxing, Markov, CP semigroup satisfying Assumption \ref{assumption3} and,
by (\ref{gnack inequality}) and Theorem \ref{Spohn},
\begin{equation*}
\lim_{\varkappa \rightarrow \infty }g(\varkappa ^{-\frac{1}{2}},1)\in
(0,\infty )\ .
\end{equation*}%
In other words, for some constants $C\in \left( 0,\infty \right) $ and
sufficiently large $c^{\prime }>c$, $g(\varkappa ^{-\frac{1}{2}},1)\geq C$
for all $\varkappa >c^{\prime }$.\hfill {}$\Box $

We now use Lemma \ref{g(0,eta)} to obtain norm estimates on the difference
of the uniformly bounded semigroups $\{\mathrm{e}^{\alpha\Lambda^{\left(%
\lambda,\eta\right)}}\}_{\alpha\geq0}$ and $\{\mathrm{e}^{\alpha
P_{0}^{\left(\lambda,\eta\right)}G^{\left(\lambda,\eta\right)}P_{0}^{\left(%
\lambda,\eta\right)}}\}_{\alpha\geq0}$.

\begin{lemma}[Semigroup estimates]
\label{corollary de g(0,eta)}\mbox{
}\newline
For sufficiently small $\lambda $, there are constants $C_{\varpi
},c_{\varpi }\in (0,\infty )$ depending on $\varpi $ but not on $\lambda $, $%
\eta $, $\varpi $ and $\alpha $ such that%
\begin{equation*}
\forall \alpha \geq 0:\qquad \Vert \mathrm{e}^{\alpha \Lambda ^{\left(
\lambda ,\eta \right) }}-\mathrm{e}^{\alpha P_{0}^{\left( \lambda ,\eta
\right) }G^{\left( \lambda ,\eta \right) }P_{0}^{\left( \lambda ,\eta
\right) }}\Vert \leq C_{\varpi }(\lambda ^{2}+\mathrm{e}^{-\alpha \lambda
^{2}c_{\varpi }})\ .
\end{equation*}
\end{lemma}

\noindent \textit{Proof. }Note that the continuous semigroup $\{\mathrm{e}%
^{\alpha \Lambda ^{\left( \lambda ,\eta \right) }}\}_{\alpha \geq 0}$ on $%
\mathfrak{H}_{\mathrm{at}}$ can be represented through the inverse Laplace
transform of the resolvent of its generator. Indeed, by Lemma \ref%
{specLambda} combined with \cite[Proof of Corollary 5.15]{Nagel-engel},
\begin{equation}
\mathrm{e}^{\alpha \Lambda ^{\left( \lambda ,\eta \right) }}-\mathrm{E}%
\mathbf{=}\underset{L\rightarrow \infty }{\lim }\left\{ \frac{1}{2\pi i}%
\int_{w-iL}^{w+iL}\mathrm{e}^{\alpha z}\left( (z-\Lambda ^{\left( \lambda
,\eta \right) })^{-1}-\frac{1}{z}\right) \mathrm{d}z\right\} +\mathbf{1}-%
\mathrm{E}  \label{inequality correct1}
\end{equation}%
for any $w>0$ and with%
\begin{equation}
\mathrm{E}:=\frac{1}{2\pi i}\oint\limits_{\left\vert z\right\vert =R^{\prime
}}\mathrm{e}^{\alpha z}(z-\Lambda ^{\left( \lambda ,\eta \right) })^{-1}%
\mathrm{d}z=\frac{1}{2\pi i}\oint\limits_{\left\vert z\right\vert =R^{\prime
}}(z-\Lambda ^{\left( \lambda ,\eta \right) })^{-1}\mathrm{d}z
\label{Kato projection lambda}
\end{equation}%
and $R^{\prime }>0$ sufficiently small. Note that $0$ is a simple eigenvalue
of $\Lambda ^{\left( \lambda ,\eta \right) }$ and the last equality follows
from the fact that the map%
\begin{equation*}
z\mapsto (1-\mathrm{e}^{\alpha z})(z-\Lambda ^{\left( \lambda ,\eta \right)
})^{-1}
\end{equation*}%
is holomorphic near $z=0$. By (\ref{kato R}) and Lemma \ref{g(0,eta)}, the
Kato projection $\mathrm{E}$ associated with the generator $\Lambda ^{\left(
\lambda ,\eta \right) }$ is well--defined for sufficiently small $R^{\prime
}\in \left( 0,c\lambda ^{2}\right) $ at fixed $\lambda \neq 0$. Using again
the spectral properties of $\Lambda ^{\left( \lambda ,\eta \right) }$ given
in Lemma \ref{g(0,eta)}, we can push the integration path of the complex
integral in (\ref{inequality correct1}) to sufficiently small, but a
strictly negative real part $w^{\prime }=-c\lambda ^{2}<0$ as follows:
\begin{eqnarray*}
&&\underset{L\rightarrow \infty }{\lim }\left\{ \frac{1}{2\pi i}%
\int_{w-iL}^{w+iL}\mathrm{e}^{\alpha z}\left( (z-\Lambda ^{\left( \lambda
,\eta \right) })^{-1}-\frac{1}{z}\right) \mathrm{d}z\right\} \\
&=&\underset{L\rightarrow \infty }{\lim }\left\{ \frac{1}{2\pi i}%
\int_{-c\lambda ^{2}-iL}^{-c\lambda ^{2}+iL}\mathrm{e}^{\alpha z}\left(
(z-\Lambda ^{\left( \lambda ,\eta \right) })^{-1}-\frac{1}{z}\right) \mathrm{%
d}z\right\} \\
&&+\frac{1}{2\pi i}\oint\limits_{\left\vert z\right\vert =R^{\prime }}%
\mathrm{e}^{\alpha z}(z-\Lambda ^{\left( \lambda ,\eta \right) })^{-1}%
\mathrm{d}z-\frac{1}{2\pi i}\oint\limits_{\left\vert z\right\vert =R^{\prime
}}\frac{\mathrm{e}^{\alpha z}}{z}\mathrm{d}z \\
&=&\underset{L\rightarrow \infty }{\lim }\left\{ \frac{1}{2\pi i}%
\int_{-c\lambda ^{2}-iL}^{-c\lambda ^{2}+iL}\frac{\mathrm{e}^{\alpha z}}{z}%
(z-\Lambda ^{\left( \lambda ,\eta \right) })^{-1}\Lambda ^{\left( \lambda
,\eta \right) }\mathrm{d}z\right\} +\mathrm{E}-\mathbf{1}\ .
\end{eqnarray*}%
Note that $c\in (0,\infty )$ is a sufficiently small constant depending on $%
\varpi $ but not on on $\lambda $, $\eta $, and $\alpha $. By (\ref%
{inequality correct1}), it follows that%
\begin{equation}
\mathrm{e}^{\alpha \Lambda ^{\left( \lambda ,\eta \right) }}-\mathrm{E}%
\mathbf{=}\underset{L\rightarrow \infty }{\lim }\left\{ \lambda ^{-2}\frac{1%
}{2\pi i}\int_{-L}^{L}\frac{\mathrm{e}^{\alpha \left( ix-c\right) \lambda
^{2}}}{\left( ix-c\right) }(\left( ix-c\right) -\lambda ^{-2}\Lambda
^{\left( \lambda ,\eta \right) })^{-1}\Lambda ^{\left( \lambda ,\eta \right)
}\mathrm{d}x\right\}  \label{laplace transform inverse}
\end{equation}%
for all $\alpha \in \mathbb{R}_{0}^{+}$. Observe that we have additionally
used an obvious change of variable in the last equation to extract the
factor $\lambda ^{-2}$. Indeed, by (\ref{eq})--(\ref{idiot}) and Assumption %
\ref{assumption important} ($|\eta |\leq C\lambda ^{2}$), we have $\Vert
\Lambda ^{\left( \lambda ,\eta \right) }\Vert =\mathcal{O}(\lambda ^{2})$
and the limit (\ref{laplace transform inverse}) yields the upper bound
\begin{equation}
\Vert \mathrm{e}^{\alpha \Lambda ^{\left( \lambda ,\eta \right) }}-\mathrm{E}%
\Vert \leq \mathrm{e}^{-c\alpha \lambda ^{2}}\underset{L\rightarrow \infty }{%
\lim }\left\{ \frac{1}{2\pi i}\int_{-L}^{L}\frac{1}{\left\Vert
ix-c\right\Vert }\left\Vert (\left( ix-c\right) -\lambda ^{-2}\Lambda
^{\left( \lambda ,\eta \right) })^{-1}\right\Vert \mathrm{d}x\right\}
\label{idiot encore0}
\end{equation}%
for all $\alpha \in \mathbb{R}_{0}^{+}$. Hence we need to bound the
integrand of the last integral by some integrable function not depending on
the parameters $\lambda ,\eta $. To this end, first observe that
\begin{equation}
\lambda ^{-2}\Lambda ^{\left( \lambda ,\eta \right) }=\mathrm{U}_{0}^{\ast
}\left( \mathcal{L}_{\mathcal{R}}+\frac{\varkappa }{2}\mathfrak{L}_{\mathrm{p%
}}\right) \mathrm{U}_{0}  \label{idiot encore}
\end{equation}%
with the ratio $\varkappa :=\eta /\lambda ^{2}\in \lbrack -\varkappa
_{0},\varkappa _{0}]$ for some fixed $\varkappa _{0}\in (0,\infty )$. See (%
\ref{lambda tilde}), Theorem \ref{lemmalongtime copy(4)} (i) and Assumption %
\ref{assumption important}. Therefore, we define the map $\tilde{g}:\mathbb{R%
}\times \lbrack -\varkappa _{0},\varkappa _{0}]\rightarrow \mathbb{R}$ by%
\begin{equation}
\tilde{g}(x,\varkappa ):=\left\Vert \left( (ix-c)-\mathcal{L}_{\mathcal{R}%
}-\varkappa \mathfrak{L}_{\mathrm{p}}\right) ^{-1}\right\Vert <\infty \ .
\label{idiot encorebis}
\end{equation}%
This function is well--defined for sufficiently small $c>0$ because of Lemma %
\ref{g(0,eta)}. Moreover, since%
\begin{equation*}
\left\vert \tilde{g}(x_{1},\varkappa _{1})-\tilde{g}(x_{2},\varkappa
_{2})\right\vert \leq \left\vert \tilde{g}(x_{1},\varkappa _{1})\right\vert
\ \left\vert \tilde{g}(x_{2},\varkappa _{2})\right\vert \left( \left\Vert
x_{1}-x_{2}\right\Vert +\left\Vert \varkappa _{1}-\varkappa _{2}\right\Vert
\left\Vert \mathfrak{L}_{\mathrm{p}}\right\Vert \right) \ ,
\end{equation*}%
the function $\tilde{g}$ is (locally Lipschitz) continuous on its domain of
definition. By compactness, for any $L\in \mathbb{R}_{0}^{+}$ there is a
constant $C\in (0,\infty )$ depending on $\varpi $ but not on $\lambda $, $%
\eta $ such that
\begin{equation}
\underset{(x,\varkappa )\in \left[ -L,L\right] \times \lbrack -\varkappa
_{0},\varkappa _{0}]}{\sup }\tilde{g}(x,\varkappa )\leq C\ .
\label{Neumann series idiot}
\end{equation}%
On the other hand, the Neumann series
\begin{equation*}
((ix-c)-\mathcal{L}_{\mathcal{R}}-\varkappa \mathfrak{L}_{\mathrm{p}})^{-1}=%
\underset{n=0}{\overset{\infty }{\sum }}(ix-c)^{-\left( n+1\right) }\left\{
\mathcal{L}_{\mathcal{R}}+\varkappa \mathfrak{L}_{\mathrm{p}}\right\} ^{n}
\end{equation*}%
implies that, for sufficiently large $x>>1$ and $\varkappa \in \lbrack
-\varkappa _{0},\varkappa _{0}]$,
\begin{equation*}
\tilde{g}(x,\varkappa )\leq \underset{n=0}{\overset{\infty }{\sum }}%
\left\Vert ix-c\right\Vert ^{-\left( n+1\right) }\left\{ \left\Vert \mathcal{%
L}_{\mathcal{R}}\right\Vert +\left\vert \varkappa _{0}\right\vert \left\Vert
\mathfrak{L}_{\mathrm{p}}\right\Vert \right\} ^{n}
\end{equation*}%
because $\mathcal{L}_{\mathcal{R}}$ and $\mathfrak{L}_{\mathrm{p}}$ are
bounded operators on $\mathfrak{H}_{\mathrm{at}}$. As a consequence, there
is a sufficiently large constant $C\in (0,\infty )$ depending on $\varpi $
but not on $\lambda $, $\eta $ such that
\begin{equation}
\forall x\in \mathbb{R},\ \varkappa \in \lbrack -\varkappa _{0},\varkappa
_{0}]:\qquad \tilde{g}(x,\varkappa )\leq \min \left\{ C,\frac{2}{\left\Vert
ix-c\right\Vert }\right\} \ .  \label{Neumann series idiotbis}
\end{equation}%
By (\ref{idiot encore0})--(\ref{Neumann series idiotbis}), we conclude the
existence of a constant $C\in (0,\infty )$ depending on $\varpi $ but not on
$\lambda $, $\eta $ such that
\begin{equation}
\forall \alpha \geq 0:\qquad \Vert \mathrm{e}^{\alpha \Lambda ^{\left(
\lambda ,\eta \right) }}-\mathrm{E}\Vert \leq C\mathrm{e}^{-c\alpha \lambda
^{2}}.  \label{final1}
\end{equation}

Meanwhile, in the same way we obtain (\ref{idiot encore0}), we derive the
upper bound%
\begin{eqnarray}
& & \left\Vert \mathrm{e}^{\alpha
P_{0}^{\left(\lambda,\eta\right)}G^{\left(\lambda,\eta\right)}P_{0}^{\left(%
\lambda,\eta\right)}}-\mathrm{\tilde{E}}\right\Vert  \label{idiot encore0bis}
\\
& \leq & \mathrm{e}^{-c\alpha\lambda^{2}}\underset{L\rightarrow\infty}{\lim}%
\left\{ \frac{1}{2\pi i}\int_{-L}^{L}\frac{1}{\left\Vert ix-c\right\Vert }%
\left\Vert
(\left(ix-c\right)-\lambda^{-2}P_{0}^{\left(\lambda,\eta\right)}G^{\left(%
\lambda,\eta\right)}P_{0}^{\left(\lambda,\eta\right)})^{-1}\right\Vert
\mathrm{d}x\right\} .  \notag
\end{eqnarray}
Here,
\begin{equation}
\mathrm{\tilde{E}}:=\frac{1}{2\pi i}\oint\limits _{\left\vert z\right\vert
=R^{\prime}}\mathrm{e}^{\alpha
z}(z-P_{0}^{\left(\lambda,\eta\right)}G^{\left(\lambda,\eta\right)}P_{0}^{%
\left(\lambda,\eta\right)})^{-1}\mathrm{d}z
\end{equation}
is an operator associated with the generator $P_{0}^{\left(\lambda,\eta%
\right)}G^{\left(\lambda,\eta\right)}P_{0}^{\left(\lambda,\eta\right)}$. The
latter is well--defined for sufficiently small $R^{\prime}\in\left(0,c%
\lambda^{2}\right)$ at fixed $\lambda\neq0$.

Indeed, for $\left\vert \lambda\right\vert <<1$\ and some constant $%
C\in(0,\infty)$ depending on $\varpi$ but not on $\lambda$, $\eta$, observe
that
\begin{equation*}
\Vert
P_{0}^{\left(\lambda,\eta\right)}G^{\left(\lambda,\eta\right)}\Vert\leq
C\lambda^{2}\quad\mathrm{and}\quad\Vert
G^{\left(\lambda,\eta\right)}P_{0}^{\left(0,0\right)}\Vert\leq C\lambda^{2},
\end{equation*}
using $P_{0}^{\left(0,0\right)}G^{\left(0,0\right)}=0$. Hence, combining
these last upper bounds with (\ref{lemmalongtime copy(1)-3}) and the
triangle inequality we get
\begin{equation}
\Vert
P_{0}^{\left(\lambda,\eta\right)}G^{\left(\lambda,\eta\right)}P_{0}^{\left(%
\lambda,\eta\right)}-\Lambda^{\left(\lambda,\eta\right)}\Vert=\Vert
P_{0}^{\left(\lambda,\eta\right)}G^{\left(\lambda,\eta\right)}P_{0}^{\left(%
\lambda,\eta\right)}-P_{0}^{\left(0,0\right)}G^{\left(\lambda,\eta%
\right)}P_{0}^{\left(0,0\right)}\Vert\leq C\lambda^{4}  \label{gnack vite 20}
\end{equation}
with $C\in(0,\infty)$ not depending on $\lambda$, $\eta$ and $\varpi$. In
particular, the spectral properties of $\Lambda^{\left(\lambda,\eta\right)}$
given in Lemma \ref{g(0,eta)} together with (\ref{gnack vite 20}) implies
that $\mathrm{\tilde{E}}$ is well--defined for a sufficiently small, but
strictly positive $R^{\prime}\in\left(0,c\lambda^{2}\right)$ and satisfy%
\begin{equation}
\Vert\mathrm{\tilde{E}}-\mathrm{E}\Vert\leq C\lambda^{4}\oint\limits
_{\left\vert z\right\vert
=R^{\prime}}\Vert(z-\Lambda^{\left(\lambda,\eta\right)})^{-1}\Vert%
\Vert(z-P_{0}^{\left(\lambda,\eta\right)}G^{\left(\lambda,\eta%
\right)}P_{0}^{\left(\lambda,\eta\right)})^{-1}\Vert\mathrm{d}z\leq
C\lambda^{2}  \label{correctbisbis}
\end{equation}
for some constant $C\in(0,\infty)$ not depending on the (sufficiently small
coupling) constants $\lambda,\eta$.

To prove the inequalities (\ref{idiot encore0bis}) and (\ref{correctbisbis}%
), note that the (non--degenerate) eigenvalue $0$ is the unique element of
the spectrum of $\Lambda ^{\left( \lambda ,\eta \right) }$ within the disc
of radius $R^{\prime }$, by Lemma \ref{g(0,eta)}. By (\ref{gnack vite 20})
and Kato's perturbation theory \cite{Kato} of discrete eigenvalues, there is
a unique (non--degenerate) eigenvalue of $P_{0}^{\left( \lambda ,\eta
\right) }G^{\left( \lambda ,\eta \right) }P_{0}^{\left( \lambda ,\eta
\right) }$ and no other element of its spectrum within the disc of radius $%
R^{\prime }$, provided that $\lambda $ is sufficiently small. In fact, this
eigenvalue is also $0$ because $G^{\left( \lambda ,\eta \right) }$ is a
closed operator and the constant function $\Omega _{0}\in \mathfrak{H}_{%
\mathrm{evo}}$ satisfies $G^{\left( \lambda ,\eta \right) \ast }\Omega
_{0}=0 $ for the adjoint $G^{\left( \lambda ,\eta \right) \ast }$ of $%
G^{\left( \lambda ,\eta \right) }$ (implying that $0$ is also an eigenvalue
of $G^{\left( \lambda ,\eta \right) }$ and thus of $P_{0}^{\left( \lambda
,\eta \right) }G^{\left( \lambda ,\eta \right) }P_{0}^{\left( \lambda ,\eta
\right) }$). Hence, \ the map%
\begin{equation*}
z\mapsto (1-\mathrm{e}^{\alpha z})(z-P_{0}^{\left( \lambda ,\eta \right)
}G^{\left( \lambda ,\eta \right) }P_{0}^{\left( \lambda ,\eta \right) })^{-1}
\end{equation*}%
is holomorphic near $z=0$. Consequently,
\begin{equation*}
\mathrm{\tilde{E}}=\frac{1}{2\pi i}\oint\limits_{\left\vert z\right\vert
=R^{\prime }}(z-P_{0}^{\left( \lambda ,\eta \right) }G^{\left( \lambda ,\eta
\right) }P_{0}^{\left( \lambda ,\eta \right) })^{-1}\mathrm{d}z
\end{equation*}%
is the Kato projection onto the eigenspace associated with the eigenvalue $0$
of $P_{0}^{\left( \lambda ,\eta \right) }G^{\left( \lambda ,\eta \right)
}P_{0}^{\left( \lambda ,\eta \right) }$ and by (\ref{Kato projection lambda}%
), the first inequality of (\ref{correctbisbis}) follows.

Now, using (\ref{gnack vite 20}), the fact that $\tilde{g}(x,\varkappa)$ is
uniformly bounded, and the second resolvent equation at sufficiently small $%
\lambda$, we get the upper bound%
\begin{equation*}
\left\Vert
\left(ix-c\right)-\lambda^{-2}P_{0}^{\left(\lambda,\eta\right)}G^{\left(%
\lambda,\eta\right)}P_{0}^{\left(\lambda,\eta\right)})^{-1}\right\Vert \leq
C\ \tilde{g}(x,\eta/\lambda^{2})
\end{equation*}
for some constant $C\in(0,\infty)$ depending on $\varpi$ but not on $\lambda$%
, $\eta$. Therefore, similarly to (\ref{final1}), we infer from (\ref{idiot
encore0bis}) and properties of the function $\tilde{g}$ that there is a
constant $C\in(0,\infty)$ depending on $\varpi$ but not on $\lambda$, $\eta$
such that
\begin{equation*}
\forall\alpha\geq0:\qquad\Vert\mathrm{e}^{\alpha
P_{0}^{\left(\lambda,\eta\right)}G^{\left(\lambda,\eta\right)}P_{0}^{\left(%
\lambda,\eta\right)}}-\mathrm{\tilde{E}}\Vert\leq C\mathrm{e}%
^{-c\alpha\lambda^{2}}.
\end{equation*}
Combined this with (\ref{final1}) and (\ref{correctbisbis}), one gets the
lemma.\hfill{}$\Box$

We now conclude by the proof of Theorem \ref{lemmalongtime copy(2)}, that
is, we show, for sufficiently small $\lambda$ and any $\varepsilon\in(0,1)$,
that%
\begin{equation}
\left\Vert
\exp\left(\alpha\Lambda^{\left(\lambda,\eta\right)}\right)-\exp\left(\alpha
P_{0}^{\left(\lambda,\eta\right)}G^{\left(\lambda,\eta\right)}P_{0}^{\left(%
\lambda,\eta\right)}\right)\right\Vert \leq C_{\varpi,\varepsilon}\left\vert
\lambda\right\vert ^{2\left(1-\varepsilon\right)},  \label{gnack vite 0}
\end{equation}
where the constant $C_{\varpi,\varepsilon}\in(0,\infty)$ depends on $%
\varpi,\varepsilon$ but not on $\lambda$, $\eta$, and $\alpha$.\bigskip{}

\noindent \textit{Proof of Theorem \ref{lemmalongtime copy(2)}.} Note that
the semigroups $\{\mathrm{e}^{\alpha\Lambda^{\left(\lambda,\eta\right)}}\}_{%
\alpha\geq0}$ and $\{\mathrm{e}^{\alpha
P_{0}^{\left(\lambda,\eta\right)}G^{\left(\lambda,\eta\right)}P_{0}^{\left(%
\lambda,\eta\right)}}\}_{\alpha\geq0}$ are uniformly bounded in $%
\lambda,\eta $ as the first one is a CP semigroup and the second one is the
restriction of $\{\mathcal{T}_{\alpha}\}_{\alpha\geq0}$, see (\ref{ninja
bound semi}). Thus, Duhamel's formula yields the inequality%
\begin{equation}
\Vert\mathrm{e}^{\alpha\Lambda^{\left(\lambda,\eta\right)}}-\mathrm{e}%
^{\alpha
P_{0}^{\left(\lambda,\eta\right)}G^{\left(\lambda,\eta\right)}P_{0}^{\left(%
\lambda,\eta\right)}}\Vert\leq\alpha C\Vert
P_{0}^{\left(\lambda,\eta\right)}G^{\left(\lambda,\eta\right)}P_{0}^{\left(%
\lambda,\eta\right)}-\Lambda^{\left(\lambda,\eta\right)}\Vert
\label{gnack vite 1}
\end{equation}
for some constant $C\in(0,\infty)$ not depending on $\lambda$, $\eta$, $%
\varpi$, and $\alpha$. By (\ref{gnack vite 20}), it follows that
\begin{equation}
\Vert\mathrm{e}^{\alpha\Lambda^{\left(\lambda,\eta\right)}}-\mathrm{e}%
^{\alpha
P_{0}^{\left(\lambda,\eta\right)}G^{\left(\lambda,\eta\right)}P_{0}^{\left(%
\lambda,\eta\right)}}\Vert\leq C_{\varpi}\alpha\lambda^{4}
\label{gnack vite 2}
\end{equation}
with $C_{\varpi}\in(C,\infty)$ depending on $\varpi$ but not on $\lambda$, $%
\eta$, and $\alpha$. We finally infer from (\ref{gnack vite 2}) and Lemma %
\ref{corollary de g(0,eta)} that
\begin{eqnarray}
& & \Vert\mathrm{e}^{\alpha\Lambda^{\left(\lambda,\eta\right)}}-\mathrm{e}%
^{\alpha
P_{0}^{\left(\lambda,\eta\right)}G^{\left(\lambda,\eta\right)}P_{0}^{\left(%
\lambda,\eta\right)}}\Vert  \label{gnack vite 5} \\
& \leq & C_{\varpi}\min\left\{ \alpha\lambda^{4},\lambda^{2}+\mathrm{e}%
^{-\alpha c_{\varpi}\lambda^{2}}\right\}  \notag \\
& \leq & C_{\varpi}\left\vert \lambda\right\vert
^{2\left(1-\varepsilon\right)}(1+\left\vert \lambda\right\vert
^{2\varepsilon}+\left\vert \lambda\right\vert ^{-2\left(1-\varepsilon\right)}%
\mathrm{e}^{-c_{\varpi}\left\vert \lambda\right\vert ^{-2\varepsilon}})
\notag
\end{eqnarray}
for any $\varepsilon\in(0,1)$ and we obtain (\ref{gnack vite 0}) for
sufficiently small $\lambda$.\hfill{}{}$\Box$

\subsection{Completely positive (CP) semigroups\label{section compl positive}%
}

In the theory of open quantum systems, one is usually interested on the
restricted dynamics of some small quantum object interacting with
macroscopic reservoirs. This restricted time--evolution is described by a
map $\mathcal{C}$ on the set of density matrices of the small system. See,
for instance, \cite[Section 1.2.1]{AlickiLendi2007}. Properties of such maps
(cf. \cite[Section 1.2.2]{AlickiLendi2007}) motivate the definition of the
class of \emph{completely positive (CP) operators}:

\begin{definition}[Completely positive (CP) maps]
\mbox{ }\newline
A positive map $\mathcal{C}\in\mathcal{B}\left(\mathcal{B}\left(\mathcal{X}%
\right)\right)$ acting on the set $\mathcal{B}\left(\mathcal{X}\right)$ of
bounded operators on a Hilbert space $\mathcal{X}$ is called completely
positive (CP) if the extended map $\mathcal{C}\otimes\mathbf{1}_{\mathcal{B}%
\left(\mathbb{C}^{n}\right)}$ remains positive for all $n\in\mathbb{N}$. If $%
\mathcal{C}\left(\mathbf{1}_{\mathcal{X}}\right)=\mathbf{1}_{\mathcal{X}}$,
then the operator $\mathcal{C}$ is called a \emph{unital} map.
\end{definition}

\noindent Completely positive (CP)\emph{\ }semigroups are defined as being
the semigroups which are CP maps for all times:

\begin{definition}[Completely positive (CP) semigroups]
\mbox{
}\newline
A semigroup $\{\mathcal{C}_{t}\}_{t\geq 0}\subset \mathcal{B}\left( \mathcal{%
B}\left( \mathcal{X}\right) \right) $, with $\mathcal{X}$ being a Hilbert
space, is CP if the map $\mathcal{C}_{t}$ is CP for all $t\in \mathbb{R}%
_{0}^{+}$. If $\mathcal{C}_{t}$ is unital for any $t\in \mathbb{R}_{0}^{+}$,
then we call $\{\mathcal{C}_{t}\}_{t\geq 0}$ unital.
\end{definition}

From now on and until the end of Section \ref{section compl positive}, $%
\mathcal{X}$ is always a $n$--dimensional Hilbert space. We denote by $%
\mathcal{B}_{2}\left( \mathcal{X}\right) \equiv \mathcal{B}\left( \mathcal{X}%
\right) $ the Hilbert space of Hilbert--Schmidt operators with scalar product%
\begin{equation*}
\left\langle A,B\right\rangle _{\mathcal{B}_{2}\left( \mathcal{X}\right) }:=%
\mathrm{Tr}_{\mathcal{X}}(A^{\ast }B),\qquad A,B\in \mathcal{B}_{2}\left(
\mathcal{X}\right) .
\end{equation*}%
In the special case where a semigroup $\{\mathcal{C}_{t}\}_{t\geq 0}\subset
\mathcal{B}\left( \mathcal{B}_{2}\left( \mathcal{X}\right) \right) $ acts on
$\mathcal{B}_{2}\left( \mathcal{X}\right) $, we can define its (unique)
\emph{adjoint semigroup} $\{\mathcal{C}_{t}^{\ast }\}_{t\geq 0}\subset
\mathcal{B}\left( \mathcal{B}_{2}\left( \mathcal{X}\right) \right) $ as
usual via the equations
\begin{equation*}
\forall t\geq 0:\quad \langle \mathcal{C}_{t}^{\ast }\left( A\right)
,B\rangle _{\mathcal{B}_{2}\left( \mathcal{X}\right) }=\langle A,\mathcal{C}%
_{t}\left( B\right) \rangle _{\mathcal{B}_{2}\left( \mathcal{X}\right)
},\qquad A,B\in \mathcal{B}_{2}\left( \mathcal{X}\right) .
\end{equation*}%
Note that if the CP semigroup $\{\mathcal{C}_{t}\}_{t\geq 0}$ is unital then
$\{\mathcal{C}_{t}^{\ast }\}_{t\geq 0}$ is CP and \emph{preserves the trace}%
. A CP semigroup $\{\mathcal{C}_{t}\}_{t\geq 0}$ is called \emph{Markov }CP
semigroup if it preserves the trace. Generators of \emph{Markov }CP
semigroups $\{\mathcal{C}_{t}\}_{t\geq 0}$ and of their adjoint groups $\{%
\mathcal{C}_{t}^{\ast }\}_{t\geq 0}$ can be characterized in the finite
dimensional case as follows (cf. \cite[Theorem 2.2]%
{GoriniKossakowskiSudarshan1976} and \cite[Theorem 2]{Lindblad 1976}):

\begin{satz}[Generators of \ finite dimensional Markov CP semigroups -- I]
\label{Theorem generator CP}\mbox{ }\newline
Let $\dim \mathcal{X}=n\in \mathbb{N}$. The operator $\mathrm{L}\in \mathcal{%
B}\left( \mathcal{B}_{2}\left( \mathcal{X}\right) \right) $ is the generator
of a Markov CP semigroup $\{\mathcal{C}_{t}\}_{t\geq 0}$ if and only if
\begin{equation*}
\mathrm{L}\left( \rho \right) =-i\left[ \mathrm{h},\rho \right] +\frac{1}{2}%
\sum\limits_{j}\left\{ \left[ \mathrm{V}_{j},\rho \mathrm{V}_{j}^{\ast }%
\right] +\left[ \mathrm{V}_{j}\rho ,\mathrm{V}_{j}^{\ast }\right] \right\}
,\qquad A\in \mathcal{B}_{2}\left( \mathcal{X}\right) ,
\end{equation*}%
where $\mathrm{h}=\mathrm{h}^{\ast }\in \mathcal{B}_{2}\left( \mathcal{X}%
\right) $ and $\left\{ \mathrm{V}_{j}\right\} \subset \mathcal{B}\left(
\mathcal{X}\right) $ is a family of operators with $\sum_{j}\mathrm{V}%
_{j}^{\ast }\mathrm{V}_{j}\in \mathcal{B}\left( \mathcal{X}\right) $.
Additionally, the adjoint semigroup $\{\mathcal{C}_{t}^{\ast }\}_{t\geq 0}$
is in this case the unital CP semigroup with generator
\begin{equation*}
\mathrm{L}^{\ast }\left( A\right) =i\left[ \mathrm{h},A\right] +\frac{1}{2}%
\sum\limits_{j}\left\{ \left[ \mathrm{V}_{j}^{\ast },A\right] \mathrm{V}_{j}+%
\mathrm{V}_{j}^{\ast }\left[ A,\mathrm{V}_{j}\right] \right\} ,\qquad A\in
\mathcal{B}_{2}\left( \mathcal{X}\right) .
\end{equation*}
\end{satz}

\noindent Generators of CP semigroups are also called \emph{Lindbladian} or
\emph{Lindblad (--Kossakowski)} generators.

A more compact characterization of generators of Markov CP semigroups is
given by \cite[Sect. 4.3]{DerezinskiFruboes2006}:

\begin{satz}[Generators of \ finite dimensional Markov CP semigroups -- II]
\label{abstract CP generator}\mbox{ }\newline
Let $\dim \mathcal{X}=n\in \mathbb{N}$. The operator $\mathrm{L}\in \mathcal{%
B}\left( \mathcal{B}_{2}\left( \mathcal{X}\right) \right) $ is the generator
of a CP semigroup $\{\mathcal{C}_{t}\}_{t\geq 0}$ if and only if there is a
completely positive map $\Xi \in \mathcal{B}\left( \mathcal{B}\left(
\mathcal{X}\right) \right) $ and an operator $\Delta \in \mathcal{B}\left(
\mathcal{X}\right) $ such that%
\begin{equation*}
\mathrm{L}=\underrightarrow{\Delta }+\underleftarrow{\Delta ^{\ast }}+\Xi \ .
\end{equation*}%
Such a CP semigroup is Markov if and only if $\mathrm{L}^{\ast }\left(
\mathbf{1}_{\mathcal{X}}\right) =0$.
\end{satz}

The \emph{relaxing} property of Markov CP semigroups $\left\{ \mathcal{C}%
_{t}\right\} _{t\geq 0}\subset \mathcal{B}\left( \mathcal{B}_{2}\left(
\mathcal{X}\right) \right) $, which is crucial for our analysis, is defined
as follows:

\begin{definition}[Relaxing semigroups]
\label{relaxing}\mbox{
}\newline
A Markov CP semigroup $\{\mathcal{C}_{t}\}_{t\geq 0}\subset \mathcal{B}%
\left( \mathcal{B}_{2}\left( \mathcal{X}\right) \right) $ is called relaxing
if there is a unique trace--one positive $\rho _{\infty }\in \mathcal{B}%
_{2}\left( \mathcal{X}\right) $, i.e., a density matrix $\rho _{\infty }$,
such that, for any density matrix $\rho \in \mathcal{B}_{2}\left( \mathcal{X}%
\right) $,
\begin{equation*}
\lim_{t\rightarrow \infty }\mathcal{C}_{t}\left( \rho \right) =\rho _{\infty
}\ .
\end{equation*}
\end{definition}

\noindent In other words, a relaxing, Markov CP semigroup has a unique
invariant equilibrium state. Moreover, this state can be approximated by the
density matrix $\mathcal{C}_{t}\left( \rho \right) $ for large times and any
initial state with density matrix $\rho $. Spohn \cite[Theorem 2]{Spohn1977}
gave in 1977 a characterization of relaxing semigroups which turns out to be
very useful in our context:

\begin{satz}[Condition for a Markov CP semigroup to be relaxing]
\label{Spohn}\mbox{
}\newline
Let $\dim\mathcal{X}=n\in\mathbb{N}$. Let $\{\mathcal{C}_{t}\}_{t\geq0}%
\subset\mathcal{B}\left(\mathcal{B}_{2}\left(\mathcal{X}\right)\right)$ be a
Markov CP and $C_{0}$ semigroup with generator $\mathrm{L}$ given by Theorem %
\ref{Theorem generator CP}. If the space spanned by the family $\left\{
\mathrm{V}_{j}\right\} $ satisfies%
\begin{equation*}
\mathrm{span}\left\{ \mathrm{V}_{j}\right\} =\mathrm{span}\left\{ \mathrm{V}%
_{j}^{\ast}\right\} \subset\mathcal{B}_{2}\left(\mathcal{X}\right)
\end{equation*}
and the bicommutant%
\begin{equation}
\left\{ \mathrm{V}_{j}\right\} ^{\prime\prime}=\mathcal{B}\left(\mathcal{X}%
\right)\ ,  \label{bicommutant}
\end{equation}
then $\{\mathcal{C}_{t}\}_{t\geq0}$ is relaxing. In particular, $0$ is a
non--degenerated eigenvalue of $\mathrm{L}$ and%
\begin{equation*}
\max\left\{ \mathrm{\mathop{\rm Re}}\left\{ w\right\} \,|\, w\in\sigma\left(%
\mathrm{L}\right)\backslash\{0\}\right\} <0.
\end{equation*}
\end{satz}

\noindent As explained after Assumption \ref{assumption3}, the condition (%
\ref{bicommutant}) is a non--commutative version of the irreducibility of
classical Markov chains.

\subsection{The operator $\mathfrak{B}_{\mathrm{p},\mathcal{R}}$ and the
conservation of positivity\label{contra exemplo}}

\noindent \textbf{1.} We first give an elementary example of an operator $%
\mathfrak{B}_{\mathrm{p},\mathcal{R}}$ defined by (\ref{balanced pump
operator0}) which does \emph{not} preserve positivity. Let $d=3$, $N=2$, $%
\varpi =1$ and%
\begin{equation*}
H_{\mathrm{at}}:=\left(
\begin{array}{lll}
0 & 0 & 0 \\
0 & 1 & 0 \\
0 & 0 & 1%
\end{array}%
\right) ,\;Q_{1}:=\left(
\begin{array}{lll}
0 & 1 & 0 \\
1 & 0 & 0 \\
0 & 0 & 0%
\end{array}%
\right) ,\;Q_{2}:=\left(
\begin{array}{lll}
0 & 0 & 1 \\
0 & 0 & 0 \\
1 & 0 & 0%
\end{array}%
\right) \
\end{equation*}%
in the canonical orthonormal basis of $\mathbb{C}^{3}$. Choose the coupling
functions $\mathrm{f}_{\ell }$ such that $\mathrm{g}_{\ell }(\epsilon )>0$
for $\ell \in \left\{ 1,2\right\} $ and $\epsilon \in \left\{ -1,0,1\right\}
$, see (\ref{function gl}). Then, the family $\{V_{j,k}^{(\ell
)}\}_{j,k,\ell }\subset \mathcal{B}(\mathbb{C}^{3})$ of operators defined by
(\ref{definition V1}) for $\ell \in \left\{ 1,2\right\} $ and $j,k\in
\left\{ 1,N\right\} $ satisfy Assumption \ref{assumption3}. With this
choice, the operators $%
W_{(1,1)}^{(N,1)},W_{(1,1)}^{(N,2)},W_{(N,1)}^{(1,1)},W_{(N,2)}^{(1,1)}\in
\mathfrak{H}_{\mathrm{at}}$ defined by (\ref{W1}) are eigenvectors of $%
\mathcal{L}_{\mathcal{R}}$ with eigenvalues $\gamma _{1},\gamma _{2},\bar{%
\gamma}_{1},\bar{\gamma}_{2}\in \mathbb{C}\backslash \{0\}$, respectively.
By setting%
\begin{equation*}
H_{\mathrm{p}}:=\left(
\begin{array}{lll}
0 & 1 & 1 \\
1 & 0 & 0 \\
1 & 0 & 0%
\end{array}%
\right) ,\
\end{equation*}%
we infer from (\ref{effective stimulated transition rate}) that
\begin{equation*}
\mathbf{B}_{N,1}\left(
\begin{array}{lll}
1 & 0 & 0 \\
0 & 0 & 0 \\
0 & 0 & 0%
\end{array}%
\right) =-\left(
\begin{array}{lll}
0 & 0 & 0 \\
0 & 2\mathop{\rm Re}\{\gamma _{1}^{-1}\} & \gamma _{1}^{-1}+\overline{\gamma
_{2}}^{-1} \\
0 & \overline{\gamma _{1}}^{-1}+\gamma _{2}^{-1} & 2\mathop{\rm Re}\{\gamma
_{2}^{-1}\}%
\end{array}%
\right)
\end{equation*}%
in the canonical orthonormal basis of $\mathbb{C}^{3}$. Indeed, note that%
\begin{equation*}
h_{\mathrm{p}}\left(
\begin{array}{lll}
1 & 0 & 0 \\
0 & 0 & 0 \\
0 & 0 & 0%
\end{array}%
\right) =W_{(1,1)}^{(N,1)}+W_{(1,1)}^{(N,2)},\text{ \ }\left(
\begin{array}{lll}
1 & 0 & 0 \\
0 & 0 & 0 \\
0 & 0 & 0%
\end{array}%
\right) h_{\mathrm{p}}^{\ast }=W_{(N,1)}^{(1,1)}+W_{(N,2)}^{(1,1)}
\end{equation*}%
and%
\begin{equation*}
W_{(1,1)}^{(N,j)}h_{\mathrm{p}}^{\ast }=W_{(N,1)}^{(N,j)}+W_{(N,2)}^{(N,j)},%
\text{ \ }h_{\mathrm{p}%
}W_{(N,j)}^{(1,1)}=W_{(N,j)}^{(N,1)}+W_{(N,j)}^{(N,1)},\text{ \ }j\in
\{1,2\}.
\end{equation*}%
Thus,%
\begin{eqnarray*}
\mathbf{B}_{N,1}\left(
\begin{array}{lll}
1 & 0 & 0 \\
0 & 0 & 0 \\
0 & 0 & 0%
\end{array}%
\right) &=&-\left( \gamma _{1}^{-1}W_{(1,1)}^{(N,1)}+\gamma
_{2}^{-1}W_{(1,1)}^{(N,2)}\right) h_{\mathrm{p}}^{\ast }-h_{\mathrm{p}%
}\left( \bar{\gamma}_{1}^{-1}W_{(N,1)}^{(1,1)}+\bar{\gamma}%
_{2}^{-1}W_{(N,2)}^{(1,1)}\right) \\
&=&-\gamma _{1}^{-1}(W_{(N,1)}^{(N,1)}+W_{(N,2)}^{(N,1)})-\gamma
_{2}^{-1}(W_{(N,1)}^{(N,2)}+W_{(N,2)}^{(N,2)}) \\
&&-\bar{\gamma}_{1}^{-1}(W_{(N,1)}^{(N,1)}+W_{(N,1)}^{(N,2)})-\bar{\gamma}%
_{2}^{-1}(W_{(N,2)}^{(N,1)}+W_{(N,2)}^{(N,2)}).
\end{eqnarray*}

Let
\begin{equation*}
D:=\det \left(
\begin{array}{ll}
2\mathop{\rm Re}\{\gamma _{1}^{-1}\} & \gamma _{1}^{-1}+\overline{\gamma _{2}%
}^{-1} \\
\overline{\gamma _{1}}^{-1}+\gamma _{2}^{-1} & 2\mathop{\rm Re}\{\gamma
_{2}^{-1}\}%
\end{array}%
\right) \in \mathbb{R}
\end{equation*}%
be the determinant of the lower $2\times 2$ block diagonal part of the above
self--adjoint matrix. One can clearly find coupling functions $\mathrm{f}%
_{\ell }$ for $\ell \in \left\{ 1,2\right\} $ such that $|\mathrm{g}%
_{1}(1)|^{2}=$ $|\mathrm{g}_{2}(1)|^{2}$, but%
\begin{equation*}
\mathcal{PP}(\mathrm{g}_{1})(1)\neq \mathcal{PP}(\mathrm{g}_{2})(1)\ .
\end{equation*}%
This implies that
\begin{equation*}
\mathop{\rm Re}\{\gamma _{1}^{-1}\}=\mathop{\rm Re}\{\gamma _{2}^{-1}\}\quad
\mathrm{and}\quad \mathop{\rm Im}\{\gamma _{1}^{-1}\}\neq \mathop{\rm Im}%
\{\gamma _{2}^{-1}\}\ .
\end{equation*}%
However, in this case
\begin{equation*}
D=-\left( \mathop{\rm Im}\{\gamma _{1}^{-1}\}-\mathop{\rm Im}\{\gamma
_{2}^{-1}\}\right) ^{2}<0
\end{equation*}%
showing that $\mathbf{B}_{N,1}$ is not a map from $\mathcal{B}^{+}(\mathcal{H%
}_{1})$ to $\mathcal{B}^{+}(\mathcal{H}_{N})$ and hence $\mathfrak{B}_{%
\mathrm{p},\mathcal{R}}$ is generally not the generator of a semigroup on $%
\mathfrak{D}$ which preserves positivity.\newline

\noindent \textbf{2.} Note that under certain conditions the averaged
stimulated transition rates%
\begin{equation*}
\mathfrak{\bar{B}}_{\mathrm{p},\mathcal{R}}:=\int_{0}^{\infty }\mathrm{e}^{-s%
\mathfrak{L}_{\mathcal{R}}}\mathfrak{L}_{\mathrm{p}}\mathrm{e}^{s\mathfrak{L}%
_{\mathcal{R}}}\mathfrak{L}_{\mathrm{p}}P_{\mathfrak{D}}\ \mathrm{d}%
s=\int_{0}^{\infty }\mathrm{e}^{-s\mathfrak{L}_{\mathcal{R}}}\mathfrak{L}_{%
\mathrm{p}}P_{\mathfrak{D}}^{\perp }\mathrm{e}^{s\mathfrak{L}_{\mathcal{R}}}%
\mathfrak{L}_{\mathrm{p}}P_{\mathfrak{D}}\,ds\in \mathcal{B}(\mathfrak{H}_{%
\mathrm{at}}),
\end{equation*}%
are well--defined. This is the case, for instance, at \emph{strong
decoherence}, i.e., if%
\begin{equation*}
\min (\mathop{\rm Re}\{\sigma (\mathfrak{L}_{\mathcal{R}}P_{\mathfrak{D}%
})\})>\max (\mathop{\rm Re}\{\sigma (\mathfrak{L}_{\mathcal{R}}P_{\mathfrak{D%
}}^{\perp })\})
\end{equation*}%
as in this situation%
\begin{equation*}
\left\Vert \mathrm{e}^{-s\mathfrak{L}_{\mathcal{R}}}\mathfrak{L}_{\mathrm{p}%
}P_{\mathfrak{D}}^{\perp }\right\Vert \left\Vert \mathrm{e}^{s\mathfrak{L}_{%
\mathcal{R}}}\mathfrak{L}_{\mathrm{p}}P_{\mathfrak{D}}\right\Vert \leq C%
\mathrm{e}^{-cs}
\end{equation*}%
for constants $C,c\in (0,\infty )$ and all $t\geq 0$. Observe that the norm $%
\left\Vert \mathrm{e}^{-t\mathfrak{L}_{\mathcal{R}}}\right\Vert $
exponentially grows as $t\rightarrow \infty $, generally. In this case we
obtain, moreover, that%
\begin{equation*}
\mathrm{e}^{t\mathfrak{\bar{B}}_{\mathrm{p},\mathcal{R}}}P_{\mathfrak{D}%
}=\lim_{\varkappa \searrow 0}P_{\mathfrak{D}}\exp (\varkappa ^{-2}t\mathfrak{%
L}_{\mathcal{R}}+\varkappa ^{-1}t\mathfrak{L}_{\mathrm{p}})P_{\mathfrak{D}}\
.
\end{equation*}%
In particular, $\mathfrak{\bar{B}}_{\mathrm{p},\mathcal{R}}$ is the
generator of a positivity preserving semi--group on $\mathfrak{D}$.

Therefore we could wonder whether or not $\mathfrak{B}_{\mathrm{p},\mathcal{R%
}}$ also generates a positive semigroup if the system shows strong
decoherence. Unfortunately, this turns out to be not the case. This can be
seen in a simple variation of the example above: By introducing a third
atom--reservoir interaction term in the model above with coupling satisfying
$|\mathrm{g}_{3}(0)|^{2}>0$ and
\begin{equation*}
Q_{3}:=\left(
\begin{array}{lll}
q_{3} & 0 & 0 \\
0 & 0 & 0 \\
0 & 0 & 0%
\end{array}%
\right) ,\text{ \ \ \ }q_{3}\in \mathbb{R},
\end{equation*}%
the diagonal part $\mathfrak{L}_{\mathcal{R}}P_{\mathfrak{D}}$ is not
changed whereas $\mathfrak{L}_{\mathcal{R}}P_{\mathfrak{D}}^{\perp }$ is
shifted by $-q_{3}^{2}|\mathrm{g}_{3}(0)|^{2}P_{\mathfrak{D}}^{\perp }$.
Hence the strong decoherence regime can be always attained for sufficiently
large $q_{3}$. As the eigenvalues $\gamma _{1},\gamma _{2},\bar{\gamma}_{1},%
\bar{\gamma}_{2}$ of $%
W_{(1,1)}^{(N,1)},W_{(1,1)}^{(N,2)},W_{(N,1)}^{(1,1)},W_{(N,2)}^{(1,1)}$ are
\ simultaneously shifted by $-q_{3}^{2}|\mathrm{g}_{3}(0)|^{2}$, $D$ remains
negative for all $q_{3}\in \mathbb{R}$.

\noindent \textbf{3.} Nevertheless, Assumption \ref{assumption Markov CP B
effect} can be verified for a large class of models. For instance, this
condition is always satisfied if the $1$st and the $N$th energy levels of
the atomic part of the model are non--degenerated. This is a special case of
the following theorem:

\begin{satz}[$\mathfrak{B}_{\mathrm{p},\mathcal{R}}$ as transition rates]
\label{suff cond CP B}\mbox{ }\newline
Assume that $P_{N,1}\mathfrak{L}_{\mathcal{R}}P_{N,1}=\xi_{N,1}P_{N,1}\ $for
some $\xi_{N,1}\in\mathbb{C}$. Then Assumption \ref{assumption Markov CP B
effect} holds.
\end{satz}

\noindent \textit{Proof}. Using
\begin{equation*}
V_{j,k}^{(\ell)}\rho_{\mathfrak{g}}=e^{-\beta\left(E_{j}-E_{k}\right)}\rho_{%
\mathfrak{g}}V_{j,k}^{(\ell)}
\end{equation*}
for every $j,k\in\{1,2,\ldots,N\}$ and any $\ell\in\{1,2,\ldots,m\}$ (cf. (%
\ref{Gibbs.init}) and (\ref{definition V1})) together with (\ref{L R}), (\ref%
{Atomic Lamb shift}) and (\ref{Effective atomic dissipationbis}), we observe
that%
\begin{equation*}
\bar{\xi}_{N,1}P_{1,N}=\left(P_{N,1}\mathfrak{L}_{\mathcal{R}%
}P_{N,1}\right)^{\ast\rho_{\mathfrak{g}}}=P_{1,N}\mathfrak{L}_{\mathcal{R}%
}P_{1,N}=:\xi_{1,N}P_{1,N}\ .
\end{equation*}
Here, $\left(\cdot\right)^{\ast\rho_{\mathfrak{g}}}$ denotes the adjoint
with respect to the scalar product induced by the Gibbs state $\rho_{%
\mathfrak{g}}$ (\ref{Gibbs.init}), that is,%
\begin{equation*}
\langle A,B\rangle_{\mathrm{at},\mathfrak{g}}:=\mathrm{Tr}_{\mathbb{C}%
^{d}}\left(A^{\ast}B\ \rho_{\mathfrak{g}}\right),\qquad A,B\in\mathcal{B}(%
\mathbb{C}^{d})\ .
\end{equation*}
Indeed, $\left(P_{N,1}\mathfrak{L}_{\mathcal{R}}P_{N,1}\right)^{\ast\rho_{%
\mathfrak{g}}}=\mathfrak{L}_{\mathcal{R}}$ corresponds to the fact that the
generator $\mathfrak{L}_{\mathcal{R}}$ satisfies a \emph{quantum detailed
balance condition} with respect to the density matrix $\rho_{\mathfrak{g}}$
of the Gibbs state $\mathfrak{g}_{\mathrm{at}}$. The inverse temperature $%
\beta$ in (\ref{Gibbs.init}) is the one of the reservoir, which is
determined by the choice of the KMS state $\omega_{\mathcal{R}}$.

Hence, $\xi :=\xi _{N,1}=\bar{\xi}_{1,N}$ and
\begin{equation*}
P_{N,1}\left( \mathfrak{C}|_{\mathrm{ran}(P_{\mathfrak{D}}^{\bot })}\right)
^{-1}P_{N,1}=\xi ^{-1}P_{N,1},\quad P_{1,N}\left( \mathfrak{C}|_{\mathrm{ran}%
(P_{\mathfrak{D}}^{\bot })}\right) ^{-1}P_{1,N}=\bar{\xi}^{-1}P_{1,N}\ .
\end{equation*}%
Thus, by (\ref{effective stimulated transition rate}),
\begin{equation*}
\mathbf{B}_{N,1}=-\xi ^{-1}P_{N,N}\underleftarrow{h_{\mathrm{p}}^{\ast }}%
P_{N,1}\underrightarrow{h_{\mathrm{p}}}P_{1,1}-\bar{\xi}^{-1}P_{N,N}%
\underrightarrow{h_{\mathrm{p}}}P_{1,N}\underleftarrow{h_{\mathrm{p}}^{\ast }%
}P_{1,1}\ .
\end{equation*}%
Using the properties of $h_{\mathrm{p}}$ and $\underleftarrow{A}%
\underrightarrow{B}=\underrightarrow{B}\underleftarrow{A}$,
\begin{equation*}
\mathbf{B}_{N,1}=-\frac{2\mathrm{Re}\left( \xi \right) }{\left\vert \xi
\right\vert ^{2}}\underrightarrow{h_{\mathrm{p}}}\ \underleftarrow{h_{%
\mathrm{p}}^{\ast }}\ .
\end{equation*}%
Similarly, one also finds%
\begin{equation*}
\mathbf{B}_{1,N}=-\frac{2\mathrm{Re}\left( \xi \right) }{\left\vert \xi
\right\vert ^{2}}\underrightarrow{h_{\mathrm{p}}^{\ast }}\ \underleftarrow{%
h_{\mathrm{p}}}\ .
\end{equation*}

On the other hand, we have
\begin{eqnarray*}
\mathbf{B}_{N,N} &=&\bar{\xi}^{-1}\underrightarrow{h_{\mathrm{p}}}%
\underrightarrow{h_{\mathrm{p}}^{\ast }}P_{N,N}+\xi ^{-1}P_{N,N}%
\underleftarrow{h_{\mathrm{p}}}\underleftarrow{h_{\mathrm{p}}^{\ast }}\ , \\
\mathbf{B}_{1,1} &=&\xi ^{-1}\underrightarrow{h_{\mathrm{p}}^{\ast }}%
\underrightarrow{h_{\mathrm{p}}}P_{1,1}+\bar{\xi}^{-1}P_{1,1}\underleftarrow{%
h_{\mathrm{p}}^{\ast }}\underleftarrow{h_{\mathrm{p}}}\ .
\end{eqnarray*}%
Therefore, by (\ref{operatorB}), we can split the operator $\mathfrak{B}_{%
\mathrm{p},\mathcal{R}}$ on the subspace $\mathcal{D}$ as follows:
\begin{equation}
\mathfrak{B}_{\mathrm{p},\mathcal{R}}=\underrightarrow{\Delta }+%
\underleftarrow{\Delta ^{\ast }}+\Xi  \label{operatorBbis}
\end{equation}%
with%
\begin{equation*}
\Delta :=\frac{i\mathrm{Im}\left( \xi \right) }{\left\vert \xi \right\vert
^{2}}\left( h_{\mathrm{p}}h_{\mathrm{p}}^{\ast }+h_{\mathrm{p}}^{\ast }h_{%
\mathrm{p}}\right) +\frac{\mathrm{Re}\left( \xi \right) }{\left\vert \xi
\right\vert ^{2}}\left( h_{\mathrm{p}}h_{\mathrm{p}}^{\ast }+h_{\mathrm{p}%
}^{\ast }h_{\mathrm{p}}\right) =\bar{\xi}^{-1}\left( h_{\mathrm{p}}h_{%
\mathrm{p}}^{\ast }+h_{\mathrm{p}}^{\ast }h_{\mathrm{p}}\right) \in \mathcal{%
B}(\mathbb{C}^{d})
\end{equation*}%
and%
\begin{equation*}
\Xi :=\mathbf{B}_{1,N}+\mathbf{B}_{N,1}\in \mathcal{B}(\mathcal{H}_{\mathrm{%
at}})\ .
\end{equation*}%
Since $\mathfrak{L}_{\mathcal{R}}$ is the generator of a relaxing Markov CP
semigroup on $\mathfrak{H}_{\mathrm{at}}$ (Remark \ref{markov CP semigroup}%
), $\mathrm{Re}(\xi )\leq 0$ and $\Xi $ is thus a completely positive map.
Clearly,%
\begin{equation*}
\left( \underrightarrow{\Delta }+\underleftarrow{\Delta ^{\ast }}+\Xi
\right) ^{\ast }(\mathbf{1})=\left( \underrightarrow{\Delta ^{\ast }}+%
\underleftarrow{\Delta }+\Xi \right) (\mathbf{1})=0\ .
\end{equation*}%
Therefore, we infer from (\ref{operatorBbis}) and Theorem \ref{abstract CP
generator} that $\mathfrak{B}_{\mathrm{p},\mathcal{R}}$ generates a Markov
CP semigroup on $\mathfrak{H}_{\mathrm{at}}$. As $\mathcal{D}$ is an
invariant subspace of this generator, $\mathfrak{B}_{\mathrm{p},\mathcal{R}}$
generates a Markov semigroup on $\mathcal{D}$ which preserves positivity.
\hfill {}$\Box $

\noindent \textit{Acknowledgments:} This work has been supported by the
grant MTM2010-16843 of the Spanish {}\textquotedblleft Ministerio de Ciencia
e Innovaci{\'{o}}n\textquotedblright {}\ and a grant of the
{}\textquotedblleft Inneruniversit{\"{a}}re Forschungsf{\"{o}rderung}%
\textquotedblright {}\ of the Johannes Gutenberg University in Mainz. We
thank Volker Bach for his support and the referees for pointing out a
mistake in the proof of Lemma \ref{lemmalongtime}.

\end{document}